\documentclass[10pt, a4paper]{article}

\usepackage[english]{babel}  
\usepackage[utf8]{inputenc}     
\usepackage[T1]{fontenc}    
\usepackage[adobe-utopia]{mathdesign}
\usepackage{frcursive}
\usepackage{amsmath} % \dfrac{}
\usepackage{mathrsfs} % \mathscr{}
\DeclareMathAlphabet{\mathdutchcal}{U}{dutchcal}{m}{n}
\SetMathAlphabet{\mathdutchcal}{bold}{U}{dutchcal}{b}{n}
\DeclareMathAlphabet{\mathdutchbcal}{U}{dutchcal}{b}{n}
\usepackage{graphicx}  
\usepackage[labelformat=simple]{subcaption} % \begin{subfigures}
\renewcommand\thesubfigure{(\alph{subfigure})}  % to have Figure 1(a) with \ref{subfig:blabla}
\usepackage{float} % to force the figure being at the exact position in the text  \begin{figure}[H]
\usepackage{tabularx}
\usepackage{multirow}  % to join rows in a table
\usepackage[dvipsnames]{xcolor}  % \textcolor{}     
\usepackage[hidelinks]{hyperref}                 
\usepackage[left=3cm,right=3cm,top=2cm,bottom=2cm]{geometry}   
\usepackage{pdflscape}
\usepackage{tikz} 
\usetikzlibrary{arrows}
\usetikzlibrary{arrows.meta} 
\usepackage[numbers, sort&compress]{natbib}
\usepackage{authblk}
\usepackage{listings}
\usepackage{algorithm}
\usepackage{algpseudocode}
\newtheorem{remark}{Remark}
\usepackage{ulem}
\newcommand{\n}[1]{n_{_\mathrm{#1}}}

\newcommand{\fint}[1]{\mathdutchcal{f}_\mathrm{#1}}
\newcommand{\g}[1]{g_{_\mathrm{#1}}}
\newcommand{\N}[1]{N_{_\mathrm{#1}}}
\newcommand{\lDf}{\lambda_{D, \ \text{co}}^{\text{front}}}
\newcommand{\lDb}{\lambda_{D, \ \text{co}}^{\text{back}}}
\newcommand{\lWb}{\lambda_{W, \ \text{co}}^{\text{back}}}
\newcommand{\lDbPlus}{\lambda_{D, \ \text{co} \ +}^{\text{back}}}
\newcommand{\lWbPlus}{\lambda_{W, \ \text{co} \ +}^{\text{back}}}
\newcommand{\lDbMinus}{\lambda_{D, \ \text{co} \ -}^{\text{back}}}
\newcommand{\lWbMinus}{\lambda_{W, \ \text{co} \ -}^{\text{back}}}
\newcommand{\lDbprim}{\frac{d\lambda_{D, \ \text{co}}^{\text{back}}}{ds}}
\newcommand{\lWbprim}{\frac{d\lambda_{W, \ \text{co}}^{\text{back}}}{ds}}
\newcommand{\lDfd}{\lambda_{D, \ \text{di}}^{\text{front}}}
\newcommand{\lDbd}{\lambda_{D, \ \text{di}}^{\text{back}}}
\newcommand{\lWbd}{\lambda_{W, \ \text{di}}^{\text{back}}}
\newcommand{\vc}{v_{\text{co}}}

\newcommand{\vd}{v_{\text{di}}}

\newcommand{\vdc}{v_{\text{di (cor)}}}
\newcommand{\vcwt}{v_{\text{co}}^{_\mathrm{\text{WT}}}}

\newcolumntype{M}[1]{>{\centering\arraybackslash}m{#1}}
\newcolumntype{C}[1]{>{\centering\arraybackslash}p{#1}}
\newcolumntype{L}[1]{>{\raggedright\let\newline\\\arraybackslash\hspace{0pt}}p{#1}}

\newcommand{\lk}[1]{{\textcolor{black}{#1}}}

\newcommand{\VC}[1]{{\textcolor{black}{#1}}}

\definecolor{backcolour}{rgb}{0.95,0.95,0.95}
\lstdefinestyle{mystyle}{
    backgroundcolor=\color{backcolour},   
    commentstyle=\color{Periwinkle},
    basicstyle=\ttfamily\footnotesize,
    breakatwhitespace=false,         
    breaklines=true,                 
    captionpos=b,                    
    keepspaces=true,                 
    numbers=left,                    
    numbersep=5pt,                  
    showspaces=false,                
    showstringspaces=false,
    showtabs=false,                  
    tabsize=2}
\title{Stochastic dynamics at the back of a gene drive eradication wave}
\author[,1]{Léna Kläy\thanks{Corresponding author: \texttt{lena.klay@mnhn.fr}}}
\author[2]{Léo Girardin}
\author[1]{Florence Débarre\thanks{Corresponding author: \texttt{florence.debarre@normalesup.org}}}
\author[3]{Vincent Calvez\thanks{Corresponding author: \texttt{vincent.calvez@math.cnrs.fr}}}
\affil[1]{\small{Institute of Ecology and Environmental Sciences Paris (IEES Paris), Sorbonne Université, CNRS, IRD, INRAE, Université Paris Est Creteil, Université de Paris, Paris Cedex 5, France.}}
\affil[2]{\small{Institut Camille Jordan, UMR 5208 CNRS and Universite Claude Bernard Lyon 1, France}}
\affil[3]{\small{CNRS, Univ Brest, UMR 6205, Laboratoire de Math\'ematiques de Bretagne Atlantique, France}}
\date{}

\begin{document}
\maketitle
%\linenumbers % To number lines

\normalsize

\section*{Abstract}

Gene drive alleles bias their own inheritance to offspring. They can fix in a wild-type population in spite of a fitness cost, and even lead to the eradication of the target population if the fitness cost is high. However, this outcome may be prevented or delayed if areas previously cleared by the drive are recolonised by wild-type individuals. Here, we investigate the conditions under which these stochastic wild-type recolonisation events are likely and when they are unlikely to occur in one spatial dimension. More precisely, we examine the conditions ensuring that the last individual carrying a wild-type allele is surrounded by a large enough number of drive homozygous individuals, resulting in a very low chance of wild-type recolonisation. To do so, we make a deterministic approximation of the distribution of drive alleles within the wave, and we split the distribution of wild-type alleles into a deterministic part and a stochastic part. Our analytical and numerical results suggest that the probability of wild-type recolonisation events increases with lower fitness of drive individuals and with smaller local carrying capacity. Numerical simulations show that these results extend to two spatial dimensions. The role of the migration rate however, is less clear but has a lower impact. We further demonstrate that, in the event of wild-type recolonization, the probability of subsequent drive reinvasion decreases with smaller values of the intrinsic growth rate of the population. Overall, our study paves the way for further analysis of wild-type recolonisation at the back of eradication traveling waves.

\vspace{0.5cm}

\noindent Keywords: gene drive, chasing, population dynamics, stochastic simulation, spatial modeling.

\newpage 

\clearpage 

\section{Introduction}

Artificial gene drive is a genetic engineering technology that could be used for the management of natural populations \cite{rode2019, burt2003, segelbacher2022}. Gene drive alleles bias their own inheritance towards a super-Mendelian rate, therefore driving themselves to spread quickly through a population despite a potential fitness cost \cite{alphey2020, burt2003, burt2018}. Homing gene drives rely on gene conversion to bias their transmission. In a heterozygous cell, the gene drive cassette located on one chromosome induces a double-strand break on the homologous chromosome. This damage is repaired through homology directed repair, which duplicates the cassette. By repeating through generations, such gene conversion favours the spread of the drive cassette in the target population. Gene conversion can theoretically take place at any point in the life cycle, either in the germline or in the zygote. Current implementations however focus on gene conversion in the germline \cite{bier2022}. 

Gene drive constructs can be designed to either spread a gene of interest in a population (an outcome called \textit{population replacement}), or to reduce the size of the target population by lowering the fitness of drive individuals (called \textit{population suppression} if the intended goal is to reduce population density, and \textit{population eradication} if the intended goal is to eradicate the population). The fitness cost carried by the drive, usually caused by the alteration of an essential fertility or viability gene, associated with the super-Mendelian propagation, can lead to the complete extinction of the target population \cite{kyrou2018, hammond2021a, burt2003,godfray2017,girardin2021}.

\begin{figure}[h]
\centering
\includegraphics[width = \textwidth]{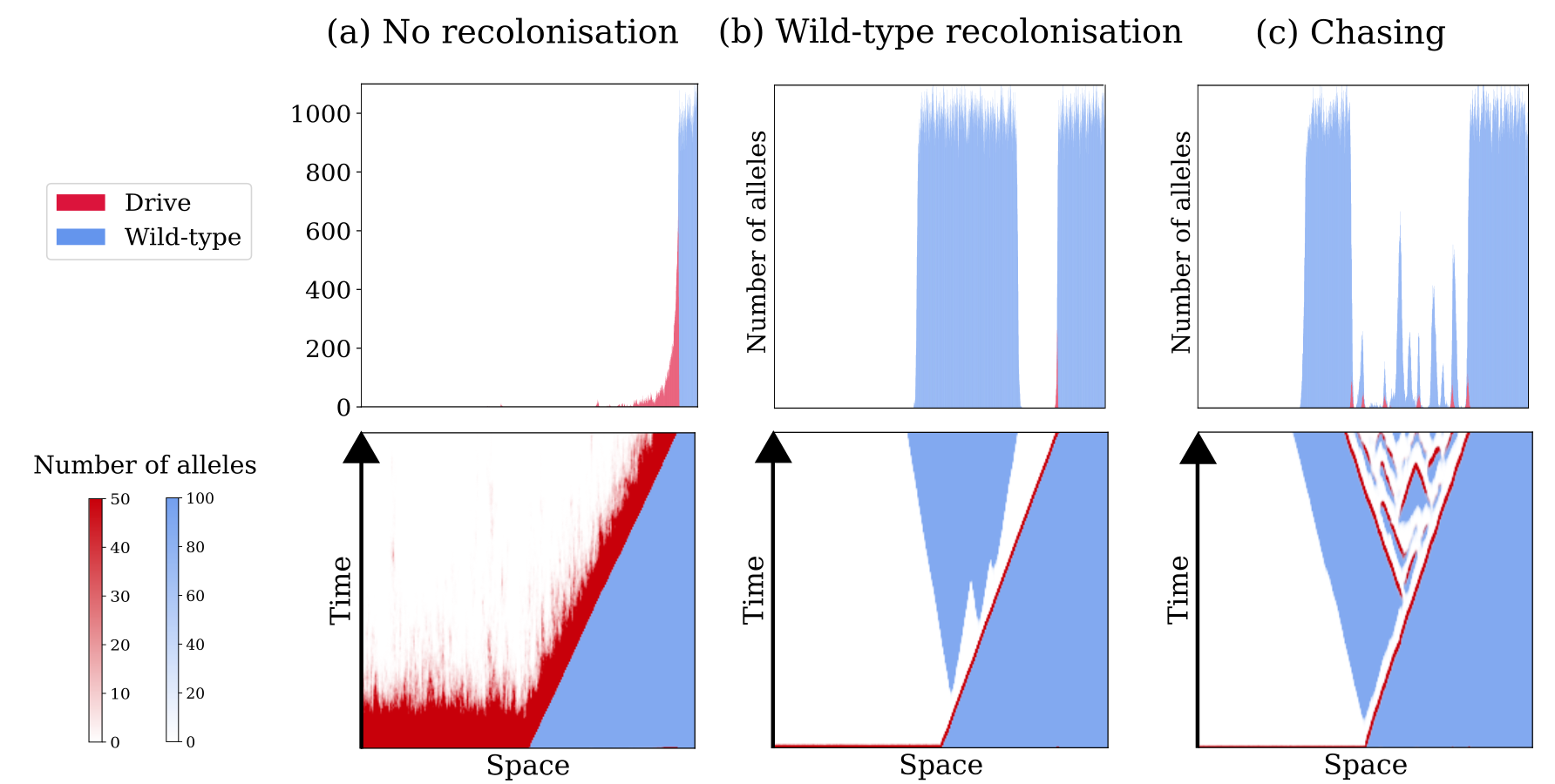}  
\caption{ \lk{Illustrations of the different outcomes of drive propagation, presented as snapshots (top row, at a fixed time $T$) and kymographs (bottom row, from time $0$ to $T$), for different fitness cost (columns). The fitness cost carried by the drive goes from low on the left column to high on the right column, and we observe different dynamics in space: (a) no recolonisation event at all (successful drive invasion), (b) wild-type recolonisation and persistence, (c) successive wild-type recolonisation and drive reinvasion (\textit{chasing}). In the bottom row, each kymograph shows the evolution of the spatial profile (horizontal axis) over time (vertical axis), where color intensity encodes number of alleles. Every simulation starts with the left half of the domain full of drive alleles (red), and the right half  full of wild-type alleles (blue) at time $t=0$, which corresponds to the lower horizontal slice in each kymograph. The higher horizontal slice represents the spatial configuration at $t=T$, whose profile is shown in the top row.} For a better visualisation, we adjusted the colormap of the kymographs to have saturation for 50 drive individuals and 100 wild-type individuals.}
\label{fig:illu_chasing}
\end{figure}

However, eradication may fail because of the recolonisation of cleared areas by wild-type individuals. Such recolonisation dynamics can prevent the total elimination of the target population. Some wild-type individuals might then stay indefinitely in the environment, or the wild-type subpopulation may be invaded again by drive individuals, leading to local extinction of the sub-population; afterwards, the resulting cleared area can be recolonised again by some wild-type individuals, and so on. These infinite dynamics have been referred to as ``colonisation-extinction'' dynamics \cite{north2020} or ``chasing'' dynamics \cite{champer2021}. In this article, we use the term ``wild-type recolonisation'' if the recolonising wild-type individuals stay indefinitely in the environment (Figure \ref{fig:illu_chasing}(b)), and ``chasing'' if at least another drive recolonisation event follows, leading to potential infinite reinvasions (Figure \ref{fig:illu_chasing}(c)).

\lk{Wild-type recolonisation events have been reported in many models of eradication gene drive, mostly discrete individual-based models \cite{north2020, north2019, birand2022, champer2021, champer2022a, liu2023a, liu2022a, paril2022, zhu2023}. This modelling approach, however, is constrained by computational resources, and simulated populations typically did not exceed several tens of thousands of individuals. To simulate a larger population, we implement our models in a population-based framework, tracking the population size at each spatial site rather than the position of each individual. In particular, this allows us to study the influence of the carrying capacity of the environment on the probability of wild-type recolonisation events.}

\lk{We also investigate the influence of migration rate and fitness on wild-type recolonisation events using an analytical approach supported by simulations. This work  complements two previous studies that relied solely on numerical simulations \cite{paril2022, champer2021}. Both studies concluded that high dispersal rates reduce the probability of recolonisation. However, Champer \textit{et al.} found that higher fitness decreases the chance of recolonisation \cite{champer2021}, whereas Paril and Phillips obtained contrasted results on X-shredder and W-shredder drives \cite{paril2022}.}

In this work, we focus on quantifying the probability that a wild-type recolonisation event does not occur within a realistic time window (Figure \ref{fig:illu_chasing}(a)). The rationale of this analysis is based on the following idea: if the last individual carrying a wild-type allele is surrounded by many drive individuals, then it is very unlikely that stochastic fluctuations can leave behind individuals carrying wild-type alleles capable of repopulating the empty place. On the contrary, if individuals carrying wild-type alleles are still present when the drive population is scarce, then it is conceivable that these individuals can recolonise with significant probability. Here, we will focus on the first facet of this rationale, using a deterministic approach to approximate the location of dense drive population. This approximation must however be complemented with a stochastic approximation to determine the position of the last individual carrying at least one wild-type allele. We explain how to handle the deterministic parts in an analytical way, then we give some heuristics of how to deal with the stochastic part. The former is based on the theory of reaction-diffusion traveling waves, which was the main focus of \cite{klay2023}, whereas the latter is based on a connection with the time of extinction in Galton-Watson branching processes.

Our results suggest that the probability of wild-type recolonisation events increases with smaller fitness of drive individuals and smaller local carrying capacity. \VC{The role of the migration rate in the model is less clear, but has a lower impact.} To check the robustness of our analytical and numerical findings in one spatial dimension, we run stochastic simulations in two spatial dimensions varying both the fitness cost and the carrying capacity. The 2D results are consistent with the 1D case, except that wild-type recolonisation events happen faster in 2D than in 1D (as they can occur in a variety of directions). We also give analytical arguments to show that, in case of a wild-type recolonisation event, the probability of a following drive reinvasion event decreases with smaller values of the intrinsic growth rate of the population.

\section{Models}

\subsection{Continuous deterministic model}
\label{sec:cont model}

\lk{We present the model step-by-step. It is an  extension from our previous works \cite{girardin2021, klay2023, klay2025}. For a genetically and spatially homogeneous population, we denote by $n(t)$ the number of individuals at time $t$. Note that unlike in \cite{girardin2021, klay2023, klay2025}, we consider the number of individuals instead of the rescaled density. We denote by $r$ the population's intrinsic growth rate and by $K$ the carrying capacity of the environment. The birth term:}
\begin{equation}
    B(n(t)) = r \ \left( 1-\frac{n(t)}{K} \right)  + 1
\end{equation}
\lk{is considered density-dependent, modulated by fitness $f$. The death term:}
\begin{equation}
    D(n(t)) = 1
\end{equation}
\lk{is considered constant. Several other modeling options have been explored in \cite{girardin2021, klay2025}. The population dynamics is given by:}
\begin{equation}
\partial_t n(t)  =     B(n(t)) f n(t) - D(n(t)) n(t) =  \big( r \ \left( 1-\frac{n(t)}{K} \right)  + 1  \big) f  n(t) - n(t)  \qquad (\forall t>0).
\end{equation}

\lk{Then, we add genetic diversity in the population. We denote by $n_i$ the number of individuals with genotype $i$. The population is diploid, with sexual reproduction, so we follow three genotypes: wild-type homozygotes ($i = WW$), drive homozygotes ($i = DD$) and heterozygotes ($i = DW$). The fitness $f_i$ depends on the genotype: wild-type homozygotes have fitness $f_{WW} = 1$, drive homozygotes have fitness $f_{DD} = 1 - s$, where $s \in (0,1)$ is the fitness cost of the drive, and drive heterozygotes have fitness $f_{DW} = 1 - s h$, where $h \in (0,1)$ is the dominance parameter.}

\lk{We assume that mating occurs at random and we do not distinguish sexes. The probability that a pair of parents with given genotypes produces offspring with a specific genotype depends on the moment at which gene conversion takes place and on the rate $c$ at which gene conversion is successful ($0 \leq c \leq 1$). Here we assume that gene conversion takes place in the germline, because this is the timing currently successfully implemented in the lab \cite{champer2017, champer2022a}. The probabilities of parental and offspring combinations are detailed in Table \ref{tab:mating_term_germline} (Appendix~\ref{app:growth_terms}). We assume that failed gene conversion leaves the wild-type allele intact. In other words, we do not consider here the emergence of resistance alleles through non-homologous end-joining. }

\lk{Finally, we assume that the individuals live in a one-dimensional space; they interact and reproduce locally, and they move in a diffusive manner. We therefore use a reaction-diffusion framework and denote by $\sigma^2$ the diffusion rate. For sake of clarity, we now omit temporal and spatial variables $t$ and $x$ in the notations ($n_i = n_i (t, x)$).}

\lk{Therefore, the population dynamics including genetic diversity, mating and movement is now given by:}

\begin{equation}
\label{eq:par_ger}
 \left\{ \small
    \begin{array}{ll}
      \partial_t \n{DD} - \sigma^2 \partial_{xx}^2 \n{DD} =  (1-s) \left(r \ \left(1-\dfrac{n}{K}\right)+1\right) \ \dfrac{ \frac{1}{4} \ (1+c)^2 \  \n{DW}^2 +  (1+c) \ \n{DW} \n{DD} + \n{DD}^2  }{n}  - \n{DD} ,\\
        \\
      \partial_t \n{DW} - \sigma^2 \partial_{xx}^2 \n{DW} =  (1-sh) \left(r \ \left(1-\dfrac{n}{K}\right)+1\right) \ \dfrac{ (1+c) \ \n{WW} \n{DW} + 2 \ \n{WW} \n{DD} + \frac{1}{2} \ (1-c^2) \ \n{DW}^2 +  (1-c)  \ \n{DW} \n{DD} }{n}  - \n{DW}, \\
        \\
      \partial_t \n{WW} - \sigma^2 \partial_{xx}^2 \n{WW} =  \left(r \ \left(1-\dfrac{n}{K}\right)+1\right) \  \dfrac{ \n{WW}^2 +  (1-c) \  \n{WW} \n{DW} + \frac{1}{4} \ (1-c)^2 \  \n{DW}^2 }{n}  -  \n{WW} .
    \end{array}
\right. 
\end{equation}

\subsubsection{From genotypes to alleles}

In \cite{klay2023}, we established that model \eqref{eq:par_ger} can be reduced to two equations instead of three, focusing on allele numbers $( \n{D},  \n{W})$ instead of genotype numbers $( \n{DD}, \n{DW}, \n{WW})$. The transformation is given by  $\n{D} = \n{DD} + \alpha \  \n{DW}$ and $\n{W} = \n{WW} + (1-\alpha) \ \n{DW}$, with $\alpha = \frac{1+c}2$ when gene conversion occurs in the germline. With this transformation, system~\eqref{eq:par_ger} is equivalent to the following one:
\begin{equation}\label{eq:par_ger_nD_nW}
   \left\{
    \begin{array}{ll}
      \partial_t \n{D}  - \sigma^2 \partial_{xx}^2 \n{D} \ = \  \n{D} \ \Big[   \g{D}( \n{D},  \n{W})  - 1 \Big], \\
        \\
      \partial_t  \n{W}  - \sigma^2 \partial_{xx}^2  \n{W} = \  \n{W} \ \Big[  \g{W}( \n{D},  \n{W})  - 1 \Big],
    \end{array}
\right. 
\end{equation}
where $\g{D}$ and $\g{W}$ are the per-capita growth rates associated with each allele, and are given by:
\begin{subequations}
\begin{equation}\label{eq:gD}
     \g{D}( \n{D},  \n{W}) =  \left(r \ \left(1-\dfrac{n}{K}\right)+1\right) \Big[ (1-s) \dfrac{\n{D}}{n} +   (1-sh) \  (1+c) \   \dfrac{\n{W}}{n} \Big], 
\end{equation} \begin{equation}\label{eq:gW}
     \g{W}( \n{D},  \n{W}) =   \left(r \ \left(1-\dfrac{n}{K}\right)+1\right)  \Big[ \  \dfrac{\n{W}}{n}   +   (1-sh)  \  (1-c) \  \dfrac{\n{D}}{n} \Big].
\end{equation}
\end{subequations}
In the following, the term ``number of drive alleles'' (resp. ``number of wild-type alleles'') refers to the quantities $n_D$ (resp. $n_W$), and represents an haploid count.

\subsubsection{Drive invasion and traveling waves} \label{sec:trav_wave_cont}

As it is classical in spatial ecology \cite{murray2007}, we investigate the spatial invasion of the drive allele by means of traveling waves. We seek stationary solutions in a reference frame moving at speed $\vc$ (``\textbf{co}'' stands for \textbf{co}ntinuous model). \lk{To this end, we perform the change of variable $z = x-\vc t$:}
\begin{equation} \label{eq:trav_waves}
\left\{
    \begin{array}{ll}
       \n{D}(t,x) = \N{D}(x-\vc t) = \N{D}(z)  \quad \quad (\forall t >0) \ (\forall x \in \mathbb{R}), \\
       \n{W}(t,x) = \N{W}(x-\vc t)  = \N{W}(z) \quad \ \  (\forall t >0) \ (\forall x \in \mathbb{R}),\\
    \end{array}
\right.
\end{equation}
Plugging this particular shape of a solution in  \eqref{eq:par_ger_nD_nW}, we deduce the following pair of equations for the traveling wave profiles ($\N{D}$, $\N{W}$):
\begin{equation} \label{eq:par_ger_wave}
\left\{
    \begin{array}{ll}
      -\vc\N{D}' - \sigma^2 \N{D}'' & =  \  \N{D} \ \left[ \left(r \ \left(1-\dfrac{N}{K}\right)+1\right) \Big[ (1-s) \dfrac{\N{D}}{N} +   (1-sh) \  (1+c) \   \dfrac{\N{W}}{N}  \Big]  - 1 \right], \\
       \\
      -\vc\N{W}' - \sigma^2 \N{W}'' & =   \  \N{W} \ \left[ \left(r \ \left(1-\dfrac{N}{K}\right)+1\right)  \Big[ \  \dfrac{\N{W}}{N}   +   (1-sh)  \  (1-c) \  \dfrac{\N{D}}{N}  \Big]  - 1 \right].
    \end{array}
\right.
\end{equation}
We impose the following conditions: 
\begin{equation} \label{eq:ass_trav_waves}
\left\{
    \begin{array}{ll}
        \vc >0, \\
       \N{D}(-\infty) = 0 \,,\quad  \N{D}(+\infty) = 0 , \\
       \N{W}(-\infty) = 0 \,,\quad  \N{W}(+\infty) = 1 . \\
\end{array}
\right.
\end{equation}
These conditions mean that we consider a drive eradication wave propagating from the left-hand side to the right-hand side of our figures. A schematic illustration of the wave is drawn in Figure \ref{fig_back_wave}(a). In this mathematical framework, the limiting values $\N{D,W}(-\infty) = 0 $ listed in system~\eqref{eq:ass_trav_waves} correspond to the eradication of the population after the drive wave has passed. The condition to attain this outcome reads $r<\frac{s}{1-s}$ \cite{girardin2021, klay2023}.

Traveling wave solutions contain important information for the biological interpretation of the results. The speed  $\vc$ is the rate of propagation of the drive population. Under assumptions \eqref{eq:ass_trav_waves}, the profile $\N{D}(z)$ has exponential decay both when $z$ goes to $+\infty$ and $-\infty$ (Figure \ref{fig:front_deter}). The profile $\N{W}(z)$ has exponential decay when $z$ goes to $-\infty$. This can be expressed as follows:
\begin{subequations}\label{eq:ansatz decay}
\begin{align}
    \text{At the front of the wave:} \quad & \N{D}(z) \approx \exp \left(\lDf z \right), \\
     \text{At the back of the wave:} \quad  & \N{D}(z) \approx \exp \left(\lDb z \right), \\
      \text{At the back of the wave:} \quad  & \N{W}(z) \approx \exp \left(\lWb z \right).
\end{align}
\end{subequations}

The rates of decay denoted by $\lambda$ play a key role in our analysis. It would be possible to derive several qualitative conclusions to some extent of generality, but we opt for simplicity, and we limit our analysis to the case of pulled traveling waves. \lk{A wave is said to be pulled if the wave speed coincides with the minimal speed of the linearized problem at low density (here low number of drive alleles). This occurs when the (drive alleles) population has its best reproductive success at low density and only these low densities contribute to the formation of the leading edge of the invasion. Conversely, a wave is said to be pushed if the wave speed is strictly larger than the minimal speed of the linearized problem. In contrast with pulled waves, the best reproductive success is at intermediate or high densities and these densities contribute significantly to the formation of the leading edge of the invasion. The speed $\vc$ is explicitly known in the case of a pulled wave:} 
\begin{equation}\label{eq:speed_lDf}
   \vc = 2 \sigma \sqrt{ (1-sh)(1+c) - 1 }
\end{equation}
(see Appendix~\ref{app:lambda_cont} for details). 

As shown in \cite{klay2023}, when gene conversion occurs in the germline, the drive wave is always pulled for $h < 0.5$. In this study, we consider $h=0.4$ for the numerical illustrations.  
We obtain formulas for the exponential rates of profiles $\n{D}$ and $\n{W}$ at the front and at the back of the wave, namely:%
\begin{subequations}\label{eq:exp_lDfd}
\begin{align}
     \lDf = & \frac{1}{\sigma} \left( - \sqrt{ (1-sh)(1+c) - 1 } \right), \label{eq:lDf} \\
    \lDb = & \frac{1}{\sigma} \left( -   \sqrt{ (1-sh)(1+c) - 1 } + \sqrt{ (1-sh)(1+c)  -  ( r+1 ) (1-s) } \right), \label{eq:lDb_cont} \\
    \lWb = & \frac{1}{\sigma} \left(  -   \sqrt{ (1-sh)(1+c) - 1 } + \sqrt{ (1-sh)(1+c)  - (r+1) (1-sh) (1-c) } \right), \label{eq:lWb_cont}
\end{align}
\end{subequations}
(see Appendix~\ref{app:lambda_cont} for details).

\subsection{Discrete stochastic model}

To observe wild-type recolonisation events, we need to consider a stochastic discrete model, based on the allele dynamics described by system \eqref{eq:par_ger_nD_nW}. \lk{In the following, we present this model in one spatial dimension, however it can easily be extended in two spatial dimensions (see Section \ref{sec:1D-2D}).} We denote the number of drive and wild-type alleles at spatial site $x$ and time $t$ as: \begin{equation}
    \n{D}(t,x) = \n{D}^{t,x}, \quad \n{W}(t, x) = \n{W}^{t, x}  \quad \text{and} \quad   n(t,x) = n^{t, x} \quad  \quad \forall  \ t \in \{ 0, \mathrm{d}t, 2 \mathrm{d}t, ...\},  \ \forall x \in \{ 0, \mathrm{d}x, 2 \mathrm{d}x,  ... \} .
\end{equation} 

In our stochastic simulations, each individual reproduces, dies and disperses independently of the reproduction, death, and dispersal of others. We alternate between two types of events: 1) allele production/disappearance in each spatial site and 2) allele migration in the neighbouring spatial sites. During one time unit, a drive allele duplicates on average $\g{D}(\n{D}^{t, x}, \n{W}^{t, x})$ times, as defined in Eq.\eqref{eq:gD}, and a wild-type allele duplicates on average $\g{W}(\n{D}^{t, x}, \n{W}^{t, x})$, as defined in Eq.\eqref{eq:gW}. One allele (drive or wild-type) disappears at rate $1$ within one time unit, corresponding to the death of an individual carrying it. To model the number of new alleles and removed alleles, we use Poisson distributions with these respective means. A Poisson distribution corresponds to the number of events observed during a time interval of given length, when the waiting time between two consecutive events is given by an exponential distribution. For one type of event, under an exponential distribution, the expected future waiting time is independent of waiting time already passed.

Second, we consider migration: at each time step, an allele migrates with probability $m$ outside its original site. It goes to one neighbouring spatial site, either on the right or on the left, with equal probabilities. To model this event, we use two Bernoulli distributions: one with probability $m$ to determine if the individual migrates, the second with probability $\frac{1}{2}$ in case of success, to determine the welcoming site (right or left).

\begin{remark}[Migration rate vs. diffusion coefficient]\label{remark_mig_diff}
Both the migration rate (denoted $m$, discrete models) and the diffusion coefficient (denoted  $\sigma^2$, continuous models) describe the ability of individuals to move in space. Their relationship is given by
\begin{equation}
    \sigma^2 = \dfrac{m (\Delta x)^2}{2 \Delta t}.
\end{equation}
\end{remark}

We implement this model in a population-based way: we follow the number of alleles in each spatial site instead of following the position of each living individual (which would then be an individual-based model). In each site, we draw independent events of production/disappearance of alleles relative to the number of alleles in this site, and the same goes for migration events.

These dynamics are summarised in Figure \ref{fig:schema_stoch} and a pseudocode is provided in Appendix \ref{app:code_stoch}. The main advantage of this approach is that it allows to simulate systems with very large number of alleles.

\begin{figure}[H]
        \centering
     \includegraphics[width=\textwidth]{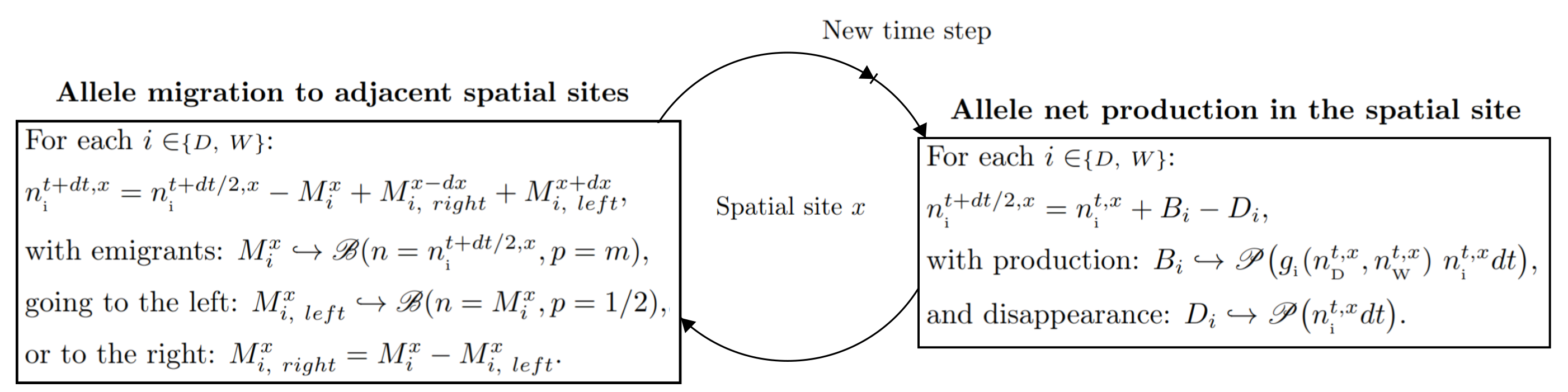}
        \caption{Consecutive steps in the stochastic discrete model.}
        \label{fig:schema_stoch}
\end{figure}

The discrete model also exhibits finite speed propagation, but the rate of propagation differs from its continuous version \eqref{eq:speed_lDf} for two reasons: firstly, the stepping-stone discretisation, and secondly, the stochastic effects due to finite population size. These two corrections of the speed are very well documented \cite{brunet1997, brunet2001}. However, for the sake of simplicity, we prefer to conduct our analysis based on the formulas of the continuous model, as in Section \ref{sec:cont model}. In Appendix \ref{subsec:corr_speed_dis}, we briefly present each correction, and we compare the outcomes of the theory and the numerical simulations, in order to support our continuous approximation.

\subsection{Initial conditions and parameters for numerical simulations}\label{subsec:para}

\begin{figure}[H]
    \centering
    \includegraphics[height=4.5cm]{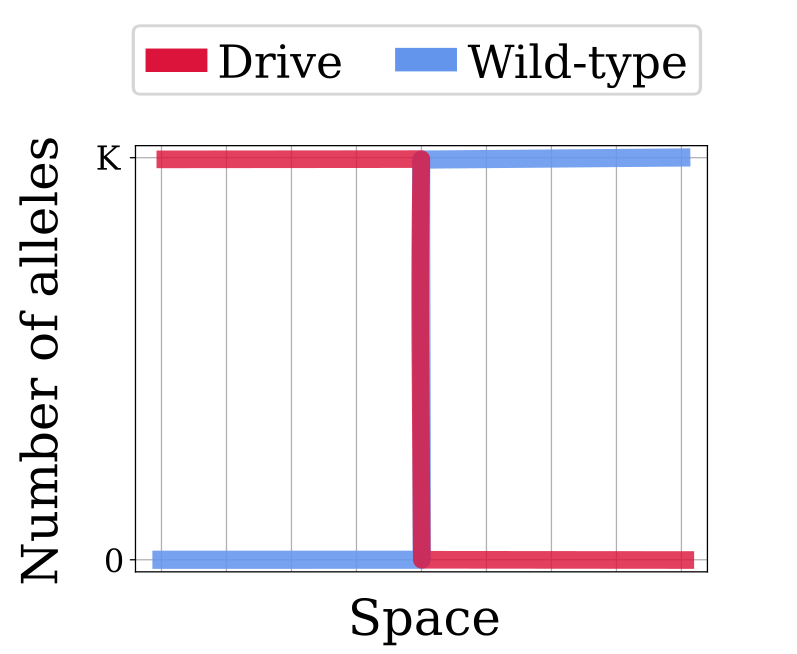}
   \caption{Initial conditions for numerical simulations in one spatial dimension.}
    \label{fig:CI_stoch}
\end{figure}

Initial conditions for our numerical simulations are illustrated in Figure \ref{fig:CI_stoch}. The left half of the domain is full of drive alleles ($\n{D}=K$), and the right half is full of wild-type alleles ($\n{W}=K$). Reference to right or left positions are made in the context of this initial state. 

Numerical simulations are performed with the same set of parameters (unless specified otherwise), summarised in Table \ref{tab:parameters_stoch}. The code is available on GitHub (\url{https://github.com/LenaKlay/Stochastic-dynamics-at-the-back-of-a-gene-drive-eradication-wave}). A citable, versioned release has also been deposited in Zenodo (\url{https://doi.org/10.5281/zenodo.17201147}). We ran our simulations in Python 3.10.12, with the Spyder 6.0.7 environment.

\begin{table}[H]
    \centering
    \begin{tabular}{llll}
    \textbf{Parameter} & \textbf{Range value} &  \textbf{Value}  & \textbf{Description}\\
$r$  & $(0,+ \infty)$ &  $0.1$ &  Intrinsic growth rate  \\
   $c$  & $[0,1]$ & $0.9$  &  Conversion rate \\
   $s$ & $(0,1)$ &  $0.3$ or $0.7$ & Fitness cost of drive homozygotes   \\
   $h$  & $[0,1]$ &  $0.4$ & Drive dominance  \\
    $m$  & $[0,1]$ &  $0.2$ & Migration rate  \\
    $K$  & $(0,+ \infty)$ & from $10^3$ to $10^8$ & Local carrying capacity \\
     $\mathrm{d}x$  & $(0,1]$  &  $1$ & Spatial step between two sites \\
    $\mathrm{d}t$  & $(0,1]$  &  $0.1$ & Temporal step \\
    $T$ & $(0,+ \infty)$  &  $1000$  & Simulation maximum time \\ 
    \end{tabular}
    \caption{Model parameters and numerical values used in the simulations.}
    \label{tab:parameters_stoch} 
\end{table}

\section{\lk{Methods}}

\subsection{\lk{A theoretical framework under the continuous approach}} \label{sec:results1}

The parameters detailed in Section \ref{subsec:para} have been chosen to verify three conditions in the deterministic model: 

i) The drive invades for all initial conditions by means of a pulled wave. This is enforced when $h<0.5$ and 
\begin{equation}
    (1-sh)(1+c) > 1, \label{eq:conditiondriveinv}
\end{equation}
which is the obvious condition for the definition of the wave speed $\vc$ in \eqref{eq:speed_lDf}. The biological interpretation goes as follows: wild-type alleles are so numerous at the front of the wave that at least one parent in each couple is wild-type homozygote (WW), so that offspring carrying drive alleles are necessarily heterozygotes. Thus, the production of drive alleles relies on heterozygotes, having a fitness of $(1-sh)$, a drive allele production rate of $(1+c)$ and a mortality rate of $1$.

ii) The final equilibrium is an empty environment, that is, we impose the following threshold on the net growth rate: 
\begin{equation}\label{eq:extinction}
    r<\frac{s}{1-s},
\end{equation}
which is equivalent to $\lDb>0$ \eqref{eq:lDb_cont} (exponential decay at the back of the wave). 

iii) \VC{We exclude the coexistence of wild-type and drive alleles having both positive frequencies at the back of the wave, see \cite{klay2023}, and discussion in Appendix \ref{app:exlu_coex}:
\begin{equation}
    (1-sh)(1-c) < (1-s). \label{eq:conditionfreq1}
\end{equation}
This is equivalent to $\lWb > \lDb$, ensuring that the drive curve is less steep than the wild-type curve as in Figure \ref{fig:zoom} (in opposition with $\lWb = \lDb$ in the case of coexistence).}

These three conditions establish a suitable framework for the observation of wild-type recolonisation events in the stochastic model (suitable but not sufficient). Figure \ref{fig:front_deter} represents the typical outcome expected in the continuous model under these assumptions, with the two profiles $N_D$ (red) and $N_W$ (blue) and a speed of propagation $v$, as described by the set of equations \eqref{eq:par_ger_wave}--\eqref{eq:ass_trav_waves}. The relative intensities of exponential decay at the back of the wave are key to our analysis, see Figure \ref{fig:zoom}.

\subsection{\lk{Wild-type recolonisation events in the stochastic model}}

\begin{figure}
    \centering
    \begin{subfigure}{0.43\textwidth}
    \centering
      \caption{Propagating wave in the continuous model}
     \includegraphics[width=0.8\linewidth]{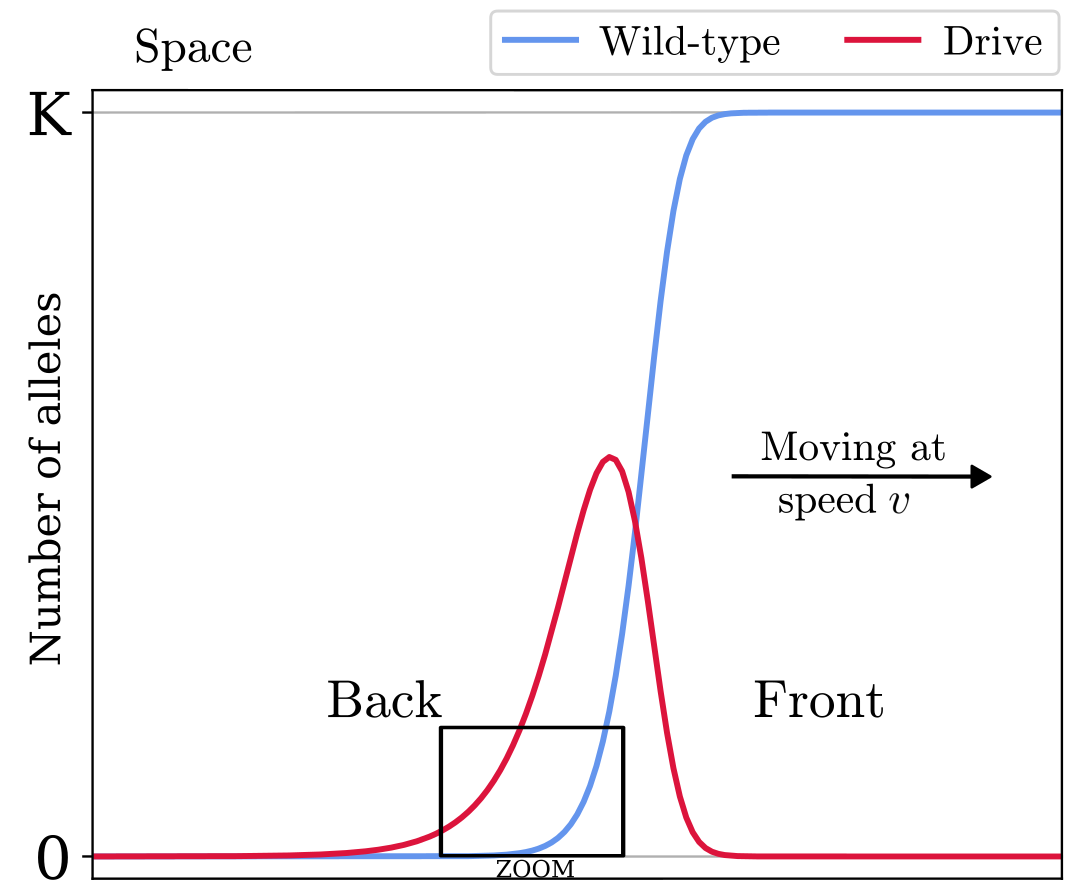}
     \label{fig:front_deter}
\end{subfigure}
\hfill
\begin{subfigure}{0.48\textwidth}
   \centering
    \caption{Zoom at the back of the wave: both densities decay exponentially fast to zero in case of eradication} 
    \includegraphics[width=0.6\linewidth]{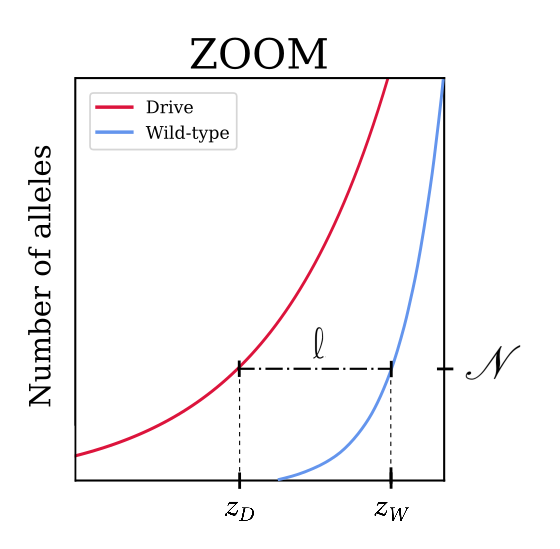}
    \label{fig:zoom}
\end{subfigure}
\\
    \begin{subfigure}[t]{0.45\textwidth}
         \centering
         \caption{$K=10^8$, $s=0.3$}        
         \includegraphics[width=\linewidth]{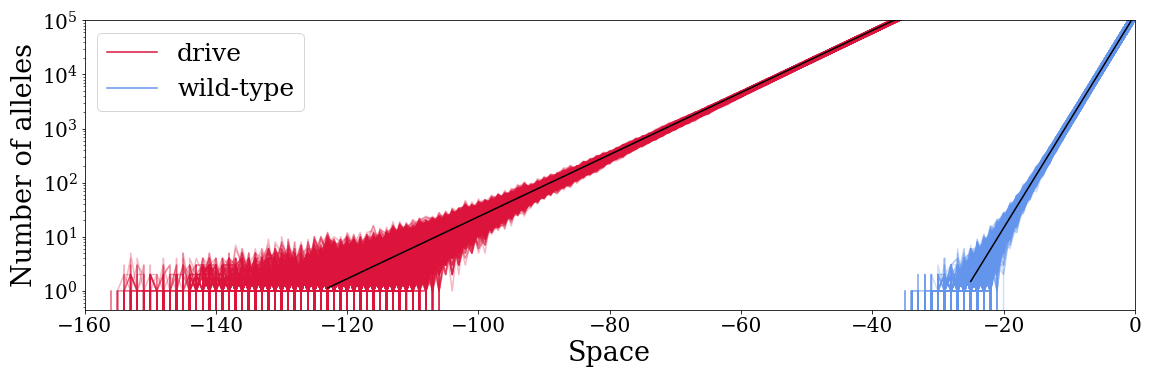}
         \label{fig_back_wave_0.3_8}
    \end{subfigure}
    \hfill
     \begin{subfigure}[t]{0.45\textwidth}
        \centering
         \caption{$K=10^8$, $s=0.7$}
         \includegraphics[width=\linewidth]{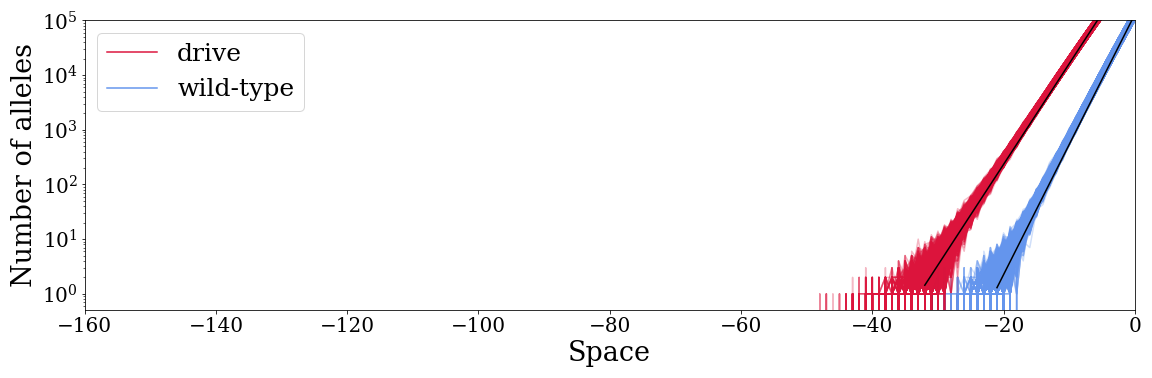}
       \label{fig_back_wave_0.7_8}
    \end{subfigure}
    \\
    \begin{subfigure}[t]{0.45\textwidth}
         \centering
         \caption{$K=10^5$, $s=0.3$}
         \includegraphics[width=\linewidth]{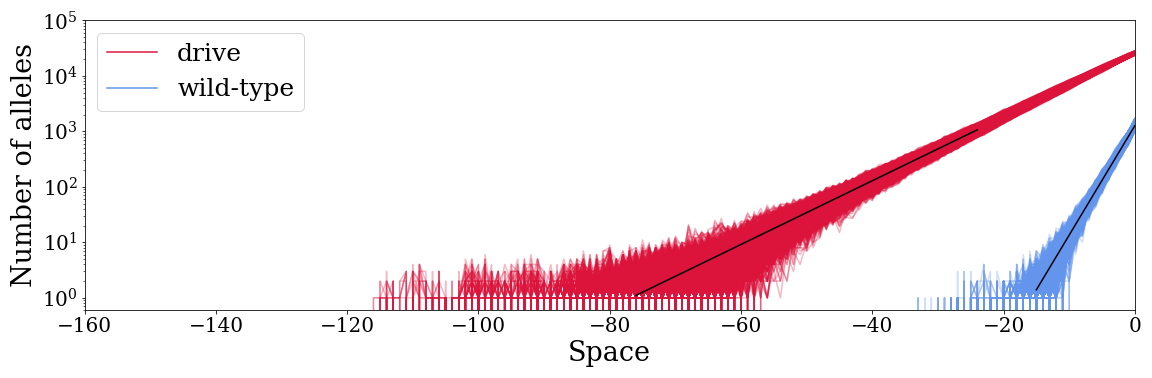}        
          \label{fig_back_wave_0.3_5} 
    \end{subfigure}
        \hfill
    \begin{subfigure}[t]{0.45\textwidth}
         \centering
          \caption{$K=10^5$, $s=0.7$ -- Wild-type recolonisation}       
         \includegraphics[width=\linewidth]{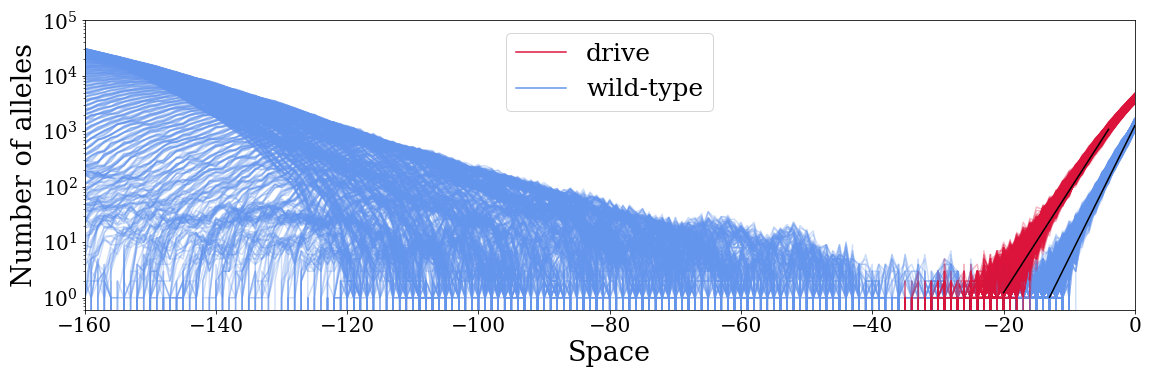}         
         \label{fig_back_wave_0.7_5}
    \end{subfigure}
\\
    \begin{subfigure}[t]{\textwidth}
         \centering    
         \caption{$K=10^8$, $s=0.3$}
         \includegraphics[width=\linewidth]{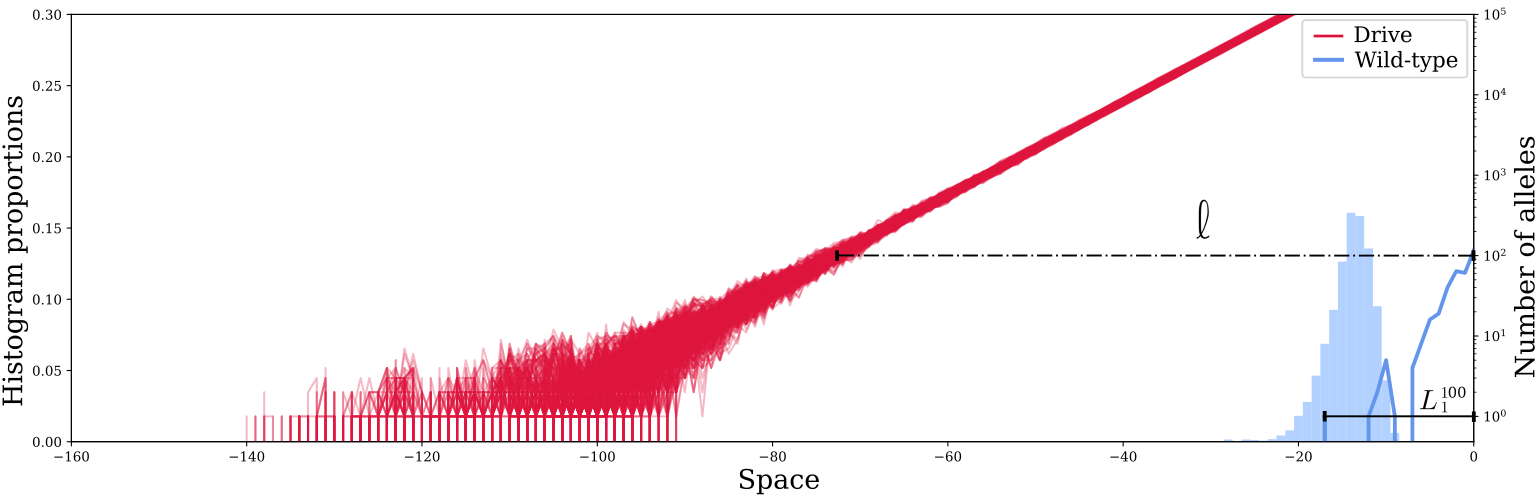}         
         \label{fig:6_histogram3}
    \end{subfigure}
     \caption{Number of drive alleles (in red) and number of wild-type alleles (in blue) at the back of the wave, for the continuous model in linear scale (panels (a, b)) and the stochastic model in log scale (panels (c, d, e, f, g)). The point of abscissa $z=0$ is arbitrary placed at the last spatial site with more than $1000$ wild-type alleles in panels (c, d, e, f) and with more than $100$ wild-type alleles in panel (g). We superimpose multiple realisations corresponding to 2000 different timepoints in the simulation, to observe the stochastic variations. \lk{We denote by $\ell$ the distance between the last position with more than $\mathscr{N}$ drive alleles and the last position with more than $\mathscr{N}$ wild-type alleles (illustrated theoretically in panel (b) and shown for one stochastic realization in panel (g)). We denote by $L_{1}^{\mathscr{N}}$, the distance between the position of the last individual carrying a wild-type allele and the last position with more than $\mathscr{N}$ wild-type alleles (shown for one stochastic realization in panel (g)). In contrast to $\ell$, this distance is genuinely stochastic.} The exponential approximations from Appendix~\ref{app:lambda_dis} are shown for each wave in black. We observe a wild-type recolonisation event for $s=0.7$ and $K=10^5$. In panel (g), for $s=0.3$ and $K=10^8$, we add in the figure a histogram showing the spatial distribution of the last wild-type allele in lighter blue, highlighting the high stochasticity of this position.}
    \label{fig_back_wave}
\end{figure}

The continuous representation fails at very low number of alleles per site: \lk{the fewer alleles there are, the greater the stochasticity}. This can be seen in Figures \ref{fig_back_wave}(c--f), showing the outcomes of stochastic simulations, and where the drive and the wild-type allelic distributions are plotted in log scale. Only the back of the wave is shown on these panels to focus on the occurrence of recolonisation events. Multiple time points are superimposed in these figures, and the curves are translated to have a constant number of wild-type alleles at the spatial origin on the right hand side (either 100 or 1000 alleles depending on the figures). The black lines are the exponential approximations for each, as described with detail in Appendix~\ref{app:lambda_dis} (they are the analogues of expressions \eqref{eq:exp_lDfd} for the stepping stone discrete approximation). For a large drive fitness cost ($s=0.7$) and a relatively small carrying capacity ($K=10^5$), we observe one wild-type recolonisation event within this window of time (Figure~\ref{fig_back_wave}\subref{fig_back_wave_0.7_5}). Note that in this example, the slopes of the wild-type and drive curves are similar, and  carrying capacity is relatively low. 

\subsection{A theoretical framework to characterise the absence of recolonisation \lk{combining both the stochastic and the deterministic approach}}

\lk{In a spatially homogeneous setting, a simulation initialized with a few wild-type individuals and many drive individuals will very likely lead to extinction of the wild-type allele \cite{champer2021}. For the investigation of the stochastic effects at the back of the wave, }
we posit that wild-type recolonisation or chasing can very likely be prevented if the last individual carrying a wild-type allele is surrounded by a sufficiently large number of drive individuals. This reference number, denoted by $\mathscr{N}$, should be large enough to avoid stochastic effects, but much smaller than the carrying capacity $K$.
We choose $\mathscr{N}=100$ in this paper. 

To get a clearer view, we superimpose in Figure \ref{fig_back_wave}(g) the distribution of drive alleles at several times ($\n{D}$ in red), the distribution of wild-type alleles at a single time ($\n{W}$ in  blue) and the histogram of the position of the last wild-type allele (light blue). We see that a clear separation exists between the distribution of the last wild-type allele and the area with few drive alleles (below 100) when $s=0.3$ preventing the wild-type recolonisation with high probability.

To determine when wild-type recolonisation or chasing is very unlikely, and evaluate the contributions of each parameter, we aim to separate the space between a deterministic zone and a stochastic zone  at the back of the wave, depending on the number of alleles per site.

We denote by $\ell$ the distance between the last position with more than $\mathscr{N}$ drive alleles and the last position with more than $\mathscr{N}$ wild-type alleles, illustrated in Figure \ref{fig_back_wave}(g). We \lk{postulate} that this distance can be well approximated by a deterministic model. This entails choosing $\mathscr{N}$ sufficiently large. Moreover, to ease the analytical calculations, we are going to use the spatially continuous model rather than the discrete one for the values of $\lambda$ characterising the curves.

The second quantity of interest is $L_{1}^{\mathscr{N}}$, the distance between the position of the last individual carrying a wild-type allele and the last position with more than $\mathscr{N}$ wild-type alleles, also illustrated in Figure \ref{fig_back_wave}(g). In contrast to $\ell$, this distance is genuinely stochastic.

\begin{figure}[H]
\centering
\begin{subfigure}{0.48\textwidth}
    \centering
    \caption{$K=10^8$, $s=0.3$}
     \includegraphics[width = \textwidth]{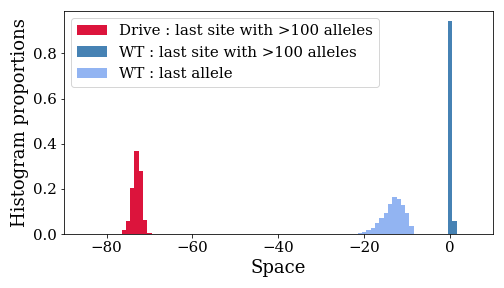}
       \label{fig:hist_deter_or_not_s3}
\end{subfigure}
\hfill
\begin{subfigure}{0.48\textwidth}
    \centering
         \caption{$K=10^8$, $s=0.7$}
    \includegraphics[width = \textwidth]{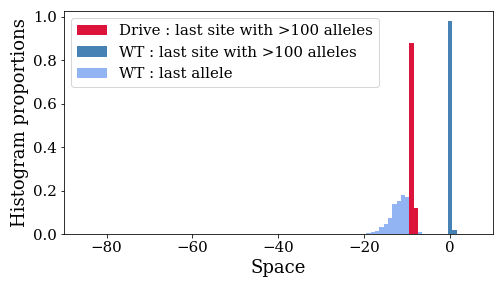}
     \label{fig:hist_deter_or_not_s7}
\end{subfigure}
 \caption{Relative positions of the last spatial site with more than 100 drive alleles at the back of the wave (in red), the last spatial site with more than 100 wild-type alleles at the back of the wave (in blue) and the last wild-type allele at the back of the wave (in light blue). We collected these positions at 2000 different time points during the simulation to plot these histograms. We arbitrarily set the last spatial site with more than $100$ wild-type alleles at the points of abscissa $z=0$. The statistical distribution of the last wild-type individuals is highly stochastic in contrast with the two other involving more than $100$ alleles. We also observe that this distribution is not symmetric, indicating rare events of last individuals carrying a wild-type allele being far away at the back of the wave. Note that deterministic distance $\ell$ (between the red and blue distributions) accounts for most of the variation when $s$ decreases, while stochastic distance $L_{1}^{\mathscr{N}}$ (between the light blue and blue distributions) remains similar.} 
   \label{fig:hist_deter_or_not}
\end{figure}

Figure \ref{fig:hist_deter_or_not} represents the distribution of the furthest spatial positions associated with at least $\mathscr{N}=100$ drive alleles (in red) and $\mathscr{N}=100$ wild-type alleles (in blue), and the furthest spatial position associated with a single wild-type allele in the site (in lighter blue), that is, the position of the last individual carrying a wild-type allele. We clearly see that the two former distributions are concentrated around a deterministic value, whereas the latter is more spread out. \VC{The two former distributions are fairly concentrated thanks to the choice of the population threshold $\mathscr{N}=100$. Their relative distance is thus characterized by the deterministic distance $\ell$.}

We can recast our mathematical problem as $L_{1}^{\mathscr{N}} < \ell$ with high probability, meaning that the last wild-type individuals is very likely surrounded by more than $\mathscr{N}$ alleles, which is clearly the case in Figure \ref{fig:hist_deter_or_not_s3} for small value of the drive fitness cost $s$ ($s = 0.3$). In contrast, when $s$ is larger ($s=0.7$), the last wild-type allele is possibly surrounded by relatively few drive alleles when $s=0.7$, because the distribution of the position of the last wild-type allele (light blue) falls beyond the distribution of the last site with more than $\mathscr{N} = 100$ alleles (Figure \ref{fig:hist_deter_or_not_s7}). Note that deterministic distance $\ell$ (between the red and blue distributions) accounts for most of the variation when $s$ decreases, while stochastic distance $L_{1}^{\mathscr{N}}$ (between the light blue and blue distributions) remains similar.

Here, the value of $\mathscr{N}$ is quite arbitrarily, but it has little influence on the mathematical argument. Our conclusions will be mainly qualitative, as we are lacking analytical predictions, in particular about the stochastic part $L_{1}^{\mathscr{N}}$. 

\section{Results}

\subsection{\lk{Analysis: characterising the deterministic distance $\ell$}}\label{subsec:deter_dist}

\begin{figure}[H]
    \centering   \includegraphics[width=0.55\linewidth]{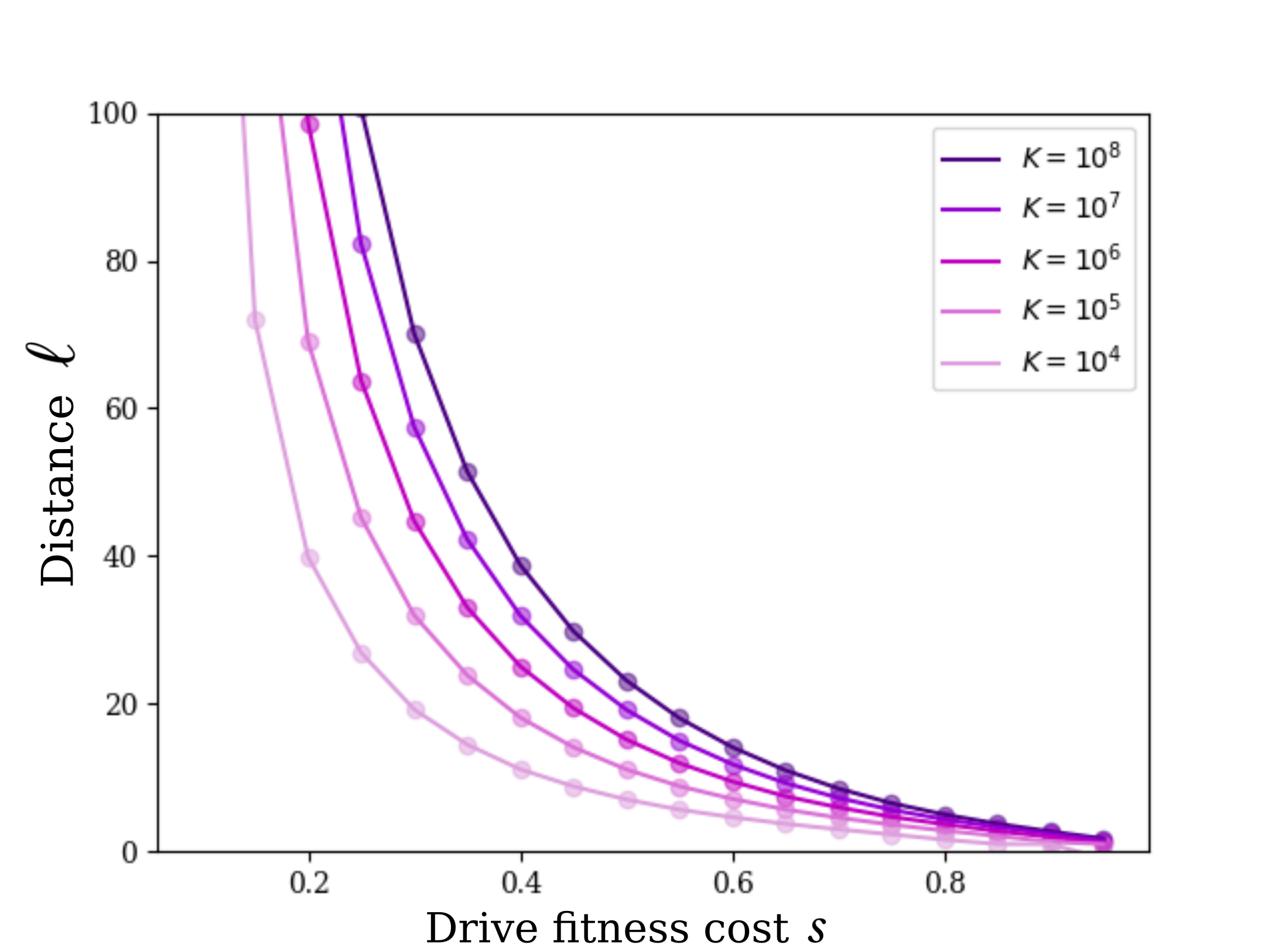}
    \caption{Distance $\ell$ for different values of the carrying capacity $K$ and the drive fitness cost $s$, computed through deterministic simulations as in \cite{klay2023}. The distance $\ell$ increases with $K$ and decreases with $s$.}
    \label{fig:dis_ref}
\end{figure}

In the deterministic continuous approximation at the back of the wave, the quantity $\ell$ corresponds to the distance between the two points of abscissa $z_D$ and $z_W$  \lk{(see Figure \ref{fig:zoom})} such that
\begin{equation}
    \rho_{D}^{\text{back}} K \exp{(\lDb z_D)} = \rho_{W}^{\text{back}} K \exp{(\lDb z_W)} = \mathscr{N}.
\end{equation}
The factor $ \rho_{D}^{\text{back}} \in (0,1)$ , resp. $ \rho_{W}^{\text{back}} \in (0,1) $, is characterised by the shape of the continuous traveling wave profile $N_D$, resp. $N_W$. We arbitrarily set $\mathcal{N}$ so that the continuous profile is accurately approximated by an exponential ramp for $z<0$, see Figure \ref{fig_back_wave}(b). This leads to the following approximation for $\ell$ with $z_D<z_W<0$:
\begin{equation} \label{eq:lc}
\text{$\ell$} = z_W - z_D = \dfrac{1}{\lWb}\log\left (\frac{\mathscr{N}}{ \rho_W^{\text{back}} K } \right) - \dfrac{1}{\lDb}\log\left (\frac{\mathscr{N}}{ \rho_D^{\text{back}} K} \right),
\end{equation}
with the decay exponents $\lWb$ \eqref{eq:lDb_cont} and $\lDb$ \eqref{eq:lDb_cont}. \lk{To obtain tractable results, we use the formula:}
\begin{equation}\label{eq:lc2}
    \ell \sim \left( \dfrac{1}{\lWb} - \dfrac{1}{\lDb} \right) \log\left (\frac{\mathscr{N}}{ K } \right)
\end{equation}
as $\rho_W^{\text{back}}<1 \ll K/\mathscr{N}$, and $\rho_D^{\text{back}}<1 \ll K/\mathscr{N}$ in logarithmic scale. The next three \lk{paragraphs} are devoted to the analytical investigation of the influence of the carrying capacity $K$, the diffusion coefficient $\sigma^2$ and the drive fitness cost $s$ on distance $\ell$. Figure \ref{fig:dis_ref} illustrates numerically distance $\ell$ with respect to the carrying capacity and the drive fitness cost.

\paragraph{The deterministic distance $\ell$ is increasing with the carrying capacity $K$}\label{subsubsec:deter_carry_capa}

The carrying capacity $K$ is a \lk{scaling factor influencing the maximum number of individuals in the population} without changing the exponential decay at the back of the wave ($\lDb$ is independent from $K$). From Eq.\eqref{eq:lc2}, we \lk{find} that the distance $\ell$ increases with respect to the carrying capacity $K$. More precisely, if the carrying capacity is multiplied by a factor $10$, then $\ell$ increases by: 

\begin{equation}\label{eq:increases_K}
    \log (10)\left( \frac{1}{\lDb} - \frac{1}{\lWb} \right).
\end{equation}

Quantity \eqref{eq:increases_K} is strictly positive under condition \eqref{eq:conditionfreq1}.

\paragraph{The deterministic distance $\ell$ is increasing with the diffusion coefficient $\sigma^2$}

The influence of the diffusion rate $\sigma^2$ can be appreciated by noting that $\lDb$ \eqref{eq:lDb_cont} and $\lWb$ \eqref{eq:lDb_cont} are inversely proportional to $\sigma$. \lk{Therefore, the distance}  $\ell$ is proportional to $\sigma$ (Eq.\eqref{eq:lc2}), \lk{and thus} it increases with the migration rate $m$ (see Remark \ref{remark_mig_diff}).

%\subsubsection{The deterministic distance $\ell$ is decreasing with the drive fitness cost $s$}\label{subsubsec:deter_fit}

\paragraph{The deterministic distance $\ell$ is decreasing with the drive fitness cost $s$}\label{subsubsec:deter_fit}
\lk{More work is required to} determine the \lk{variation} of $\ell = z_W - z_D$ \lk{with respect to} the drive fitness cost $s$. Indeed, it is not complicated to see that both $z_W$ and $z_D$ are decreasing with respect to the fitness cost $s$ in Eq.\eqref{eq:lc2}, but establishing the monotonicity of the difference is more involved.

Below we are going to establish that 
\begin{equation}\label{eq:result353}
    \dfrac{d}{ds}\left( \dfrac{1}{\lWb} - \dfrac{1}{\lDb} \right) > 0 ,
\end{equation}
leading to the conclusion that $\ell$ is decreasing with $s$ in the approximation \eqref{eq:lc2} because $\log\left (\frac{\mathscr{N}}{ K } \right) <0$.  
Instead of comparing the two explicit solutions $\lDb$ and $\lWb$, we start from the characteristic equations satisfied by each component. The following equations can be deduced from \eqref{eq:par_ger_wave}, and lead to the formulas \eqref{eq:exp_lDfd}, see Section \ref{app:lambda_cont_back} for details.   
\begin{align}
    &     0 =  \sigma^2 (\lDb)^2  + \vc \lDb  + \Big[  ( r+1 ) (1-s)  - 1 \Big], \label{eq:lambda_eq1}\\
    &    0 = \sigma^2 (\lWb)^2  + \vc \lWb  +  \Big[ (r+1) (1-sh)   (1-c) -1   \Big]. \label{eq:lambda_eq2}
\end{align}
For the sake of simplicity, we introduce a notation for the zeroth order coefficients:
\begin{equation}\label{eq:coeff q}
    q_D(s) = ( r+1 ) (1-s)  - 1 \quad \text{and}   \quad q_W(s) = (r+1) (1-sh)   (1-c) -1,  
\end{equation}
and the (common) velocity wave for which we recall the formula:
\begin{equation}\label{eq:common v}
    v(s) =  2 \sigma \sqrt{ (1-sh)(1+c) - 1 }.
\end{equation}
We differentiate each equation Eq.\eqref{eq:lambda_eq1} and Eq.\eqref{eq:lambda_eq2} with respect to $s$, and we find:
\begin{align}
    &     0 =   \sigma^2 2\lDb(s) \lDbprim(s)  + v(s) \lDbprim(s)  + \frac{d v}{ds}(s) \lDb(s) + \frac{d q_D}{ds}(s),  \\
    &    0 =   \sigma^2 2\lWb(s) \lWbprim(s)  + v(s) \lWbprim(s)  + \frac{d v}{ds}(s)  \lWb(s) + \frac{d q_W}{ds}(s), 
\end{align}
from which we deduce:
\begin{align}
    &  \frac{d}{ds} \left( \dfrac{1}{\lDb} \right)(s)   = \dfrac{ v'(s) \lDb(s)  + q_D'(s)}{ \lDb(s)^2 ( \sigma^2 2\lDb(s)  + v(s))} = F\left(q_D', \lDb\right), \\
    &   \frac{d }{ds}\left( \dfrac{1}{\lWb} \right)(s)  =  \dfrac{ v'(s) \lWb(s)  + q_W'(s)}{ \lWb(s)^2 (  \sigma^2 2\lWb(s)  + v(s))}  = F\left(q_W', \lWb\right).
\end{align}
We are going to use the monotonicity properties of $F$ combined with some useful properties about $q_D, q_W$ and $\lDb, \lWb$ in order to establish the chain of inequalities leading to the result 
\begin{equation*}
    F\left(q_D', \lDb\right) < F\left(q_D', \lWb\right) < F\left(q_W', \lWb\right). 
\end{equation*}
Firstly, the function $F$ is increasing with respect to the variable $\lambda$:
\begin{align}
\begin{split}
   \frac{\partial F}{\partial \lambda} (q, \lambda) & = \dfrac{v'  \lambda ^2  ( \sigma^2 2\lambda   + v ) - (  v'  \lambda   + q' ) (  \sigma^2 6 \lambda^2  +2  v \lambda )}{ \lambda^4 ( \sigma^2 2\lambda  + v)^2} \\
  &  = \dfrac{- 4  \sigma^2 v' \lambda^3 - v v' \lambda^2 - 6  \sigma^2 q' \lambda^2 - 2 v q' \lambda}{\lambda^4 ( \sigma^2 2\lambda  + v)^2} > 0.
\end{split}
\end{align}
In fact, each term in the above numerator is positive: $-v'(s)>0$ \eqref{eq:common v}, and $-q'(s)>0$, both for $q_D'$ and $q_W'$ \eqref{eq:coeff q}. When this partial monotonicity is combined with the aforementioned fact that $\lWb > \lDb$ (condition (16)), we get that
\begin{equation}\label{eq:ineq1}
    F\left(q_D', \lDb\right) < F\left(q_D', \lWb\right)
\end{equation}
Secondly, the function $F$ is clearly increasing with respect to the variable $q'$. Moreover, we have $q_D'(s) = - (r+1)< - (1-c)h (r+1) =  q_W'(s)<0$, because $(1-c) h < 1$. Therefore, we can deduce that
\begin{equation}\label{eq:ineq2}
    F\left(q_D', \lWb\right) < F\left(q_W', \lDb\right). 
\end{equation}
Bringing inequalities \eqref{eq:ineq1} and \eqref{eq:ineq2} together, we obtain:
\begin{equation}
    \frac{d}{ds} \left( \dfrac{1}{\lDb} \right)   =   F\left(q_D', \lDb\right) <  F\left(q_W', \lWb\right) = \frac{d }{ds}\left( \dfrac{1}{\lWb} \right)   ,
\end{equation}
from which we deduce \eqref{eq:result353} and the fact that $\ell$ is decreasing with respect to $s$.

\begin{remark}[Intrinsic fitness vs. fitness cost]
In the literature, the intrinsic fitness $\fint{D}$ is sometimes considered instead of the fitness cost $s$. The drive intrinsic fitness is the overall ability of drive alleles to increase in frequency when the drive proportion is close to zero. Its value is closely linked with $s$, and given by first line of system \eqref{eq:par_ger_nD_nW}:
\begin{equation}
    \fint{D} = (1-sh)(1+c)-1 \quad \iff \quad s = \dfrac{c-\fint{D} }{h(1+c)} .
\end{equation}
\lk{Increasing the fitness cost $s$ reduces the fitness $\fint{D}$,} so that any conclusion on the influence of $s$ can be immediately transferred to $\fint{D}$.  
\end{remark}

\subsection{\lk{Heuristics: Characterising the stochastic distance $L_{1}^{\mathscr{N}}$}}

We now consider the position of the last individual carrying a wild-type allele, for which stochastic effects are predominant (Figure \ref{fig:hist_deter_or_not}). The aim of this section is to draw a parallel with spatial Galton-Watson processes in a fixed frame, for which there exists a mathematical theory -- see for instance \cite{bertacchi2018}.

Galton-Watson processes describe a single isolated population in time, based on the assumption that individuals give birth and die independently of each other, and follow the same distribution for each of these events. It appears that Galton-Watson processes with migration are very sensitive to spatial parameters: as we work on a limited spatial domain, we focus on extinction time instead of distance to the last wild-type allele, less affected by this restriction.

We need to verify two assumptions before heuristically reducing our stochastic problem to a spatial Galton-Watson process: i) first, the stochastic distance $L_{1}^{\mathscr{N}}$ ($L^{100}_1$ in the following) can be transformed into an extinction time problem and ii) second, the dynamics of the wild-type alleles at the back of the wave can be approximated by a stochastic process of an ideal population consisting of wild-type alleles only in a isolated space.

In the following, the positions of the last wild-type allele and the extinction times are all computed conditionally to the absence of wild-type recolonisation in the global simulation: as we try to characterise the conditions preventing wild-type recolonisation in a realisable time window, we assume that recolonisation has not happened yet.

\subsubsection{Converting distances to extinction times} \label{subsubsec:dist_time}

We denote by $L_{0}^{100}$ the distance between the empty adjacent site on the left of the last wild-type allele, and the last position with more than $100$ wild-type alleles at the back of the wave. We have:

\begin{equation}
    L_{0}^{100} = L_{1}^{100} + \mathrm{d}x,
\end{equation}
with $\mathrm{d}x$ the size of a spatial step. We also define $T^{100}_0$ as the time that the last spatial site with more than $100$ wild-type alleles (at the back of the wave) takes to go extinct. 

\begin{figure}[H]
\centering
\begin{subfigure}{0.48\textwidth}
    \centering
    \caption{$s=0.3$}
    \includegraphics[width=\textwidth]{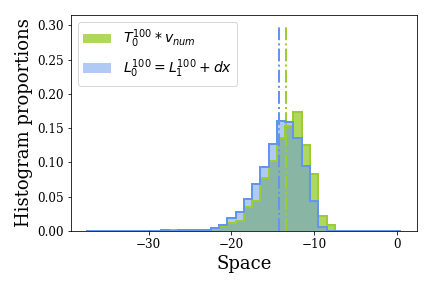}
\end{subfigure}
\hfill
\begin{subfigure}{0.48\textwidth}
    \centering
    \caption{$s=0.7$}
    \includegraphics[width=\textwidth]{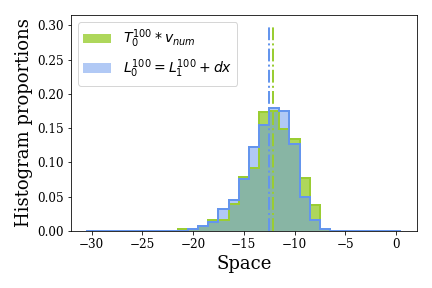}
\end{subfigure}
\caption{Superimposition of the statistical distributions of distance $L_{0}^{100}$ (in light blue) and extinction time $T^{100}_{0}$ multiplied by the speed of the wave $v_{\text{num}}$ (in light green). These distributions are calculated for $K=10^8$, over 500 different time points. The two distributions are very close for $s=0.3$, and for $s=0.7$. The distribution of $L_{1}^{100}$, the relative position of the last wild-type individual is asymmetric, like already observed in Figure~\ref{fig:hist_deter_or_not}; this asymmetry is preserved in the distribution of extinction time $T^{100}_{0}$ times speed $v_{\text{num}}$.}
\label{fig:distance_time}
\end{figure}

\lk{We take advantage of the propagation at a constant speed to convert the extinction time $T^{100}_0$ into a distance $v_{\text{num}} T^{100}_0$, where $v_{\text{num}}$ stands for the speed of the wave (as estimated from stochastic simulations). This conversion is equivalent to looking at the kymographs on Figure \ref{fig:illu_chasing} (bottom line) either from the horizontal or from the vertical perspective, with the corresponding multiplicative factor $v_{\text{num}}$. We hypothesize that $v_{\text{num}} T^{100}_0$ and $L_{0}^{100}$ follow similar distributions. To support such heuristics, we show in} Figure~\ref{fig:distance_time} the two distributions for two values of the drive fitness cost $s=0.3$ and $s=0.7$. In both cases, they appear very close from each other.

\subsubsection{From a global dynamics to an isolated population}\label{subsubsec:glob_to_iso}

We now ask whether the time $T^{100}_{0}$ (time that the last spatial site with more than 100 wild-type individuals at the back of the wave takes to go extinct) can be approximated by the extinction time $T^{100}_{0 \ \text{gw}}$ of a single isolated population modelled with an auxiliary stochastic stepping stone process. We consider a spatial Galton-Watson process in a fixed domain, where the population is distributed with an exponential spatial profile at initial time. At the back of the wave and in the absence of wild-type recolonisation, we assume that: \begin{equation}
    \dfrac{\n{D}^{t, x}}{n^{t,x}} \sim 1 \ , \ \dfrac{\n{W}^{t,x}}{n^{t,x}} \ll  1 \ \ \text{and} \ \ n^{t,x} \ll  1.
\end{equation}

We know from \eqref{eq:gW} that a wild-type allele at the back of the wave produces on average $\g{W}(\n{D}^{t,x}, \n{W}^{t,x})$ alleles (in offspring) during one time unit, with \begin{equation}
    \g{W}(\n{D}^{t,x}, \n{W}^{t,x}) = \left( r \ (1-\frac{n^{t,x}}{K})+1 \right)  \left[ \  \dfrac{\n{W}^{t,x}}{n^{t,x}}   +   (1-sh)  \  (1-c) \  \dfrac{\n{D}^{t,x}}{n^{t,x}} \right] \approx  (r+1)(1-sh)(1-c),
\end{equation}
and disappears on average at rate $1$. Thanks to this approximation, the dynamics do not depend on $\n{D}$ any more, and we can simulate a single isolated wild-type population. As before, at each time step, a wild-type allele migrates to the adjacent site on the right with probability $\frac{m}{2}$ and to the left with probability $\frac{m}{2}$.  

To be consistent in the comparison between $T^{100}_{0}$ (global dynamics) and $T^{100}_{0 \ \text{gw}}$ (single population approximation), we initiate the simulation with the profile $ \exp(z \ \lWbd)$ approximating the exponential decay of wild-type allele numbers at the back of the wave (Figure \ref{fig:zoom}). We also observe that migration from dense areas to less dense areas in the exponential profile plays a major role and significantly increases the extinction time $T^{100}_{0 \ \text{gw}}$ (Appendix \ref{app:migration}). Thus, we had to consider a large exponential initial condition with a maximum number of individuals per site being $10^6$, to fully capture the wild-type dynamics at the end of the wave.

\VC{In order to support our heuristical argument that the extinction time $T_0^{100}$ can be captured by a Galton-Watson process in a fixed domain, we ran 500 replicates for two values of the fitness cost, resp.  $s=0.3$ and $s=0.7$. We superimposed the two distributions of extinction times $T_0^{100}$ and $T^{100}_{0 \ \text{gw}}$ (both multiplied by the same $v_{\text{num}}$ in order to compare spatial distributions as above). The histograms are very close for $s=0.3$ and $s=0.7$. This preliminary result paves the way for further analysis; if one was able to characterise analytically extinction time $T^{100}_{0 \ \text{gw}}$ of this spatial Galton-Watson process, we could also have access to an analytical approximation of distance $L^{100}_{0}$ and consequently, an analytical condition ensuring the absence of wild-type recolonisation event with high probability in a realisable time window.}

\begin{figure}[H]
   \begin{subfigure}{0.48\textwidth}
    \centering
    \caption{$s=0.3$} 
    \includegraphics[width = 0.9\textwidth]{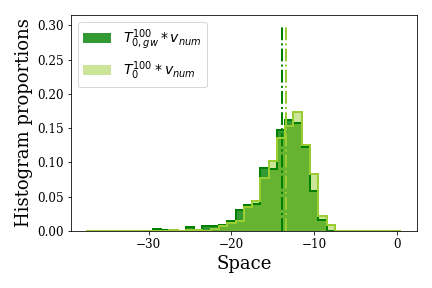}      
\end{subfigure}
\hfill
\begin{subfigure}{0.48\textwidth}
    \centering
     \caption{$s=0.7$}
    \includegraphics[width = 0.9\textwidth]{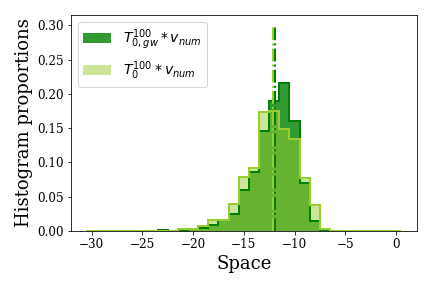}      
\end{subfigure}
    \caption{Superimposition of the statistical distributions of extinction times $T^{100}_{0}$ (in light green) and $T^{100}_{0 \ \text{gw}}$ (in dark green), multiplied by the speed of the wave $v_{\text{num}}$ to obtain distances. The extinction time $T^{100}_{0}$ is collected over 500 different time points in the global dynamics, while $T^{100}_{0 \ \text{gw}}$ is obtained running the spatial Galton-Watson process detailed in the main text 500 times. The distribution of $T^{100}_{0 \ \text{gw}}$ (single population approximation) fits very well the distribution of $T^{100}_{0}$ (global dynamics) and the asymmetry is again preserved.}
    \label{fig:gw_time}
\end{figure}

\subsection{Numerics}

\subsubsection{The influence of the fitness cost $s$ and the carrying capacity $K$}

In Figure \ref{fig_back_wave}(g), the probability to observe one wild-type recolonisation event in a realisable time window is extremely low when $s=0.3$: the furthest away the last individual carrying a wild-type allele might be is a spatial site with more than $10^4$ drive alleles. However, when $s=0.7$, this very last position corresponds to a number of drive alleles below $100$ (Figure \ref{fig:hist_deter_or_not} (b)), and consequently, a non-negligible chance of wild-type recolonisation. 

\begin{figure}[H]
\centering
\includegraphics[width=0.6\textwidth]{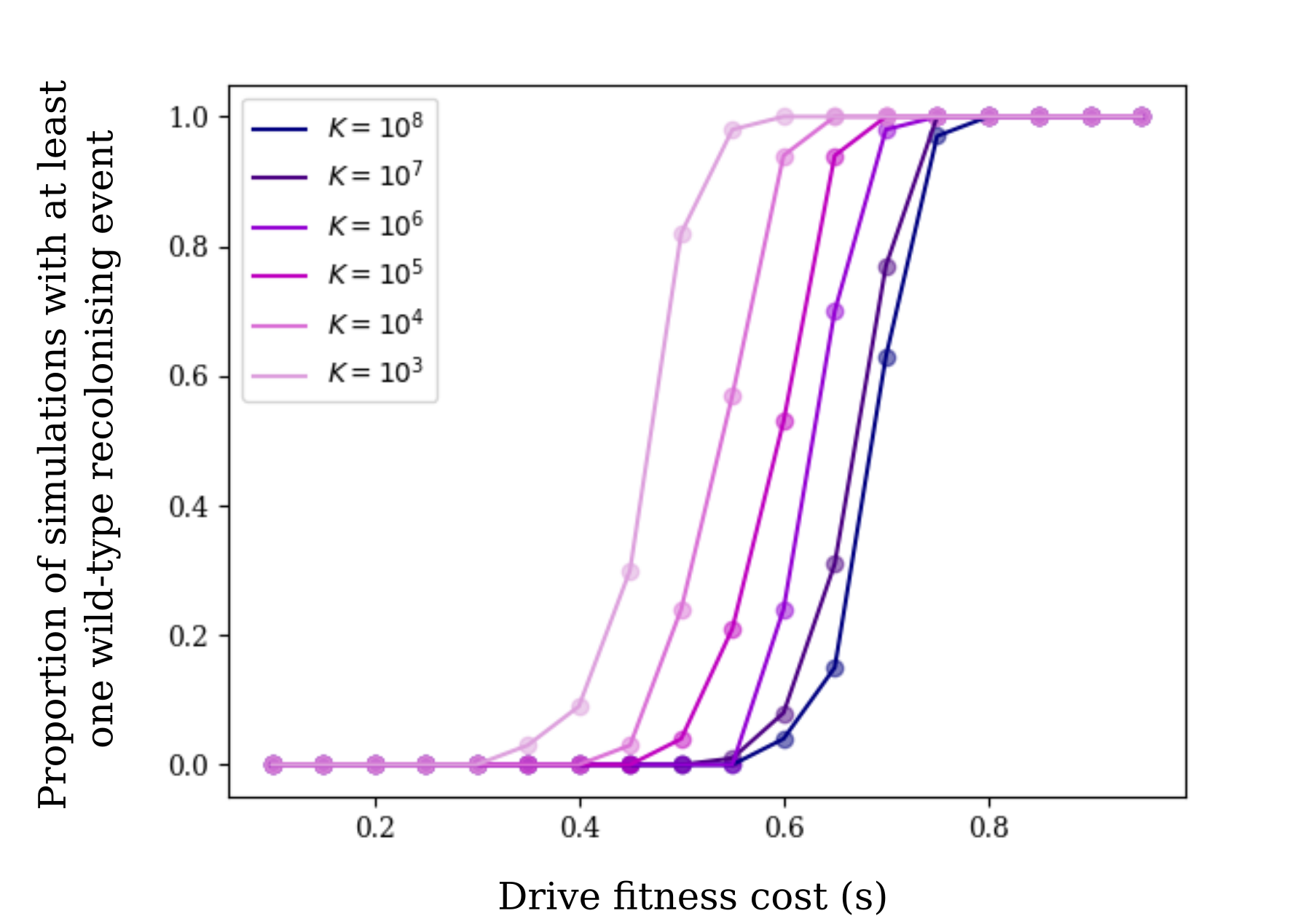}
\caption{Proportion of simulations encountering at least one wild-type recolonisation event within $1000$ units of time, as a function of $s$, the fitness disadvantage for drive, and $K$, the carrying capacity on one spatial step. For each point, we ran $100$ simulations. When $K=10^8$, the probability to observe wild-type recolonisation over $1000$ units of time is close to $0$ for $s=0.3$, and approximately $0.63$ when $s=0.7$.}
\label{fig:chasing_s_K}
\end{figure}

In Figure \ref{fig:chasing_s_K}, we numerically approximate the probability to observe wild-type recolonisation within $1000$ units of time, for different values of $s$ the drive fitness cost and $K$ the local carrying capacity. Each point of the graph is determined by the proportion of replicates where we observe wild-type recolonisation, over $100$ replicates. This probability increases with $s$ and decreases with $K$. Noticeably for a given local carrying capacity $K$, the transition between very low ($<10\%$) and very high ($>90\%$) chances of wild-type recolonisation within $1000$ units of time when the fitness cost $s$ varies, is relatively restricted: these two extreme conditions can be reached at two different $s$ values within a range of $0.2$.

\subsubsection{\VC{The role of the migration rate $m$}}

\begin{figure}[H]
\centering
\includegraphics[width=0.9\textwidth]{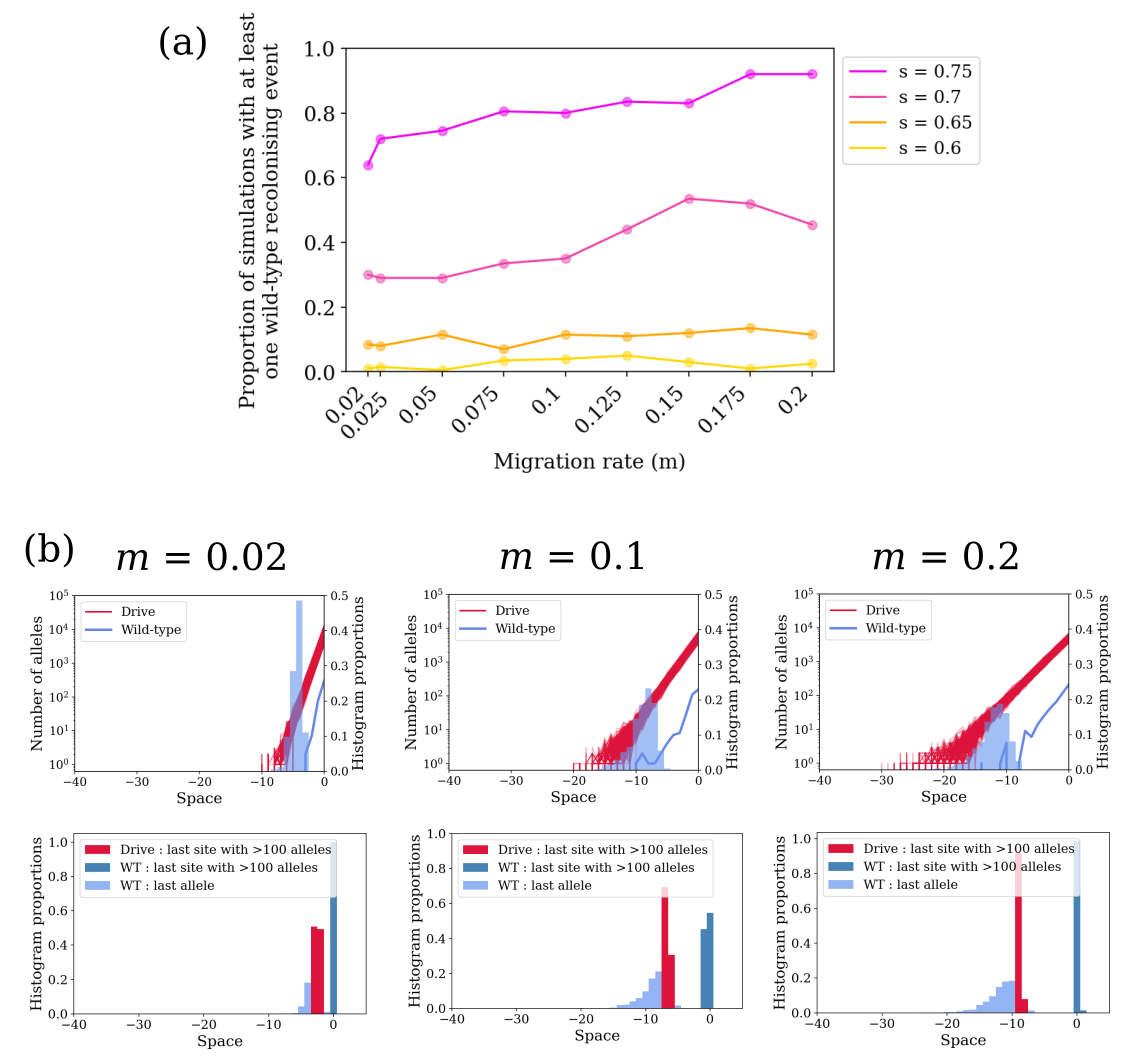}
\caption{\lk{Panel (a) is the analog of Figure \ref{fig:chasing_s_K} when the migration rate $m$ and the drive fitness cost $s$ vary. It shows the proportion of simulations encountering at least one wild-type recolonisation event within $1000$ units of time. For each point, we ran $200$ simulations. We observe a slight increase in the probability of wild-type recolonisation as $m$ increases; however, this effect is less clear and weaker than the variation induced by the drive fitness cost $s$, and appears rather sensitive to stochasticity. Panel (b) shows the back of the wave (in the absence of wild-type recolonising events) for a migration rate $m=0.02$, $0.1$ and $0.2$, and a drive fitness cost $s = 0.7$. The first row of panel (b) is the analog of Figure \ref{fig:6_histogram3}; it shows the superimposition of $2000$ drive allele waves (in red), a unique wild-type allele wave (in blue) and the histogram of the spatial distribution of the last wild-type allele (in lighter blue). The second row of panel (b) is the analog of Figure \ref{fig:hist_deter_or_not}, and shows the relative positions of the last spatial site with more than $100$ drive alleles at the back of the wave (in red), the last spatial site with more than $100$ wild-type alleles at the back of the wave (in blue) and the last wild-type allele at the back of the wave (in lighter blue, identical to the one in the first row). When the migration rate $m$ increases, the distance $\ell$ between the last spatial site with more than $100$ drive alleles and the last spatial site with more than $100$ wild-type alleles also increases, as proven in Section \ref{subsec:deter_dist}. However, this effect is counteracted by the fact that the distribution of the position of the last wild-type allele becomes more dispersed as the migration rate $m$ increases (enhanced stochastic effects). These two effects combined, we observe an overall less clear (slightly increasing) probability of wild-type recolonisation event when the migration rate $m$ increases (panel (a)).} }
\label{fig:chasing_m_s}
\end{figure}

\VC{Figure \ref{fig:chasing_m_s} (a) is the analog of Figure \ref{fig:chasing_s_K} when the migration rate $m$ and the fitness cost $s$ vary. We observe a slight increase in the probability of recolonisation as $m$ increases, but with a lower amplitude than under the variation of the fitness cost. This slight increase contrasts with our analytical conclusions predicting a decrease of the probability of recolonisation. Indeed, the analysis above informs us that the deterministic distance $\ell$ is increasing with the diffusion coefficient $\sigma^2$ (hence, the migration rate $m$).}

\VC{Therefore, the deterministic components of the wild-type and drive sub-populations are further separated when migration increases, suggesting a smaller probability of recolonisation. However, stochastic effects are enhanced under higher migration, leading to a larger spreading of the position of the last wild-type individual, suggesting a larger probability of recolonisation. These opposite trends can be viewed on the distribution of individuals in the stochastic simulations (see \ref{fig:chasing_m_s} (b)). Numerically, we find that the two effects counteract each other, leading to a slowly varying probability of recolonisation.}

\subsection{Comparison with the two-dimensional case} \label{sec:1D-2D}

We extend the 1D stochastic model to a 2D version, using the same probabilistic framework. Traveling waves now move in multiple directions at the same time, instead of one.

\begin{figure}[H]
\begin{subfigure}[t]{0.99\textwidth}
\centering
\caption{$K =10^8, s=0.3$} \label{fig:1D-2D_K8s3}
\includegraphics[width=0.18\textwidth]{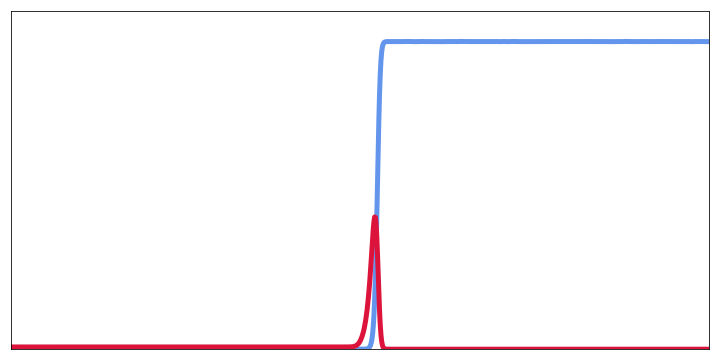} % 1D  K=10^8 et s=0.3
\includegraphics[width=0.18\textwidth]{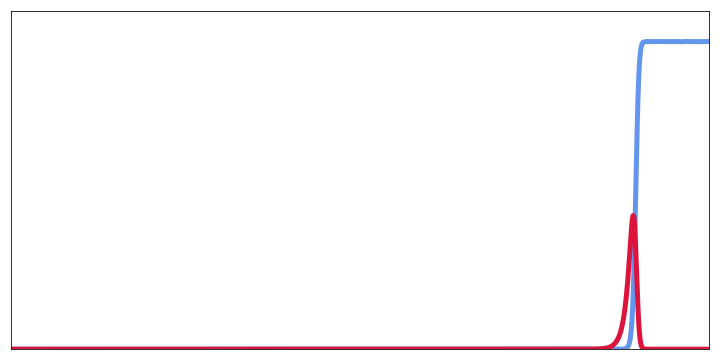}
\includegraphics[width=0.18\textwidth]{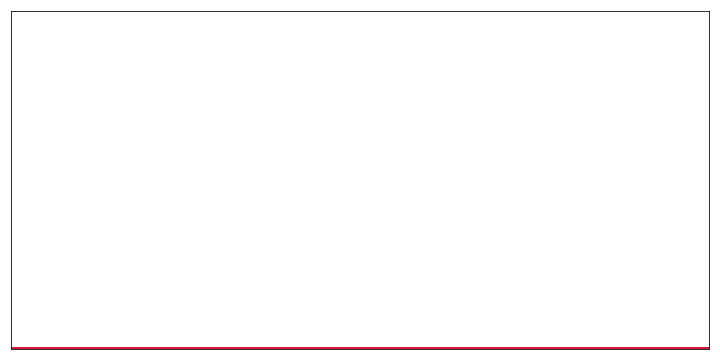}
\includegraphics[width=0.18\textwidth]{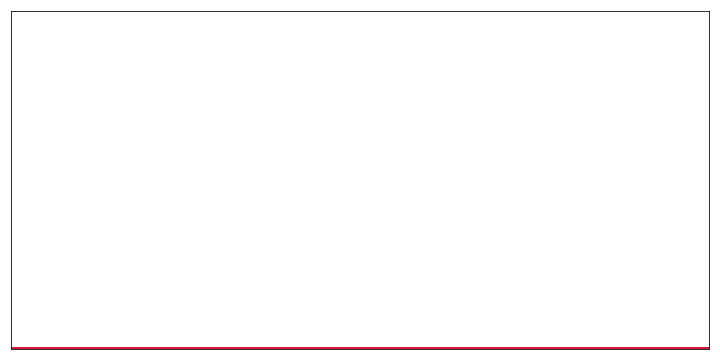}
\includegraphics[width=0.18\textwidth]{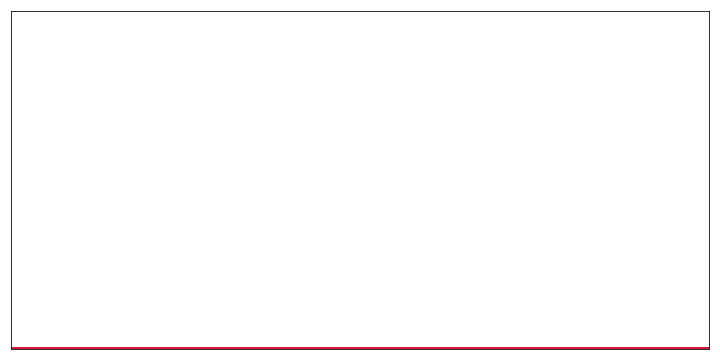}\\
\includegraphics[width=0.18\textwidth]{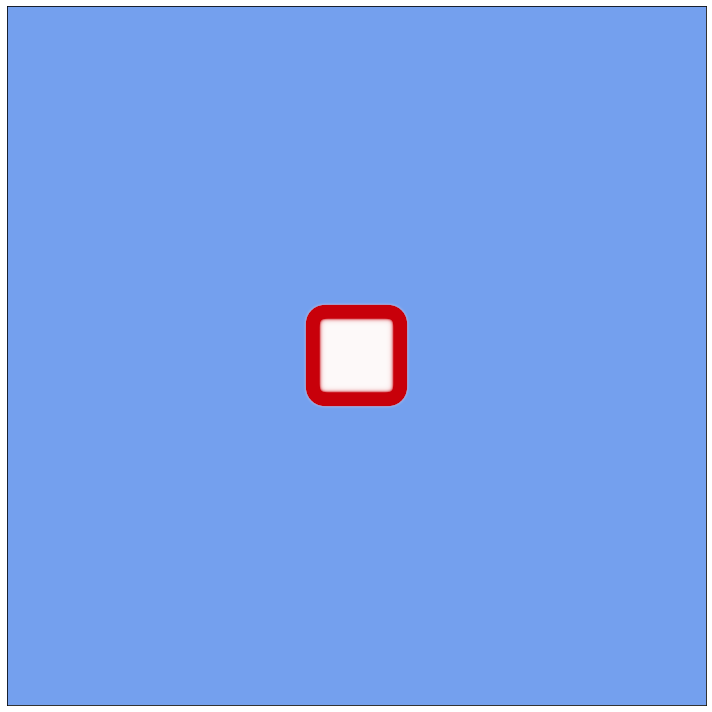} % 2D  K=10^8 et s=0.3
\includegraphics[width=0.18\textwidth]{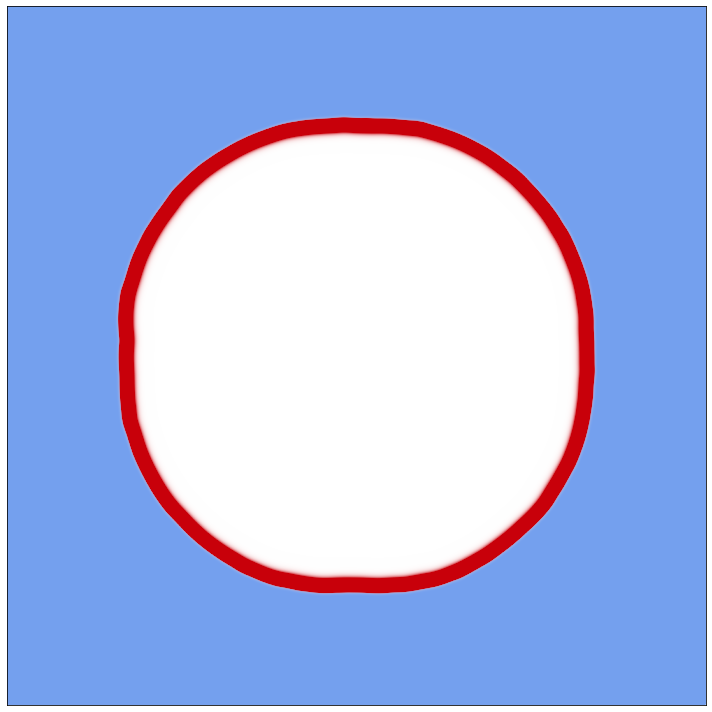}
\includegraphics[width=0.18\textwidth]{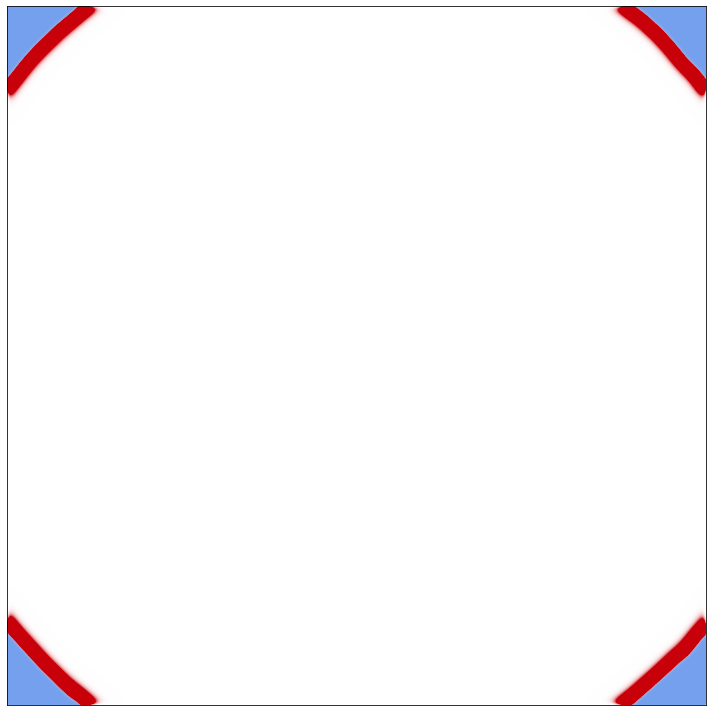}
\includegraphics[width=0.18\textwidth]{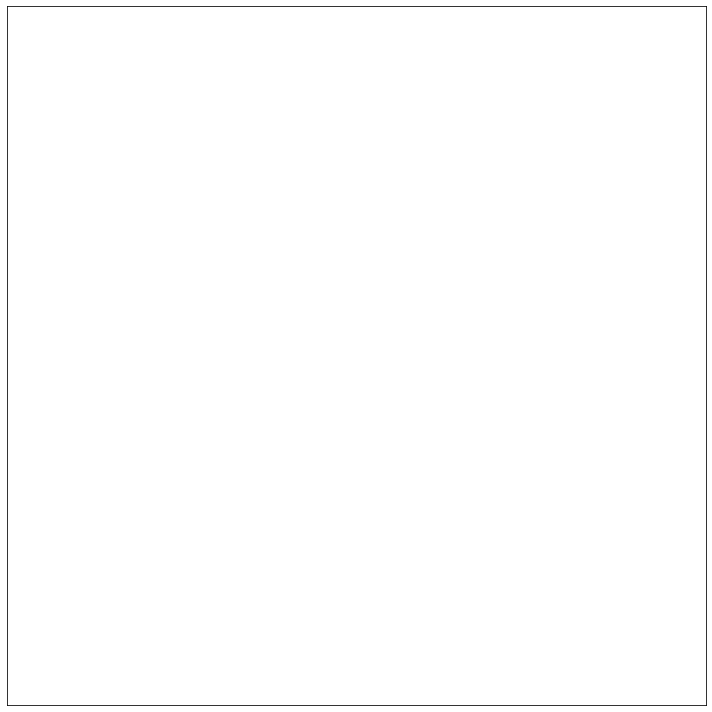}
\includegraphics[width=0.18\textwidth]{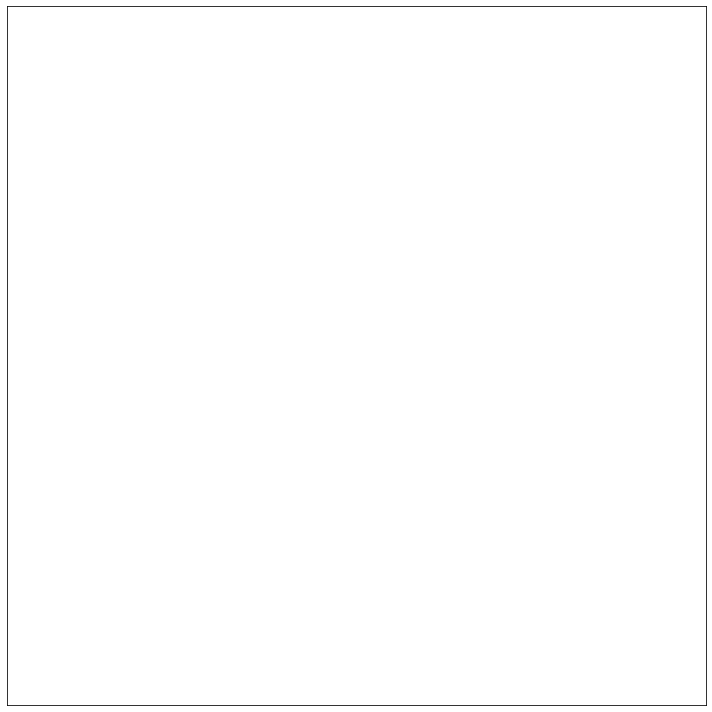}\\
\end{subfigure}
\begin{subfigure}[t]{0.99\textwidth}
\centering
\caption{$K=10^5, s=0.3$} \label{fig:1D-2D_K5s3}
\includegraphics[width=0.18\textwidth]{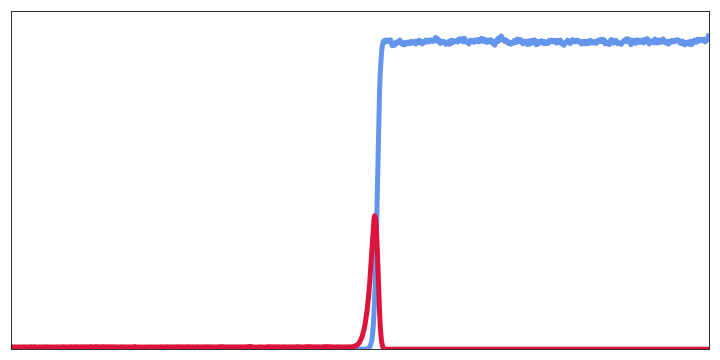} % 1D  K=10^5 et s=0.3
\includegraphics[width=0.18\textwidth]{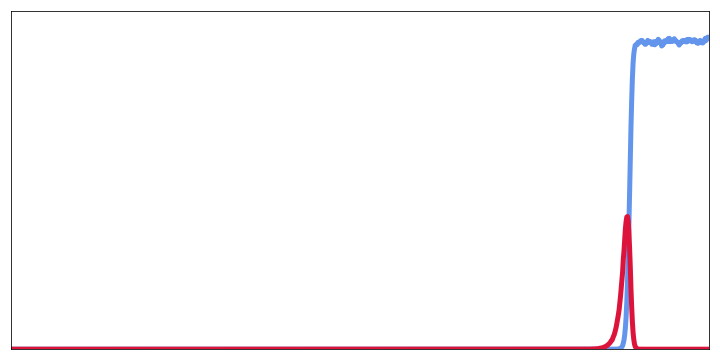}
\includegraphics[width=0.18\textwidth]{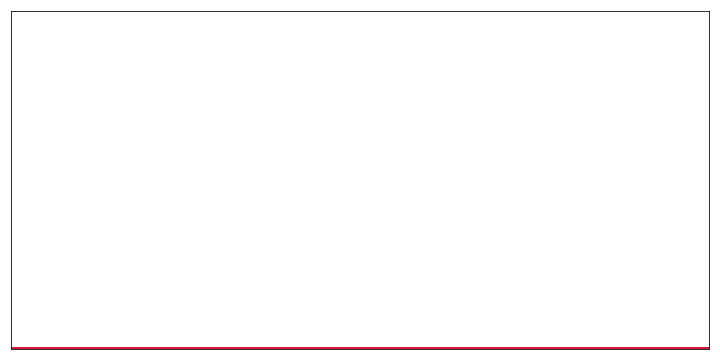}
\includegraphics[width=0.18\textwidth]{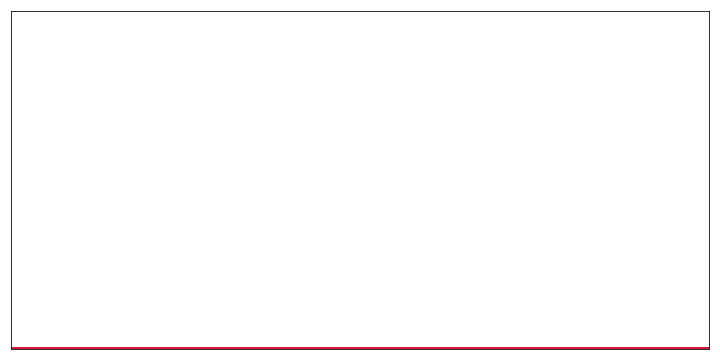}
\includegraphics[width=0.18\textwidth]{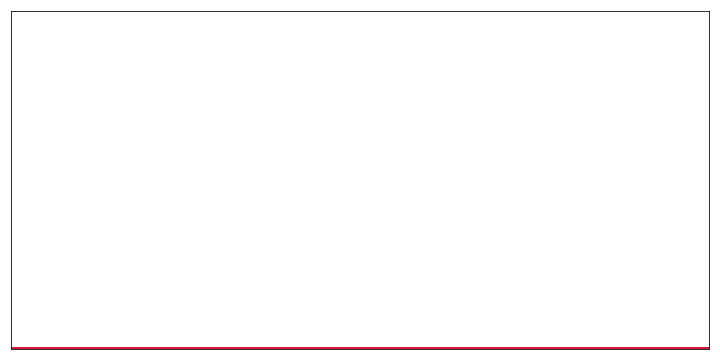}\\
\includegraphics[width=0.18\textwidth]{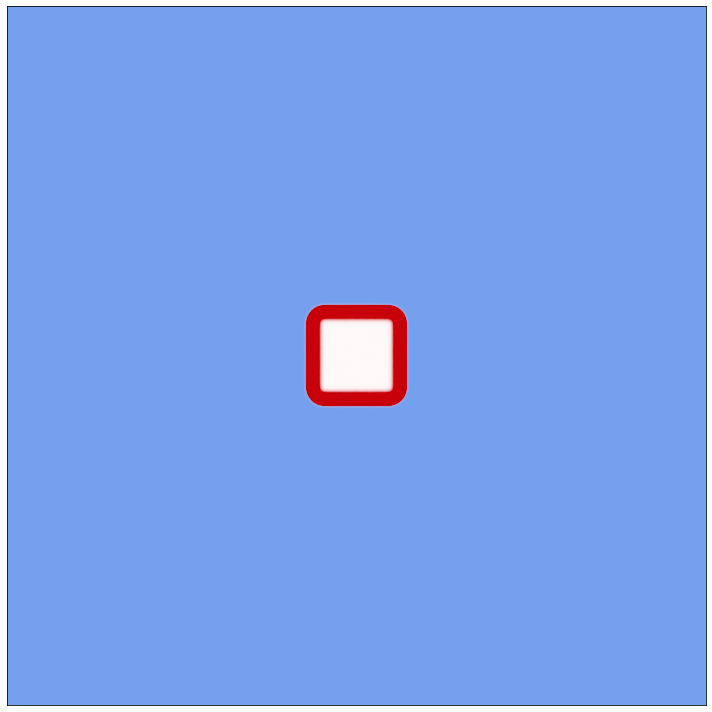} % 2D  K=10^5 et s=0.3
\includegraphics[width=0.18\textwidth]{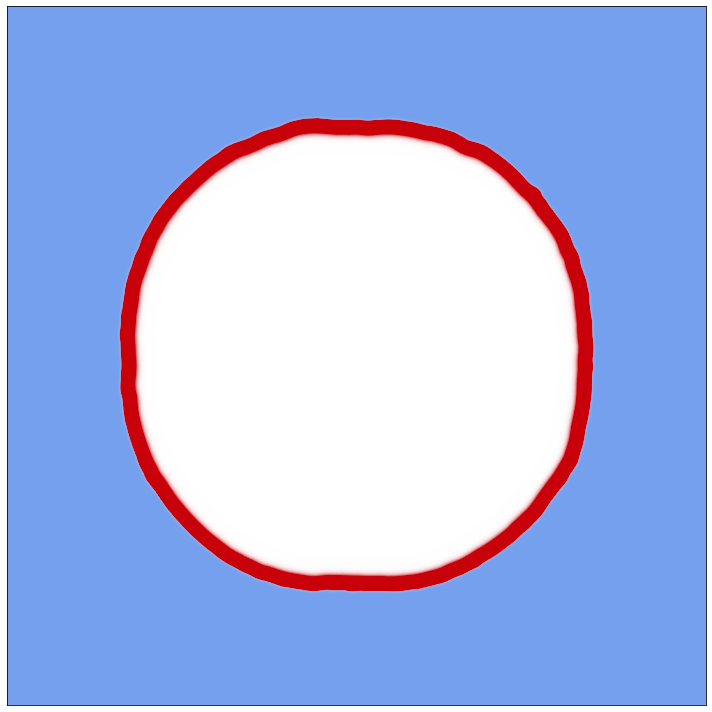}
\includegraphics[width=0.18\textwidth]{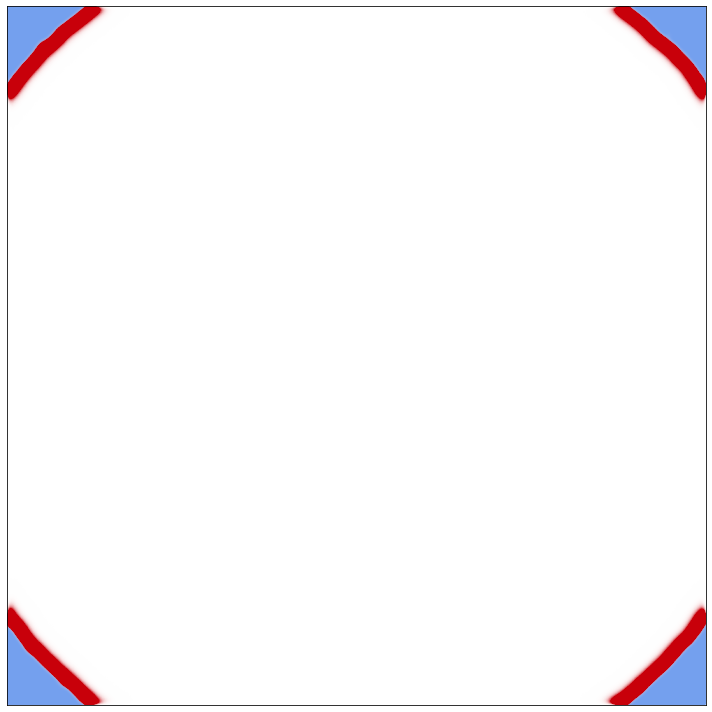}
\includegraphics[width=0.18\textwidth]{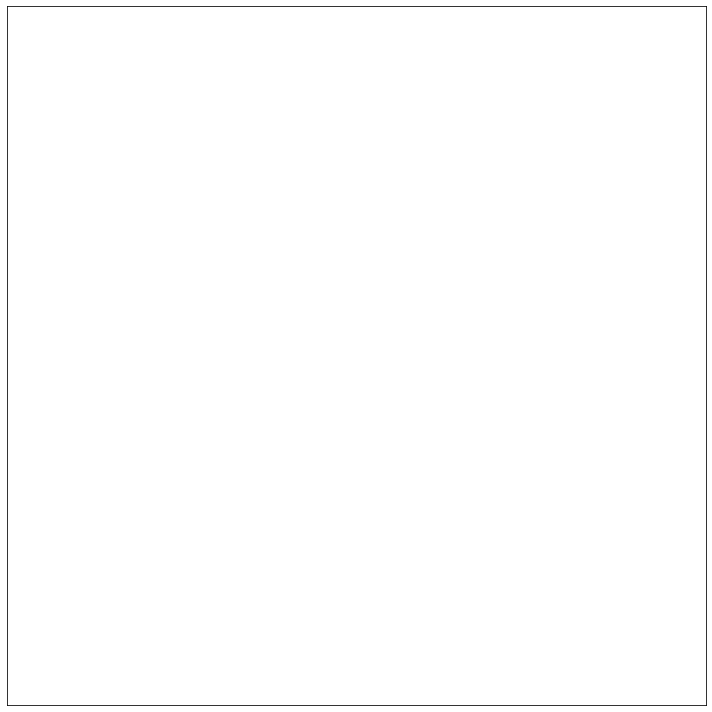}
\includegraphics[width=0.18\textwidth]{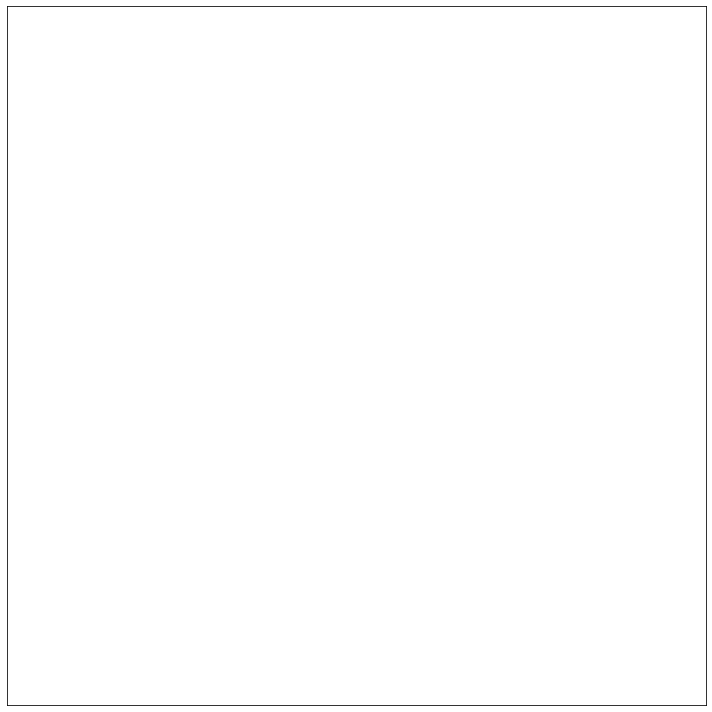}\\
\centering
\caption{$K=10^8, s=0.7$} \label{fig:1D-2D_K8s7}
\includegraphics[width=0.18\textwidth]{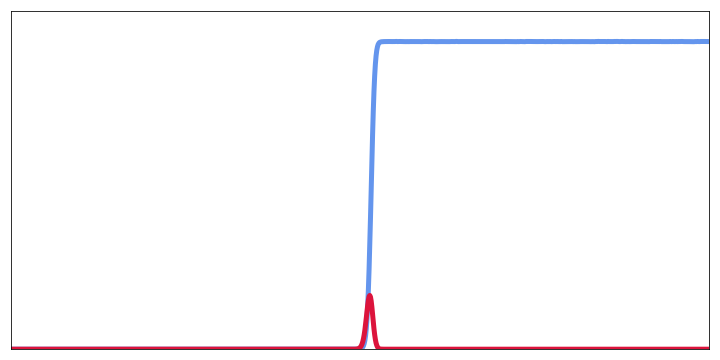} % 1D  K=10^8 et s=0.7
\includegraphics[width=0.18\textwidth]{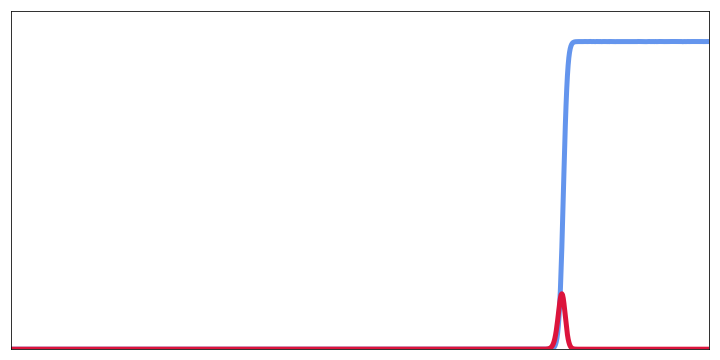}
\includegraphics[width=0.18\textwidth]{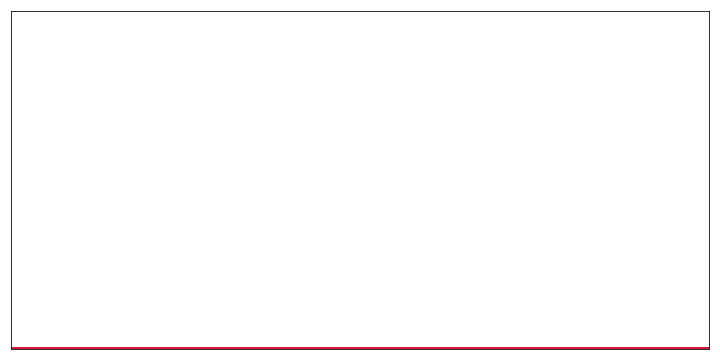}
\includegraphics[width=0.18\textwidth]{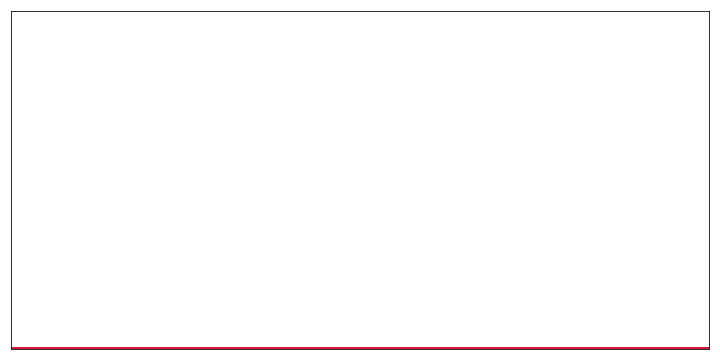}
\includegraphics[width=0.18\textwidth]{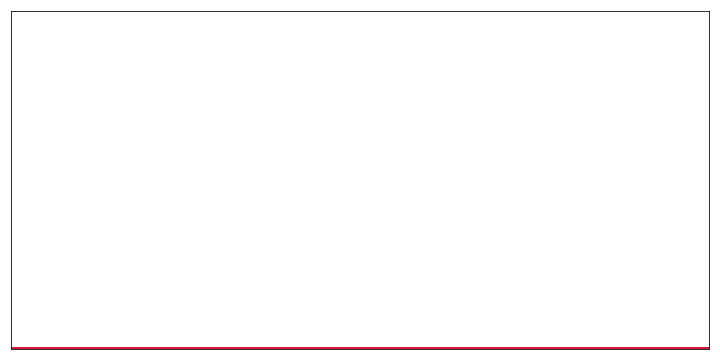}\\
\includegraphics[width=0.18\textwidth]{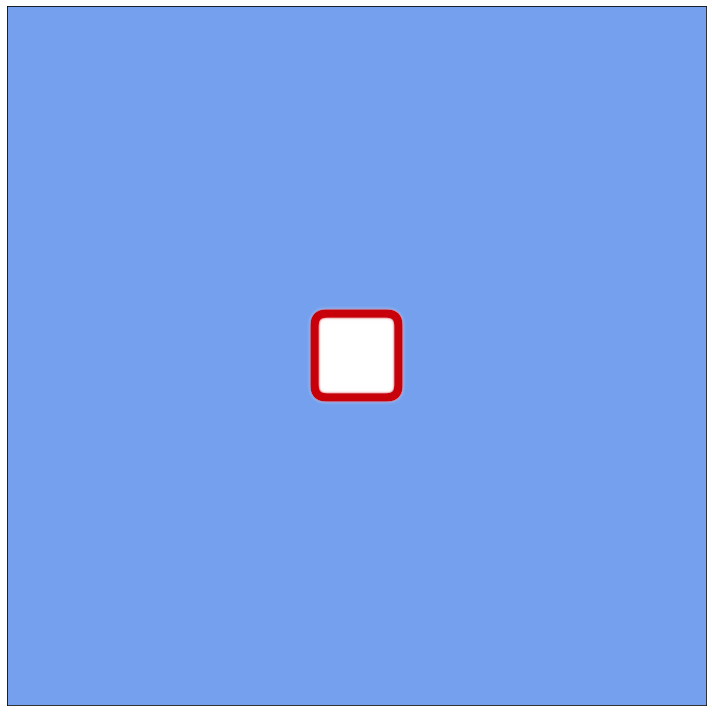} % 2D  K=10^8 et s=0.7
\includegraphics[width=0.18\textwidth]{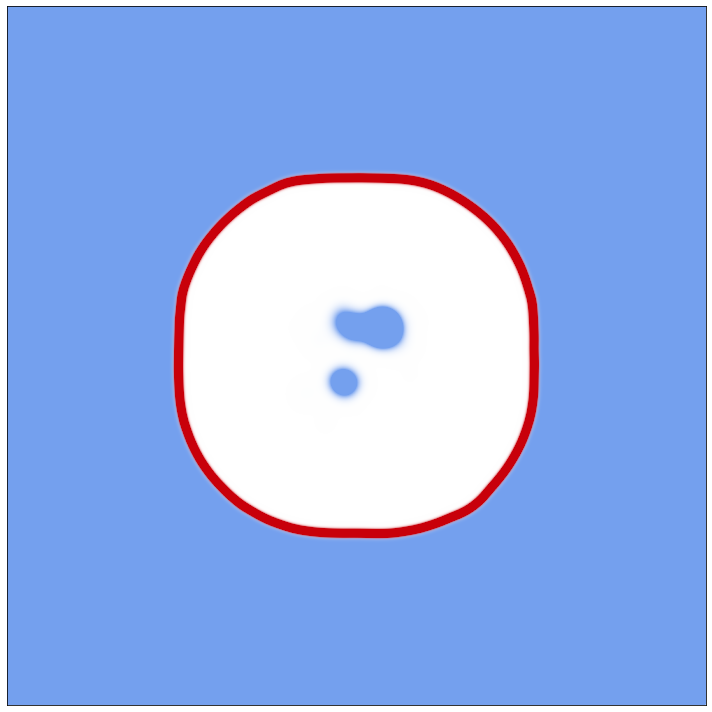}
\includegraphics[width=0.18\textwidth]{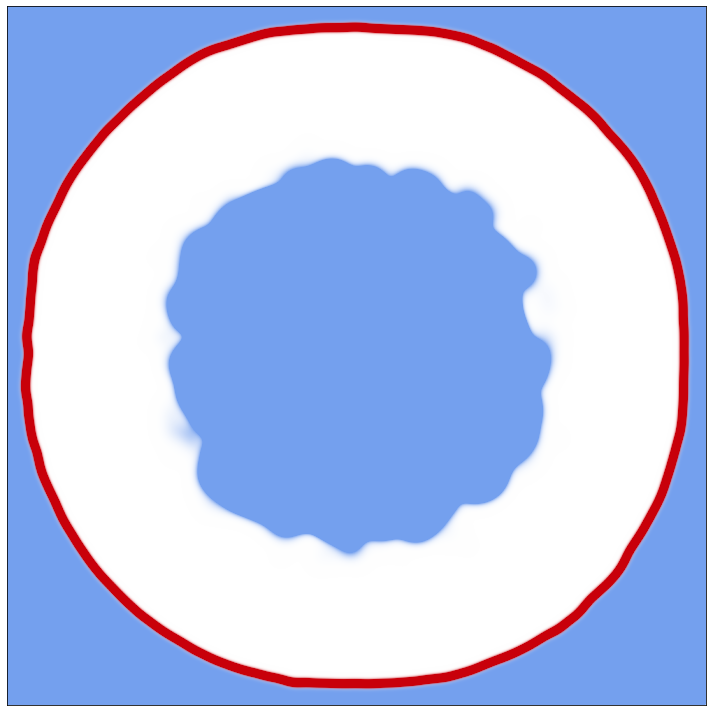}
\includegraphics[width=0.18\textwidth]{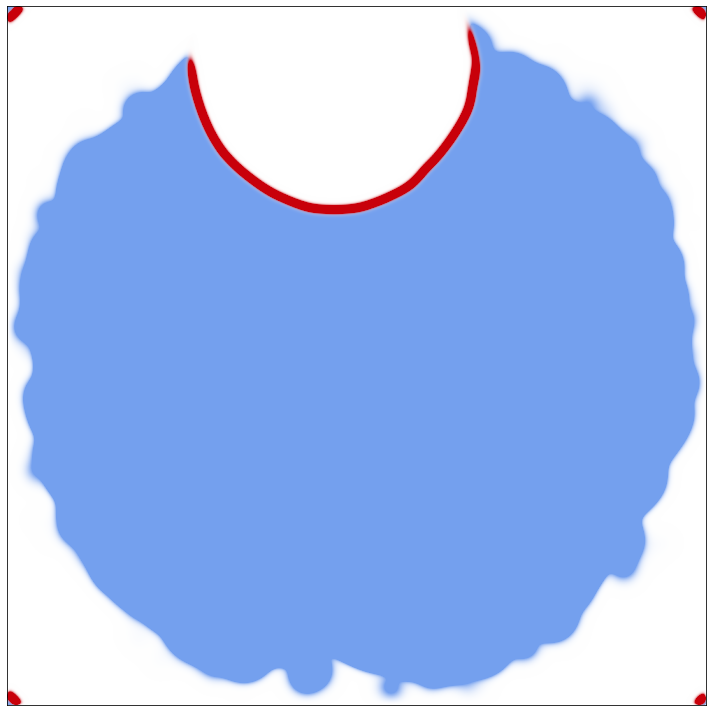}
\includegraphics[width=0.18\textwidth]{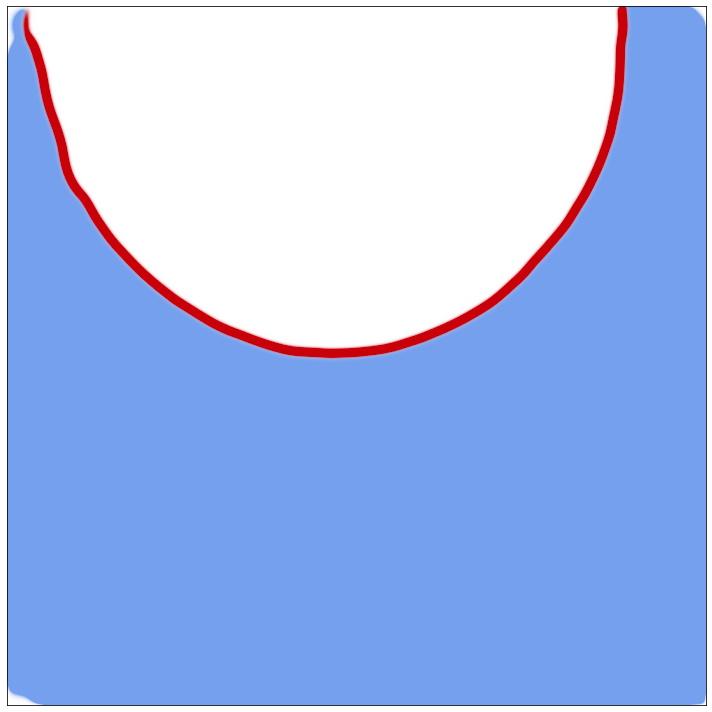}\\
\end{subfigure}
\begin{subfigure}[t]{0.99\textwidth}
\centering
\caption{$K=10^5, s=0.7$} \label{fig:1D-2D_K5s7}
\includegraphics[width=0.18\textwidth]{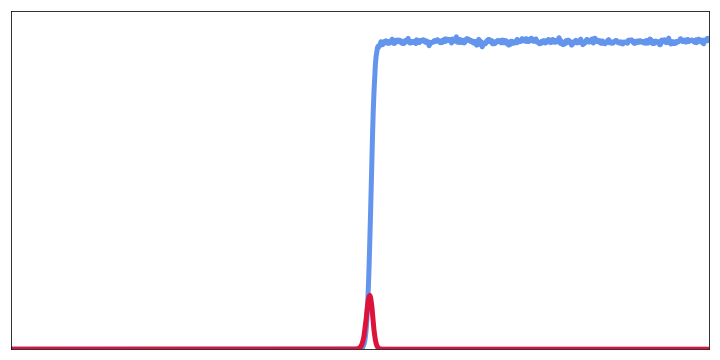} % 1D  K=10^5 et s=0.3
\includegraphics[width=0.18\textwidth]{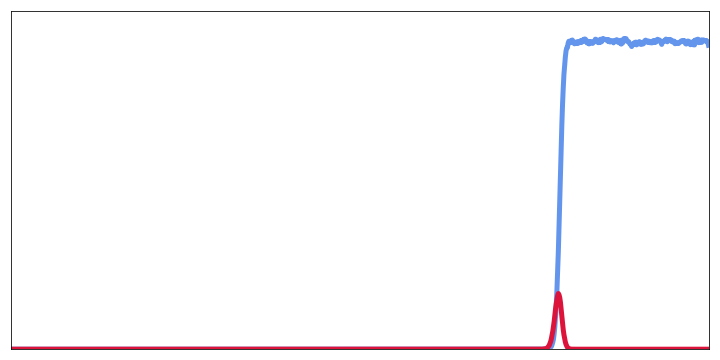}
\includegraphics[width=0.18\textwidth]{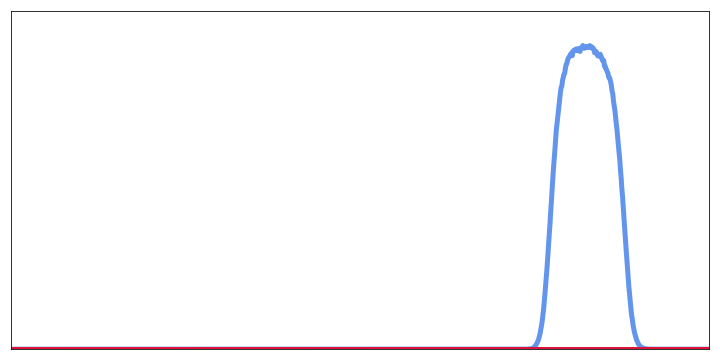}
\includegraphics[width=0.18\textwidth]{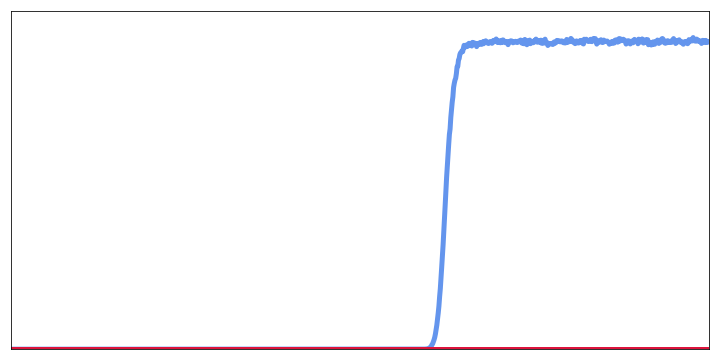}
\includegraphics[width=0.18\textwidth]{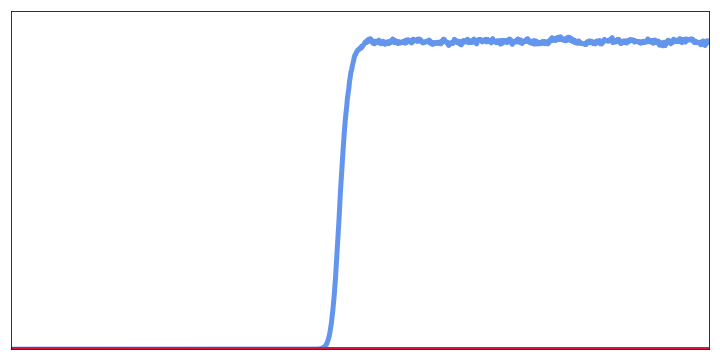}\\
\includegraphics[width=0.18\textwidth]{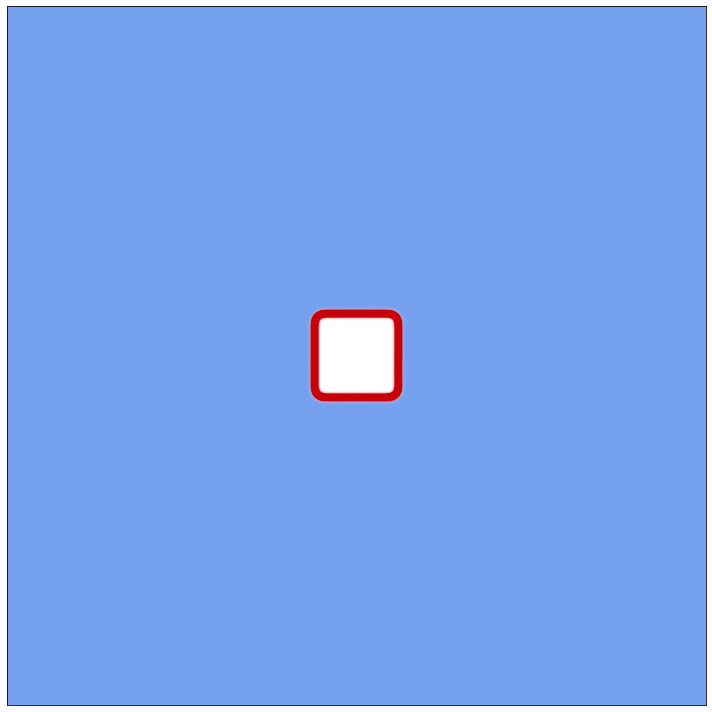} % 2D  K=10^5 et s=0.3
\includegraphics[width=0.18\textwidth]{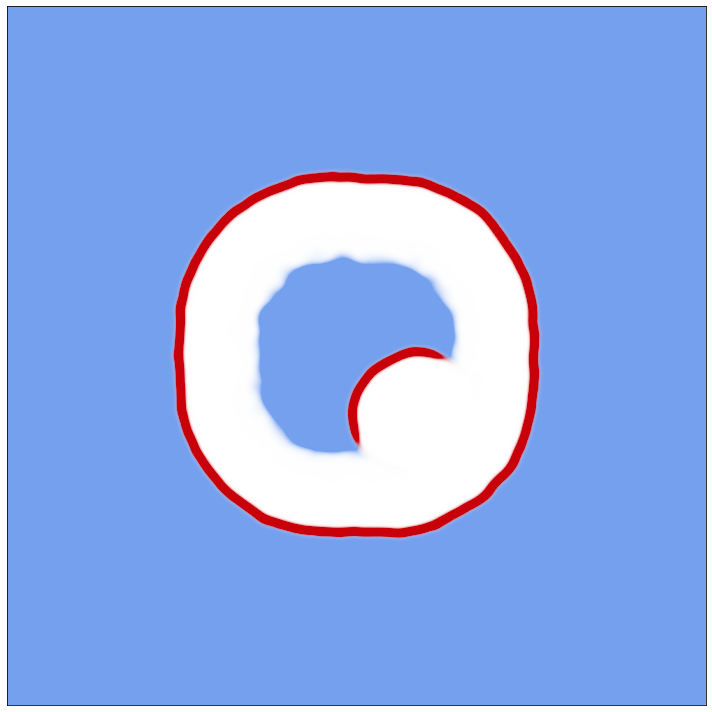}
\includegraphics[width=0.18\textwidth]{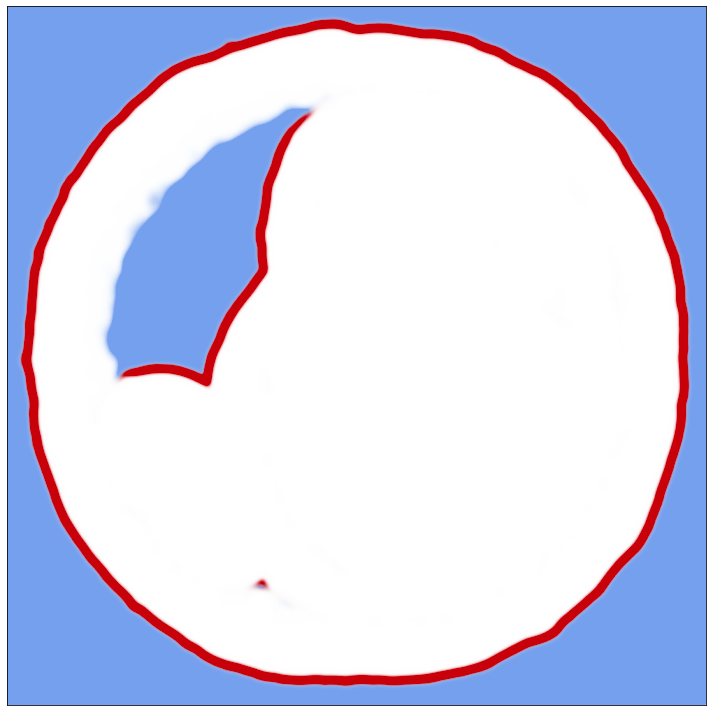}
\includegraphics[width=0.18\textwidth]{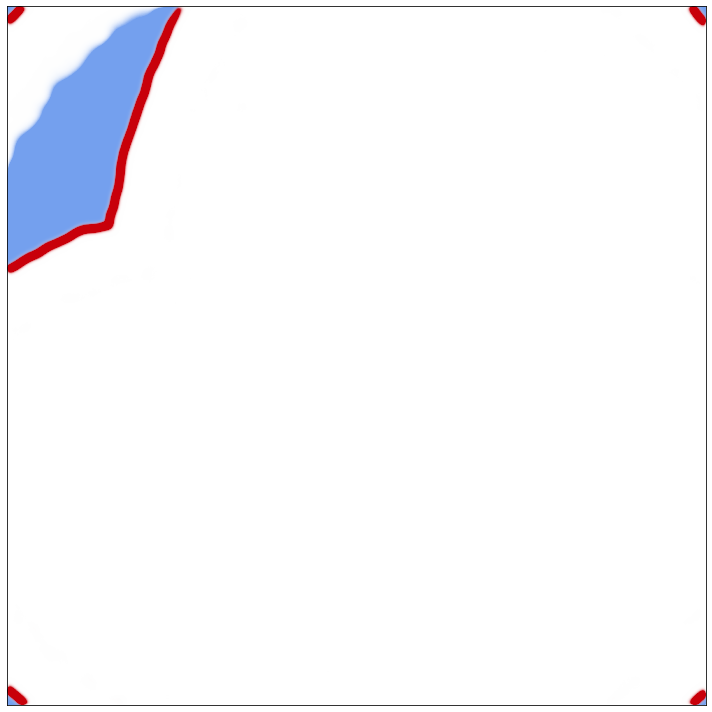}
\includegraphics[width=0.18\textwidth]{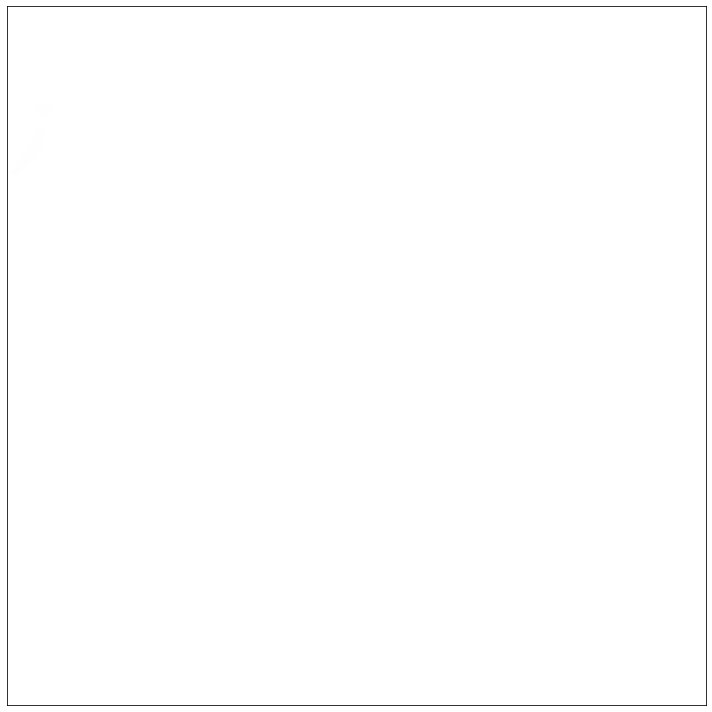}\\
\end{subfigure}
\caption{Comparison between the 1D and the 2D stochastic models, using the same numerical framework, with drive fitness cost $s=0.3$ or $s=0.7$ and carrying capacity $K=10^3$ ou $K=10^5$. Wild-type alleles are shown in blue and drive alleles in red. In 1D, the y-axis represents the number of alleles, while in 2D, the number of alleles is shown in shades of colour: the darker the colour, the more alleles of the corresponding type on the spatial site. For small drive fitness cost $s=0.3$, both 1D and 2D simulations show similar outcomes, i.e. the full eradication of the population. For high drive fitness cost $s=0.7$, the system is more sensitive to wild-type recolonisation events. The 2D domain amplifies this discrepancy: it facilitates the emergence of rare events as they can occur in a variety of directions each time. As expected, a larger carrying capacity reduces the stochasticity and consequently the probability of wild-type recolonisation event, both in 1D and 2D.}
\label{fig:1D-2D}
\end{figure}

In Figure \ref{fig:1D-2D}, we compare the numerical outcomes for medium versus large carrying capacity ($K=10^5$ and $K=10^8$), and small versus high fitness cost ($s=0.3$ and $s=0.7$). Similarly to the one-dimensional case, we observe in 2D that the probability of wild-type recolonisation events increases with higher drive fitness cost and smaller carrying capacity. Note that $\lDb$ and $\lWb$ formulas stay unchanged in 2D, meaning that all analytical results from Section \ref{subsec:deter_dist} still hold. 

The main discrepancy between 1D and 2D simulations relies on the fact that rare stochastic events tend to happen on average faster in 2D than in 1D, as they can occur in a variety of directions each time. For small drive fitness cost, the probability of wild-type recolonisation event is so low that 1D and 2D simulations result in the same outcome of complete extinction (Figure \ref{fig:1D-2D}(a,b)). However for higher fitness cost, wild-type recolonisation events occur faster in 2D than in 1D (Figure \ref{fig:1D-2D}(c,d)). Similarly, we only observe drive reinvasion event(s) in the 2D simulations within this time window (Figure \ref{fig:1D-2D}(c,d)). Additional simulations in line with these observations are shown in Appendix \ref{ann:2D_smallK}, with a smaller carrying capacity $K=10^3$.  

\subsection{From wild-type recolonisation events to chasing dynamics}\label{sec:WT_VS_chasing}

\lk{So far, we have focused on a condition under which wild-type recolonisation is unlikely. This corresponds to the outcomes of Figure \ref{fig:illu_chasing}(a), or Figures \ref{fig:1D-2D_K8s3}-\ref{fig:1D-2D_K5s3}. Interestingly, in the 2D case, Figure \ref{fig:1D-2D_K5s7} exhibits a different scenario of eradication occurring after at least one recolonisation event, and subsequent drive re-invasion. }

\begin{figure}[H]
    \centering
    \includegraphics[width=0.8\textwidth]{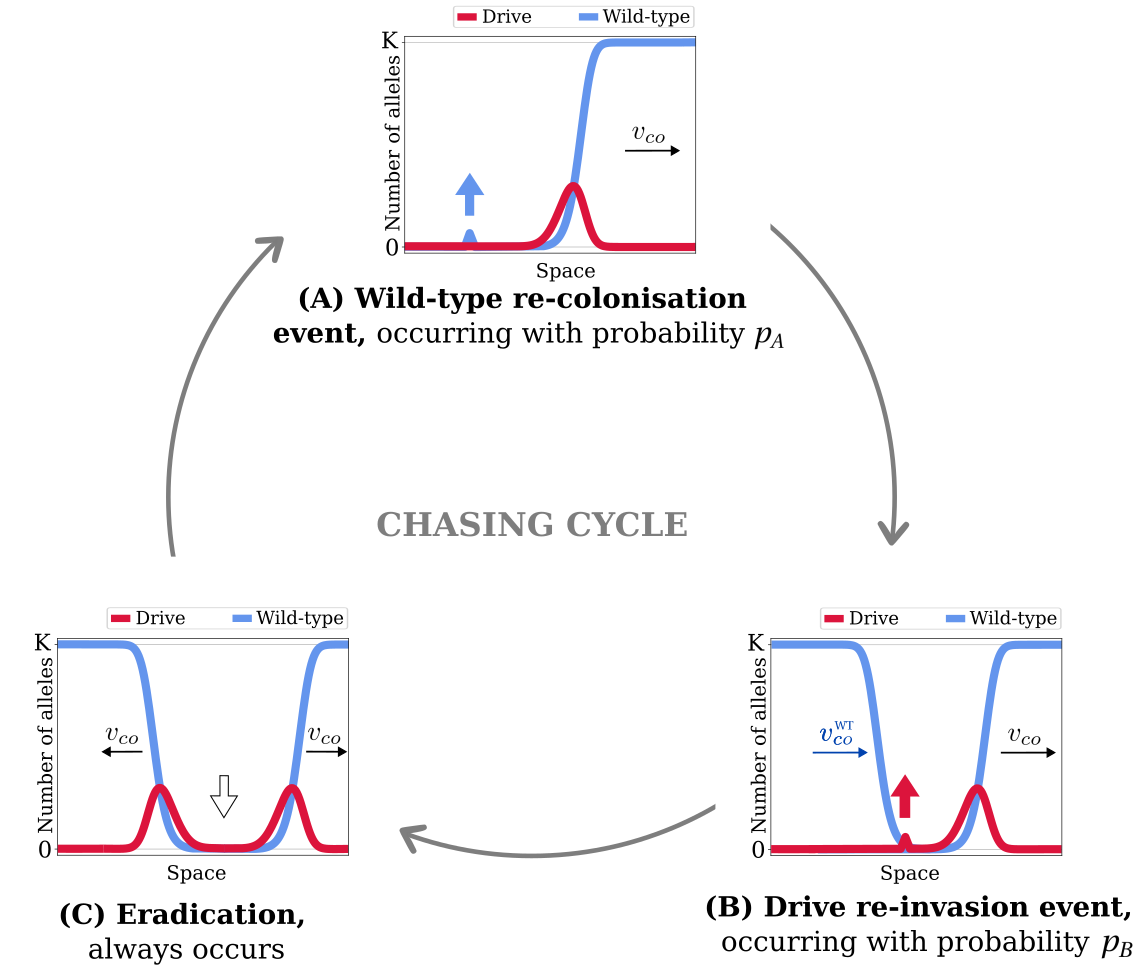}
    \caption{Illustration of the the three-step cycle underlying chasing dynamics. Wild-type recolonisation event (A) and drive reinvasion event (B) are stochastic events occurring with probabilities $p_A, p_B \in (0,1)$, while the eradication following the drive (re)invasion (C) occurs with probability $p_C = 1$ under Condition \eqref{eq:extinction}. If $p_A \gg p_B$, we would expect a full wild-type recolonisation, whereas if $p_B \gg p_A$, we would expect the population to be fully eradicated. Infinite chasing dynamics might emerge when $p_A$ and $p_B$ get closer.}
    \label{fig:schema_chasing}
\end{figure}

\lk{To analyse this scenario, we decompose it into a  three-step cycle as in Figure \ref{fig:schema_chasing}. }

\lk{Step $A$ (wild-type recolonisation, see Figure~\ref{fig:schema_chasing}) is followed by a phase where two waves propagate in the same direction at different speeds (see Figure \ref{fig:illu_chasing}(b)): a pure wild-type wave propagating at speed $\vcwt$, and a composite wild-type/drive wave propagating at speed $\vc$, where 
\begin{equation}\label{eq:speed_comparison}
  \vcwt = 2 \sigma \sqrt{r}  \quad \quad \text{and} \quad \quad  \vc =  2 \sigma \sqrt{ (1-sh)(1+c) - 1 }.
\end{equation}
The speed value $\vcwt = 2 \sigma \sqrt{r}$ is the one of a single population with intrinsic growth rate $r$ invading an empty environment (in the absence of drive individuals). The comparison of the two speeds sheds light on the role of the wild-type growth rate $r$. When $r$ is small, it is expected that the two fronts increasingly move apart, preventing drive re-invasion (step B). We tested this theoretical conclusion by reducing the value of $r$ from $r=0.1$ to $r=0.02$ (resp. $\vcwt = 0.63$ and $\vcwt = 0.28$). We observed numerically a full wild-type recolonisation (Figure \ref{fig:r_varying_r002}). Several other cases have been investigated in Appendix \ref{ann:chasing}, all supporting this intuition. }

\begin{figure}[H]
\begin{subfigure}[t]{0.99\textwidth}
\centering
\caption{$r=0.02$} 
\label{fig:r_varying_r002}
\includegraphics[width=0.18\textwidth]{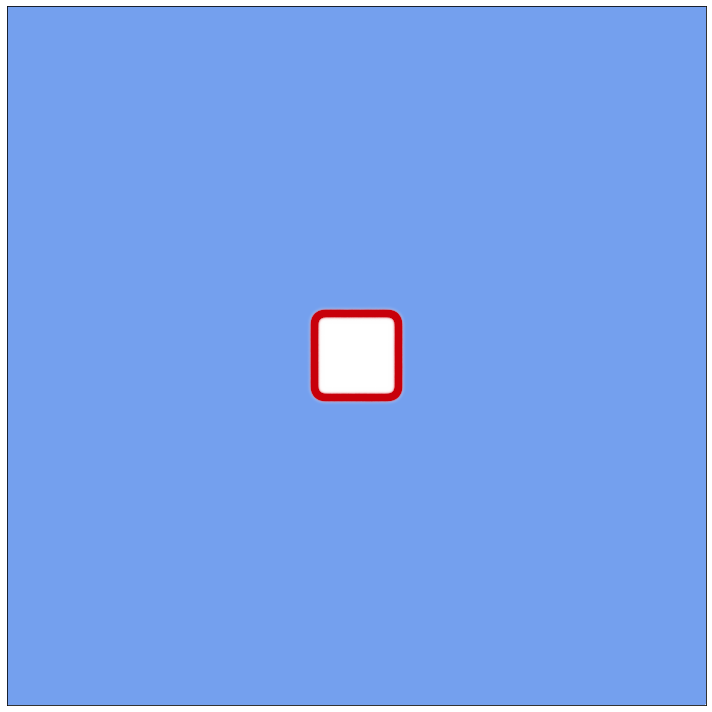} % 2D  K=10^5 et s=0.3
\includegraphics[width=0.18\textwidth]{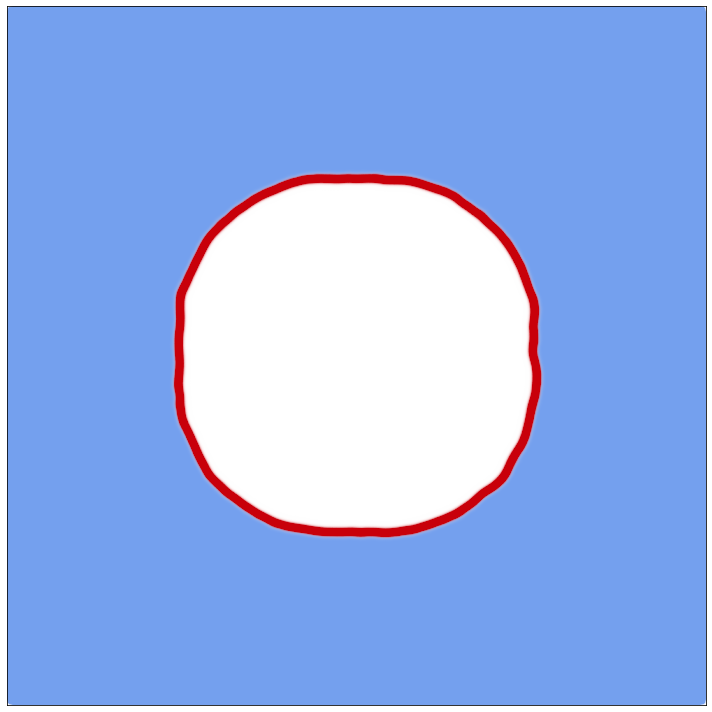}
\includegraphics[width=0.18\textwidth]{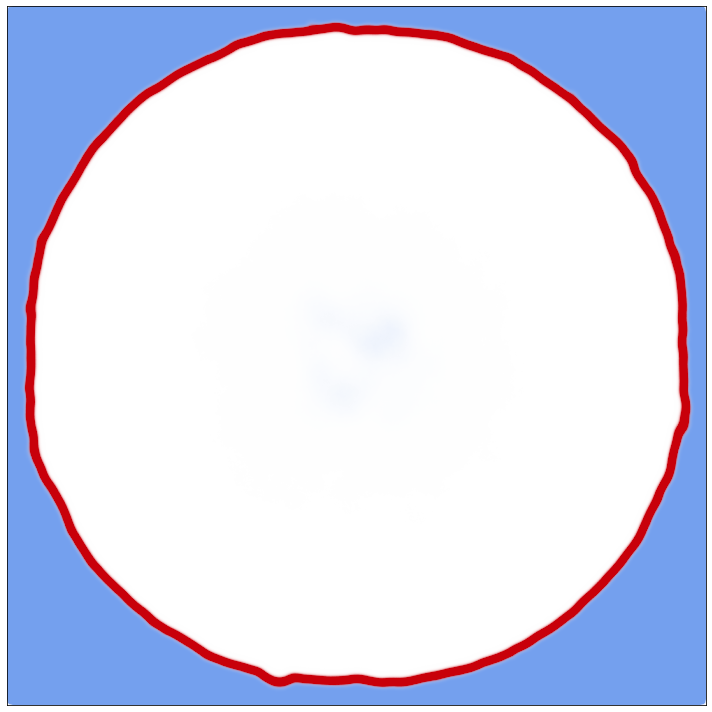}
\includegraphics[width=0.18\textwidth]{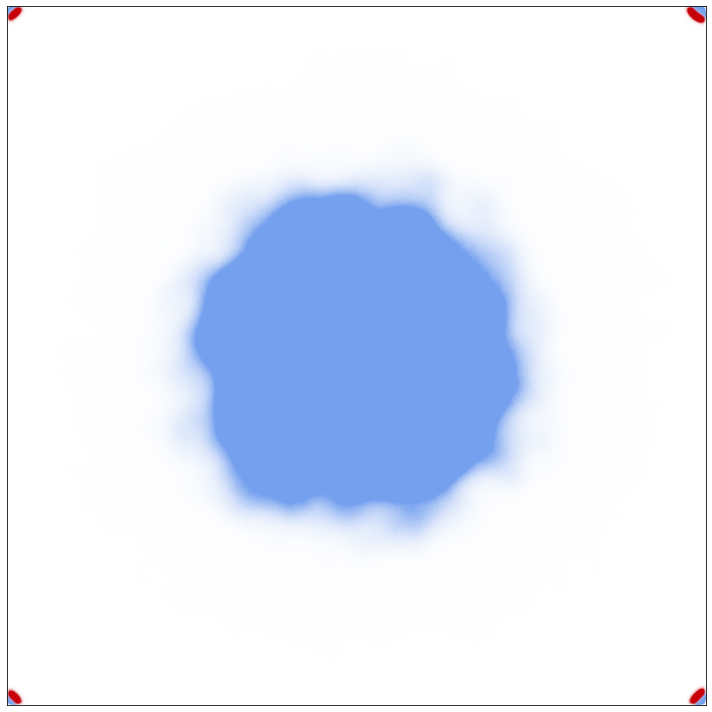}
\includegraphics[width=0.18\textwidth]{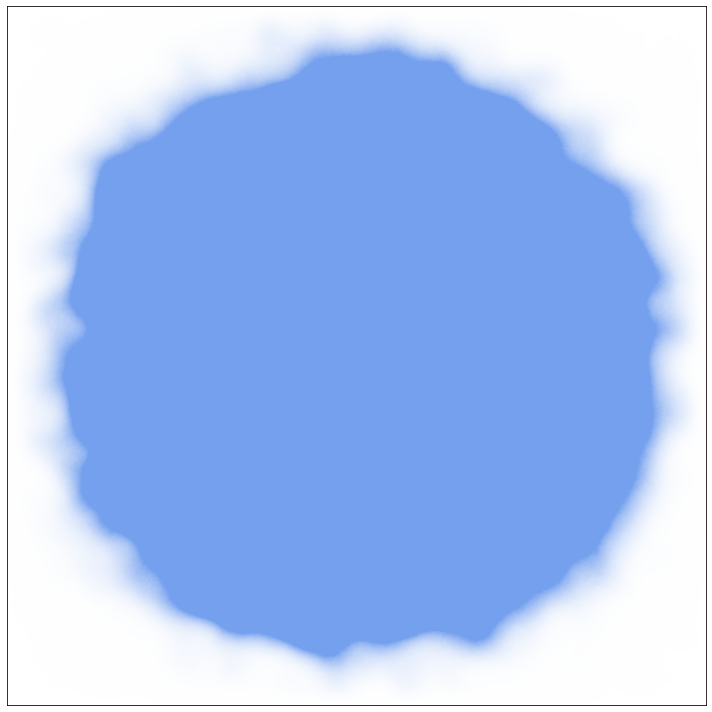}\\
\end{subfigure}
\begin{subfigure}[t]{0.99\textwidth}
\centering
\caption{$r=0.1$} 
\includegraphics[width=0.18\textwidth]{2D_K_5_s_0.7_t_20.0.png} % 2D  K=10^5 et s=0.3
\includegraphics[width=0.18\textwidth]{2D_K_5_s_0.7_t_250.0.png}
\includegraphics[width=0.18\textwidth]{2D_K_5_s_0.7_t_500.0.png}
\includegraphics[width=0.18\textwidth]{2D_K_5_s_0.7_t_750.0.png}
\includegraphics[width=0.18\textwidth]{2D_K_5_s_0.7_t_1000.0.png}\\
\end{subfigure}
\caption{Comparison of 2D numerical simulations for two different values of the intrinsic growth rate $r=0.02$ or $0.1$, with carrying capacity $K=10^5$ and drive fitness cost $s=0.7$. The case $r=0.1$ corresponds to Figure \ref{fig:1D-2D_K5s7}. On one hand, when $r=0.02$, the wild-type wave remains far behind the drive eradication wave, disabling drive reinvasion events; we witness a full wild-type recolonisation. On the other hand, when $r=0.1$, the wild-type recolonisation event is shortly followed by at least three drive reinvasion events, which lead to the complete eradication of the population.} \label{fig:r_varying}
\end{figure}

\section{Discussion}

Given both the promise and risk of eradication drives, their dynamics are worthy of careful examination. In this work, we introduced a theoretical framework to characterise the conditions leading to successful extinction of a population targeted by an eradication gene drive. We sought to delimit the conditions for the absence of wild-type recolonisation in a pulled drive eradication wave in a realisable time window. More precisely, we considered that if the last individual carrying a wild-type allele is surrounded by a large number of drive individuals (set to $ \mathscr{N} = 100$ alleles in this work) then wild-type recolonisation is very unlikely. To attest if this is the case, we determined the distance between the last individual carrying a wild-type allele and the last spatial site with more than $100$ drive alleles, at the back of the wave. This value is not straightforward to find, so we proceeded in two steps.

First, we determined the distance $\ell$ between the drive and the wild-type wave at the level line corresponding to $100$ alleles. This distance is almost deterministic, as we chose $\mathscr{N} = 100$ large enough for stochastic fluctuations to be negligible. We deduced this distance from a deterministic simulation and analytically proved that it increases with the carrying capacity $K$, increases with the migration rate $m$ and decreases with the drive fitness cost $s$. We also found that if the carrying capacity $K$ is multiplied by a factor $10$, the distance  $\ell$ increases by $ \left( \frac{1}{\lDbd} - \frac{1}{\lWbd} \right) \log(10)$, with $\lDbd $, resp.\ $\lWbd$, the exponent of the exponential profile at the back of the drive, resp.\ wild-type, wave. All these analytical results still hold in 2D.

Second, we determined the distance $L^{100}_1$ between the position of the last wild-type allele and the last position with more than $100$ wild-type alleles. This distance highly depends on the stochastic fluctuations of the last individual, and its distribution is difficult to characterise. To simplify the problem, we considered an isolated wild-type population, and we reduced heuristically the distance distribution to an extinction time distribution of a spatial Galton-Watson process, with the appropriate initialisation. Multiplying this time by the speed of the wave, we obtained a good approximation of the distance $L^{100}_1$. We leave open the quality of this approximation in mathematical terms, as well as the analysis of the distribution of the extinction time. 

Combining the two distances $\ell$ and $L^{100}_1$ allowed us to determine precisely the conditions under which wild-type recolonisation is highly unlikely. Numerically, we observed that the larger the carrying capacity is, the fewer recolonisation events we observe numerically. And the fitter the drive, the smaller the chance of wild-type recolonisation, in agreement with the results obtained in \cite{champer2021}. \lk{The role of the migration, however, is less clear: distances $\ell$ and $L^{100}_1$ seems to increase similarly when the migration rate $m$ increases, compensating one another. Numerically we observe a slight increase in the probability of wild-type recolonisation as $m$ increases.} Noticeably in our study, the transition between very low ($<10\%$) and very high ($>90\%$) chances of wild-type recolonisation when the drive fitness cost varies is relatively restricted: for $s$ the drive fitness cost varying between $0$ (no drive fitness cost) and $1$ (no possible survival for drive individuals), this happens within a range of $0.2$. Focusing on the drive intrinsic fitness, i.e. the overall ability of drive to increase in frequency when the drive proportion is close to zero, we agree on the observations made by \cite{paril2022}: the fitter the drive, the sharper the drive wave at the front \eqref{eq:lDf}, and the faster the speed \eqref{eq:speed_lDf} as initially showed in \cite{fisher1937}. In contrast at the back of the wave, the fitter the drive, the wider the drive wave, reducing the risk of wild-type recolonisation (see Section \ref{subsec:deter_dist}). We also demonstrated that if we consider the same migration rate for all genotypes, high migration rates would reduce the chance of wild-type recolonisation. This result is in agreement with the numerical observations made in \cite{paril2022, champer2021}.

The outcomes of our 2D simulations are consistent with our 1D conclusions: the probability of wild-type recolonisation events increases with higher drive fitness cost and smaller carrying capacity. Nevertheless, we observe that stochastic events (both wild-type recolonisation and drive reinvasion events) emerge in average more frequently in 2D than in 1D, leading to faster dynamics. This is due to the fact that stochastic events can occur in a variety of directions each time in 2D, instead of only one in 1D.

Finally, we explored the drive reinvasion dynamics after one (or a few) hypothetical wild-type recolonisation event(s). We demonstrated that the probability of drive reinvasion depends on the speed of both the wild-type recolonising wave and the drive eradication wave, and decreases with smaller values of the intrinsic growth rate $r$. If the probability of drive reinvasion is way higher than the probability of wild-type recolonisation, we expect the population to go extinct. 

The models that we used in this study are generalist: they could be applied to different species and gene drive constructs. They provide general conclusions but also come with necessary simplifications. For instance, we assumed a uniform landscape with random movement, which is extremely rare in the wild. Realistic migration patterns over both small and large scales would need to be taken into account to obtain better predictions. In mosquito populations, gene drive propagation can be accelerated by long distance migrations: mosquitoes can benefit from fast air currents (transporting them for hundreds of kilometres in a few hours) \cite{chapman2011, hu2016, huestis2019} or human-based modes of travel such as cars \cite{egizi2016} or planes \cite{eritja2017}. This could make it easier for wild-type individuals to permeate the drive wave, and initiate the recolonisation of empty areas.

\lk{Our generalistic models also miss a variety of other ecological parameters that need to be taken into account before any field release. The presence of a competing species or predator in the ecosystem has been shown to make eradication substantially easier than anticipated, and prevent wild-type recolonisation \cite{liu2023a}. High levels of inbreeding have also been proven to significantly increase the likelihood of recolonisation \cite{champer2021}. Overall, the life cycle of the species, the mating system, the ecological differences between males and females, the seasonal population fluctuations are all biological characteristics that might influence outcomes.}

\lk{Finally, some simplifications in our models are directly related to the type of gene drive constructs involved. The fitness component on which the disandvantage operates (fecundity or survival; in our models it operates on fecundity) might affect the speed of drive invasion, and accentuate consequences of potential Allee effects \cite{klay2025}. Emerging resistances that might alter the gene conversion ability of the drive (not included in our models) may also influence extinction outcomes \cite{rode2019, beaghton2019, hammond2017, price2020}.}

Because of its generic nature, our model cannot be used directly for risk assessment, but informs, qualitatively, on the potential failure of an eradication drive.

\clearpage

%%%%%%%%%%%%%%%%%%%%%%%%%%%%%%%%%%%%%%% Appendix %%%%%%%%%%%%%%%%%%%%%%%%%%%%%%%%%%%%%%%%%%%%%%%%%%%%%

\newpage

\Huge \textbf{Appendix} \normalsize

\appendix

% FD renumerotation des tables et des figures
\setcounter{table}{0}
\renewcommand{\thetable}{S\arabic{table}}
\setcounter{figure}{0}
\renewcommand{\thefigure}{S\arabic{figure}}

\section{Growth term details} \label{app:growth_terms}

\begin{table}[H]
\begin{tabular}{|M{.13\textwidth}|M{.43\textwidth}|M{.08\textwidth}|M{.3\textwidth}|}
\hline
  \textbf{Parents} & $ \ $ \hspace{2cm} \textbf{Gametes} \hspace{1.5cm} \textbf{Adult}  &  \textbf{Fitness} &  \textbf{Growth term} \\
 \hline
    WW + WW & \begin{tikzpicture}
\begin{scope}[every node/.style={circle}]
    \node (A) at (0,-0){};
    \node (B) at (3,0) {W,W + W,W};
    \node (C) at (6,0) {WW};
\end{scope}
\begin{scope}[>={Stealth[black]},
              every node/.style={fill=white},
              every edge/.style={draw=black, thick}]
    \path [->] (A) edge node {$1$} (B);
    \path [->] (B) edge node {$1$} (C);
\end{scope}
\end{tikzpicture} &  $1$ & $ \dfrac{\n{WW} \n{WW}}{n} $ \\
\hline %%%%%%%%%%%%%%%%%%%%%%%%%%%%%%%%%%%%%%%%%%%%%%%%%%%%%%%%%%%%%%%%%%%%%%%%%%%
  WW + WD  \vspace{0.2cm}  & \vspace{-0.2cm}  \begin{tikzpicture}
\begin{scope}[every node/.style={circle}]
    \node (A) at (0,-0.75){};
    \node (B) at (3,0) {W,W + D,D};
    \node (C) at (6,0) {WD};%
    \node (D) at (6,-1) {WW};%
    \node (E) at (3,-1.5) {W,W + W,D};
    \node (F) at (6,-2) {WD};%
\end{scope}
\begin{scope}[>={Stealth[black]},
              every node/.style={fill=white},
              every edge/.style={draw=black, thick}]
    \path [->] (A) edge node {$c$} (B);
    \path [->] (B) edge node {$1$} (C);
    \path [->] (E) edge node {$\frac{1}{2}$} (D);
    \path [->] (A) edge node {$1-c$} (E);
    \path [->] (E) edge node {$\frac{1}{2}$} (F);
\end{scope}
\end{tikzpicture} 
 &   \begin{tikzpicture}
\begin{scope}
    \node (1) at (0,0){$  1-sh $};
    \node (2) at (0,-1) {$ 1$};
    \node (3) at (0,-2) {$ 1-sh $};
\end{scope}
\end{tikzpicture} 
&   \begin{tikzpicture}
\begin{scope}
    \node (1) at (0,0){$  c \ \ (1-sh)    \ \dfrac{2 \n{WW} \n{DW}}{n} $};
    \node (2) at (0,-1) {$  (1-c) \ \frac{1}{2} \    \ \dfrac{2 \n{WW} \n{DW}}{n} $};
    \node (3) at (0,-2) {$ (1-c) \ \frac{1}{2} \  \ (1-sh)   \ \dfrac{2 \n{WW} \n{DW}}{n} $};
\end{scope}
\end{tikzpicture} \\
 \hline %%%%%%%%%%%%%%%%%%%%%%%%%%%%%%%%%%%%%%%%%%%%%%%%%%%%%%%%%%%%%%%%%%%%%%%%%
    WW + DD &  \begin{tikzpicture}
\begin{scope}[every node/.style={circle}]
    \node (A) at (0,-0){};
    \node (B) at (3,0) {W,W + D,D};
    \node (C) at (6,0) {WD};
\end{scope}
\begin{scope}[>={Stealth[black]},
              every node/.style={fill=white},
              every edge/.style={draw=black, thick}]
    \path [->] (A) edge node {$1$} (B);
    \path [->] (B) edge node {$1$} (C);
\end{scope}
\end{tikzpicture} &  $1-sh$ & $  (1-sh)   \ \dfrac{2 \n{WW} \n{DD}}{n} $ \\
\hline %%%%%%%%%%%%%%%%%%%%%%%%%%%%%%%%%%%%%%%%%%%%%%%%%%%%%%%%%%%%%%%%%%%%%%%%%%%
    WD + WD  \vspace{1.4cm} & \vspace{-0.6cm}  \begin{tikzpicture}
\begin{scope}[every node/.style={circle}]
    \node (A) at (-0.5,-1.5){};
    \node (B) at (3,0.2) {D,D + D,D};
    \node (C) at (5.5,0.2) {DD};
    \node (D) at (3,-1.5) {D,D + W,D};
    \node (E) at (5.5,-1) {DD};
    \node (F) at (5.5,-2) {WD};
    \node (G) at (3,-4) {W,D + W,D};
    \node (H) at (5.5,-3) {DD};
    \node (I) at (5.5,-4) {WD};
    \node (J) at (5.5,-5) {WW};
\end{scope}

\begin{scope}[>={Stealth[black]},
              every node/.style={fill=white},
              every edge/.style={draw=black, thick}]
    \path [->] (A) edge node {$c^2$} (B);
    \path [->] (B) edge node {$1$} (C);
    \path [->] (A) edge node {$2c(1-c)$} (D);
    \path [->] (D) edge node {$\frac{1}{2}$} (E);
    \path [->] (D) edge node {$\frac{1}{2}$} (F);
    \path [->] (A) edge node {$(1-c)^2$} (G);
    \path [->] (G) edge node {$\frac{1}{4}$} (H);
    \path [->] (G) edge node {$\frac{1}{2}$} (I);
    \path [->] (G) edge node {$\frac{1}{4}$} (J);
\end{scope}
\end{tikzpicture} &  
\begin{tikzpicture}
\begin{scope}
    \node (1) at (0,-0){$1-s $};
    \node (2) at (0,-1) {$ 1-s  $};
    \node (3) at (0,-2) {$ 1-sh $};
    \node (4) at (0,-3) {$ 1-s $};
    \node (5) at (0,-4) {$ 1-sh$};
    \node (6) at (0,-5) {$ 1 $};
\end{scope}
\end{tikzpicture} &  
\begin{tikzpicture}
\begin{scope}
    \node (1) at (0,0){$ c^2 \ \ (1-s)  \ \dfrac{\n{DW} \n{DW}}{n} $};
    \node (2) at (0,-1) {$ c \ (1-c) \ \ (1-s)      \ \dfrac{\n{DW} \n{DW}}{n}  $};
    \node (3) at (0,-2) {$ c \ (1-c)  \ \ (1-sh)      \ \dfrac{\n{DW} \n{DW}}{n} $};
    \node (4) at (0,-3) {$ (1-c)^2 \  \frac{1}{4} \ \ (1-s)     \  \dfrac{\n{DW} \n{DW}}{n}  $};
    \node (5) at (0,-4) {$ (1-c)^2 \  \frac{1}{2} \ \ (1-sh)    \ \dfrac{\n{DW} \n{DW}}{n}   $};
    \node (6) at (0,-5) {$ (1-c)^2 \  \frac{1}{4} \    \  \dfrac{\n{DW} \n{DW}}{n}  $};
\end{scope}
\end{tikzpicture}\\
\hline %%%%%%%%%%%%%%%%%%%%%%%%%%%%%%%%%%%%%%%%%%%%%%%%%%%%%%%%%%%%%%%%%%%%%%%%%%%
  WD + DD  \vspace{0.2cm}  & \vspace{-0.2cm}  \begin{tikzpicture}
\begin{scope}[every node/.style={circle}]
    \node (A) at (0,-0.75){};
    \node (B) at (3,0) {D,D + D,D};
    \node (C) at (6,0) {DD};%
    \node (D) at (6,-1) {WD};%
    \node (E) at (3,-1.5) {W,D + D,D};
    \node (F) at (6,-2) {DD};%
\end{scope}
\begin{scope}[>={Stealth[black]},
              every node/.style={fill=white},
              every edge/.style={draw=black, thick}]
    \path [->] (A) edge node {$c$} (B);
    \path [->] (B) edge node {$1$} (C);
    \path [->] (E) edge node {$\frac{1}{2}$} (D);
    \path [->] (A) edge node {$1-c$} (E);
    \path [->] (E) edge node {$\frac{1}{2}$} (F);
\end{scope}
\end{tikzpicture}  &   \begin{tikzpicture}
\begin{scope}
    \node (1) at (0,0){$ 1-s $};
    \node (2) at (0,-1) {$ 1-sh $};
    \node (3) at (0,-2) {$ 1-s $};
\end{scope}
\end{tikzpicture}  &   \begin{tikzpicture}
\begin{scope}
    \node (1) at (0,0){$  c \ \ (1-s)    \ \dfrac{2 \n{DW} \n{DD}}{n} $};
    \node (2) at (0,-1) {$ (1-c) \ \frac{1}{2} \  \  (1-sh)    \ \dfrac{2 \n{DW} \n{DD}}{n} $};
    \node (3) at (0,-2) {$ (1-c) \ \frac{1}{2} \ \ (1-s)   \ \dfrac{2 \n{DW} \n{DD}}{n} $};
\end{scope}
\end{tikzpicture} \\
 \hline %%%%%%%%%%%%%%%%%%%%%%%%%%%%%%%%%%%%%%%%%%%%%%%%%%%%%%%%%%%%%%%%%%%%%%%%%
  DD + DD & \begin{tikzpicture}
\begin{scope}[every node/.style={circle}]
    \node (A) at (0,-0){};
    \node (B) at (3,0) {D,D + D,D};
    \node (C) at (6,0) {DD};
\end{scope}
\begin{scope}[>={Stealth[black]},
              every node/.style={fill=white},
              every edge/.style={draw=black, thick}]
    \path [->] (A) edge node {$1$} (B);
    \path [->] (B) edge node {$1$} (C);
\end{scope}
\end{tikzpicture} &  $1-s$ &
\begin{tikzpicture}
\begin{scope}
    \node (1) at (0,0){$ (1-s)    \ \dfrac{\n{DD} \n{DD}}{n} $};
\end{scope}
\end{tikzpicture}\\
\hline
\end{tabular}
\caption{Growth term details when conversion occurs in the germline. Since the underlying model is the same, this Table is the same as Table 8 in \cite{klay2023}}
\label{tab:mating_term_germline}
\end{table}

\clearpage

\section{Implementation of the stochastic model}  \label{app:code_stoch}

\begin{algorithm}[H]
\caption{Main steps of the stochastic model, in one spatial dimension.}
\begin{algorithmic}

\State \textbf{Initialization:}
\State Divide space into $nb\_sites$.
\State Set $n_D(i) \gets K  \mathrm{d}x$ for $i \leq nb\_sites/2$.
\State Set $n_W(i) \gets K  \mathrm{d}x$ for $i > nb\_sites/2$.
\State Set $f_D(i), f_W(i) \gets 0$ for all sites.

\Statex

\For{$t = 0$ \textbf{to} $T$ \textbf{step} $dt$}

    \Statex
    \State \textbf{Stopping condition:}
    \If{ $\max \{ i : n_D(i) \} \geq nb\_sites - 10$ }
        \State \textbf{break loop}
    \EndIf

    \Statex
    \State \textbf{Birth and Death:}
    \For{each site $i$}
        \State $P(i) \gets n_D(i) + n_W(i)$
        \If{ $P(i) = 0$ }
            \State $f_D(i) \gets 0$, $f_W(i) \gets 0$
        \Else
            \State $f_D(i) \gets \Big(1 + r \big(1 - \tfrac{P(i)}{K  \mathrm{d}x}\big)\Big)  \dfrac{ (1-s) \, n_D(i) + (1-s h)(1+c) \, n_W(i)}{P(i)}$
            \State $f_W(i) \gets \Big(1 + r \big(1 - \tfrac{P(i)}{K  \mathrm{d}x}\big)\Big)  \dfrac{ (1-c)(1-s h) \, n_D(i) + n_W(i)}{P(i)}$
        \EndIf
        \State $n_D(i) \gets n_D(i) + \text{Poisson}(f_D(i)  n_D(i)  \mathrm{d}t) - \text{Poisson}(n_D(i)  \mathrm{d}t)$
        \State $n_W(i) \gets n_W(i) + \text{Poisson}(f_W(i)  n_W(i)  \mathrm{d}t) - \text{Poisson}(n_W(i)  \mathrm{d}t)$
        \State $n_D(i) \gets \max(n_D(i),0)$ 
        \State $n_W(i) \gets \max(n_W(i),0)$
    \EndFor

    \Statex
    \State \textbf{Migration:}
    \For{each site $i$}
        \State \textbf{Allele outflow:}
        \State $M_D(i) \sim \text{Binomial}(n_D(i), m)$, \quad $n_D(i) \gets n_D(i) - M_D(i)$, 
        \State $M_W(i) \sim \text{Binomial}(n_W(i), m)$, \quad $n_W(i) \gets n_W(i) - M_W(i)$
        \State \textbf{with equal probability to the left or right:}
        \State $M_D^L(i) \sim \text{Binomial}(M_D(i), 0.5)$, \quad $M_D^R(i) \gets M_D(i) - M_D^L(i)$
        \State $M_W^L(i) \sim \text{Binomial}(M_W(i), 0.5)$, \quad $M_W^R(i) \gets M_W(i) - M_W^L(i)$
    \EndFor

    \State \textbf{Allele inflow at the boundaries:}
    \State $n_D(1) \gets n_D(1) + M_D^L(1)$, \quad $n_W(1) \gets n_W(1) + M_W^L(1)$
    \State $n_D(N) \gets n_D(N) + M_D^R(N)$, \quad $n_W(N) \gets n_W(N) + M_W^R(N)$

    \State \textbf{Allele inflow on interior sites:}
    \For{each interior site $i$}
        \State $n_D(i) \gets n_D(i) + M_D^R(i-1) + M_D^L(i+1)$
        \State $n_W(i) \gets n_W(i) + M_W^R(i-1) + M_W^L(i+1)$
    \EndFor

\EndFor

\end{algorithmic}
\end{algorithm}

\clearpage

\section{Speed and exponential approximations of the wave} \label{app:lambda}

\subsection{Continuous model}\label{app:lambda_cont}

Since the wave is pulled, we deduce the speed $\vc $ and the exponential approximations at the back and at the front of the wave from system \eqref{eq:par_ger_wave} as detailed in the following sections. \lk{The terms ``decreasing section'' and ``increasing section'' refer to the wave profile at a fixed time $t$, illustrated in Figure \ref{fig:front_deter}.}

\subsubsection{At the front of the wave}

At the front of the wave, we assume that: \begin{equation}\label{eq:approx_}
    \dfrac{\N{D}}{N} \ll 1 \ , \ \dfrac{\N{W}}{N} \sim 1 \ \  \text{and} \  \ N \sim 1.
\end{equation}

Combined with system \eqref{eq:par_ger_wave}, we know the solution of Eq.\eqref{eq:approx_ND} is an approximation of $\N{D}$ at the front of the wave:\begin{equation}\label{eq:approx_ND}
  0  =  \sigma^2 \N{D}'' +  \vc \N{D}' + \N{D} \ \Big[  (1-sh) \  (1+c)  - 1 \Big]. 
\end{equation}

Since the wave is pulled, the speed $\vc$ is given by the minimal speed of the problem linearised at low drive allele numbers at the front of the wave. We approximate the decreasing drive section at the front of the wave by an exponential function of the form $ \N{D}(z) \approx  \rho_D^{front} \exp ( \lDf z ) $: \begin{equation}
    0 = \sigma^2 (\lDf)^2 + \vc \ \lDf +  \Big[  (1-sh) \  (1+c)  - 1 \Big],
\end{equation}     

with the determinant $\Delta = (\vc)^2 - 4 \sigma^2 \Big[  (1-sh) \  (1+c)  - 1 \Big]$. The minimal speed is given by $\Delta =0$, thus: \begin{equation} 
   \vc =  2 \sigma \sqrt{ (1-sh)(1+c) - 1 },
\end{equation}

and the corresponding exponent is given by:

\begin{equation}
    \lDf = -\dfrac{\vc}{2  \sigma^2} = - \dfrac{1}{\sigma} \sqrt{ (1-sh)(1+c) - 1 }.
\end{equation}

The quantity $(1-sh)(1+c) - 1 $ is always strictly positive for a pulled wave in case of drive invasion \cite{klay2023}.

\subsubsection{At the back of the wave} \label{app:lambda_cont_back}

At the back of the wave, we assume that: \begin{equation} \label{eq:approx_cb}
    \dfrac{\N{D}}{N} \sim 1 \ , \ \dfrac{\N{W}}{N} \ll 1 \ \ \text{and} \ \ N \ll 1.
\end{equation}

By definition of a traveling wave, the speed at the front and at the back of the wave is equal.

\subsubsection*{Drive increasing section}

Using \eqref{eq:approx_cb} in system \eqref{eq:par_ger_wave}, we know the solution of the following Eq.\eqref{eq:NDcb} is an approximation of $\N{D}$ at the back of the wave:

\begin{equation}\label{eq:NDcb}
  0 = \sigma^2 \N{D}'' +  \vc \N{D}' +  \N{D} \ \Big[  ( r+1 ) (1-s)  - 1 \Big].
\end{equation}

We approximate the increasing drive section at the back of the wave by an exponential function of the form $ \N{D}(z) \approx \rho_D^{\text{back}} \exp ( \lDb z )$ and deduce from \eqref{eq:NDcb}:

\begin{equation}
   0 = \sigma^2 (\lDb)^2  + \vc \lDb  + \Big[  ( r+1 ) (1-s)  - 1 \Big] r .
\end{equation}

\lk{The two solutions are}:
\begin{equation} \label{eq:lDb-}
    \begin{aligned}
  \lDbMinus = & \dfrac{1}{2 \sigma^2 } \left( -  \vc - \sqrt{ (\vc)^2 - 4 \sigma^2 \Big[  ( r+1 ) (1-s)  - 1 \Big] } \  \right), \\
   = & \dfrac{1}{2 \sigma^2} \left( -  2 \sigma \sqrt{ (1-sh)(1+c) - 1 } \pm \sqrt{  4  \sigma^2 \Big[  (1-sh)(1+c) - 1 \Big] - 4 \sigma^2 \Big[  ( r+1 ) (1-s)  - 1 \Big] } \  \right),\\
  = & \frac{1}{\sigma} \left( -   \sqrt{ (1-sh)(1+c) - 1 } - \sqrt{   (1-sh)(1+c) - 1  + 1  -  ( r+1 ) (1-s)   } \right) .
    \end{aligned}
\end{equation}

\begin{equation} \label{eq:lDb+}
    \begin{aligned}
  \lDbPlus = & \dfrac{1}{2 \sigma^2 } \left( -  \vc + \sqrt{ (\vc)^2 - 4 \sigma^2 \Big[  ( r+1 ) (1-s)  - 1 \Big] } \  \right), \\
   = & \dfrac{1}{2 \sigma^2} \left( -  2 \sigma \sqrt{ (1-sh)(1+c) - 1 } + \sqrt{  4  \sigma^2 \Big[  (1-sh)(1+c) - 1 \Big] - 4 \sigma^2 \Big[  ( r+1 ) (1-s)  - 1 \Big] } \  \right),\\
  = & \frac{1}{\sigma} \left( -   \sqrt{ (1-sh)(1+c) - 1 } + \sqrt{   (1-sh)(1+c) - 1  + 1  -  ( r+1 ) (1-s)   } \right) .
    \end{aligned}
\end{equation}

Since we study an eradication drive, we have $ r < \dfrac{s}{1-s} \iff 1 - (r+1)(1-s) > 0 $ \cite{klay2023}. Therefore, $\lDbMinus$ and $\lDbPlus$ are of opposite sign. At the back of the wave, the number of drive individuals tends to zero when $ z \to - \infty $ therefore we only conserve the positive solution $  \lDbPlus $.

\subsubsection*{Wild-type increasing section} 

Using \eqref{eq:approx_cb} in system \eqref{eq:par_ger_wave}, we know the solution of the following Eq.\eqref{eq:NWcb} is an approximation of $\N{W}$ at the back of the wave:

\begin{equation}\label{eq:NWcb} 
  0  = \sigma^2 \N{W}''  + v\N{W}'  + \N{W}  \Big[ (r+1) (1-sh)   (1-c) -1   \Big].
\end{equation}

We approximate the increasing wild-type section at the back of the wave by an exponential function of the form $ \N{W}(z) \approx \rho_W^{\text{back}} \exp ( \lWb z )$ and deduce from \eqref{eq:NWcb}:

\begin{equation}
   0 =  \sigma^2 (\lWb)^2  + \vc \lWb  +  \Big[ (r+1) (1-sh)   (1-c) -1   \Big]   .
\end{equation}

The two solutions are: \begin{equation}
    \begin{aligned} \label{eq:lWb-}
    \lWbMinus = &  \dfrac{1}{2 \sigma^2} \left( - \vc - \sqrt{ (\vc)^2 - 4  \sigma^2 \Big[ (r+1) (1-sh)   (1-c) -1   \Big]  } \  \right), \\
   = & \dfrac{1}{2 \sigma^2} \left( -  2 \sigma \sqrt{ (1-sh)(1+c) - 1 } - \sqrt{  4 \sigma^2 \Big[  (1-sh)(1+c) - 1 \Big] - 4 \sigma^2 \Big[ (r+1) (1-sh)   (1-c) -1   \Big]  } \ \right) ,\\
  = & \frac{1}{\sigma} \left( -   \sqrt{ (1-sh)(1+c) - 1 } - \sqrt{   (1-sh)(1+c)  -  1 + 1  - (r+1) (1-sh)   (1-c)     } \right) .
    \end{aligned}
\end{equation}

 \begin{equation}
    \begin{aligned} \label{eq:lWb+}
    \lWbPlus = &  \dfrac{1}{2 \sigma^2} \left( - \vc + \sqrt{ (\vc)^2 - 4  \sigma^2 \Big[ (r+1) (1-sh)   (1-c) -1   \Big]  } \  \right), \\
   = & \dfrac{1}{2 \sigma^2} \left( -  2 \sigma \sqrt{ (1-sh)(1+c) - 1 } + \sqrt{  4 \sigma^2 \Big[  (1-sh)(1+c) - 1 \Big] - 4 \sigma^2 \Big[ (r+1) (1-sh)   (1-c) -1   \Big]  } \ \right) ,\\
  = & \frac{1}{\sigma} \left( -   \sqrt{ (1-sh)(1+c) - 1 } + \sqrt{   (1-sh)(1+c)  -  1 + 1  - (r+1) (1-sh)   (1-c)     } \right) .
    \end{aligned}
\end{equation}

Since we study an eradication drive, we have $ r < \dfrac{s}{1-s} \iff r+1  < \dfrac{1}{1-s}$ \cite{klay2023} and therefore : 

\begin{equation}\label{eq:51}
    (r+1) (1-sh) (1-c) < \dfrac{1-sh}{1-s} (1-c) < 1-c < 1
\end{equation}

\lk{Since $1-(r+1)(1-sh)(1-c) > 0$ \eqref{eq:51}, $\lWbMinus$ and $\lWbPlus$ are of opposite sign}. At the back of the wave, the number of wild-type individuals tends to zero when $ z \to - \infty $: we only conserve the positive solution $\lWbPlus$.

\subsection{Discrete model} \label{app:lambda_dis}

To determine the speed of the wave and the exponential approximations in the discrete stochastic model, we focus on the mean dynamics \lk{which can be deduced in the limit of large carrying capacity $K\to +\infty$.} For drive alleles, we have:

\begin{equation} \label{eq:discretD}
 \left\{ \small
\begin{array}{ll}
 \n{D}^{t+\frac{dt}{2}, x} &=  \left( ( \g{D}(\n{D}^{t,x}, \n{W}^{t,x})    -  1) \ \mathrm{d}t + 1 \right) \  
  \n{D}^{t,x}, \\
  \\
\n{D}^{t+\mathrm{d}t,x} &= (1-m) \  \n{D}^{t+\frac{dt}{2},x} + \dfrac{m}{2} \ ( \n{D}^{ t+\frac{dt}{2}, x+\mathrm{d}x} + \n{D}^{t+\frac{dt}{2},x-\mathrm{d}x}),
\end{array}
\right.
\end{equation}

and for wild-type alleles: 

\begin{equation} \label{eq:discretW}
 \left\{ \small
\begin{array}{ll}
 \n{W}^{t+\frac{dt}{2}, x} &=  \left(  (  \g{W}(\n{D}^{t,x}, \n{W}^{t,x})   - 1) \mathrm{d}t  + 1  \right) \ \n{W}^{t,x}, \\ 
 \\
\n{W}^{t+\mathrm{d}t,x} &= (1-m) \  \n{W}^{t+\frac{dt}{2},x} + \dfrac{m}{2} \ ( \n{W}^{t+\frac{dt}{2}, x+\mathrm{d}x} + \n{W}^{ t+\frac{dt}{2}, x-\mathrm{d}x,}).
\end{array}
\right.
\end{equation}

where the first lines correspond to the birth and death dynamics, and the second lines correspond to the migration. Combining the two lines in each system, we obtain: 

\begin{equation} \label{eq:discretDWT}
 \left\{ \small
\begin{array}{ll}
 \n{D}^{t+\mathrm{d}t,x} &= \left( ( \g{D}(\n{D}^{t,x}, \n{W}^{t,x})   -  1) \ \mathrm{d}t + 1 \right) (1-m) \  \n{D}^{t,x} + \left( ( \g{D}(\n{D}^{t,x+\mathrm{d}x}, \n{W}^{t,x+\mathrm{d}x})   -  1) \ \mathrm{d}t + 1 \right)  \dfrac{m}{2} \  \n{D}^{ t,x+\mathrm{d}x} \\ \\  &+   \left( ( \g{D}(\n{D}^{t,x-\mathrm{d}x}, \n{W}^{t,x-\mathrm{d}x})   -  1) \ \mathrm{d}t + 1 \right) \dfrac{m}{2} \   \n{D}^{t,x-\mathrm{d}x},  \\ 
 \\
 \n{W}^{t+\mathrm{d}t, x} &=  \left(  (  \g{W}(\n{D}^{t,x}, \n{W}^{t, x})   - 1) \mathrm{d}t  + 1  \right)   (1-m) \  \n{W}^{ t, x} +  \left(  (  \g{W}(\n{D}^{t,x+\mathrm{d}x}, \n{W}^{t, x+\mathrm{d}x})   - 1) \mathrm{d}t  + 1  \right) \dfrac{m}{2} \  \n{W}^{t, x+\mathrm{d}x}\\ \\  & + \left(  (  \g{W}(\n{D}^{t,x-\mathrm{d}x}, \n{W}^{t, x-\mathrm{d}x})   - 1) \mathrm{d}t  + 1  \right) \dfrac{m}{2} \ \n{W}^{ t, x-\mathrm{d}x}.
\end{array}
\right.
\end{equation}

with:  \begin{equation}
     \g{D}( \n{D},  \n{W}) =   \left( r \ (1-n)+1 \right) \Big[ (1-s) \dfrac{\n{D}}{n} +   (1-sh) \  (1+c) \   \dfrac{\n{W}}{n} \Big],
\end{equation} \begin{equation}
     \g{W}( \n{D},  \n{W}) =   \left( r \ (1-n)+1 \right)  \Big[ \  \dfrac{\n{W}}{n}   +   (1-sh)  \  (1-c) \  \dfrac{\n{D}}{n} \Big].
\end{equation}

The speed of the wave in the discrete model is denoted $\vd $. \lk{$\vd$ differs from its continuous version $\vc$ due to the finite size of the population, however both speeds coincide in the limit $d\!t, d\!x\to 0$.}

\subsubsection{At the front of the wave}

At the front of the wave, we assume that: \begin{equation}\label{eq:approx_df}
    \dfrac{\n{D}^{t,x}}{n^{t,x}} \ll 1 \  ,  \  \dfrac{\n{W}^{t,x}}{n^{t,x}} \sim 1 \ \ \text{and} \ \ n^{t,x} \sim 1.
\end{equation} 

These approximations are also true around $x$, at the spatial sites $x+\mathrm{d}x$ and $x-\mathrm{d}x$. Using \eqref{eq:approx_df} in the first line of system \eqref{eq:discretDWT}, we know the solution of the following Eq.\eqref{eq:NDdf} is an approximation of $\n{D}$ at the front of the wave:

\begin{equation}\label{eq:NDdf}
    \n{D}^{t+\mathrm{d}t, x} = \left( \left(  (1-sh)   (1+c)   -  1 \right) \ \mathrm{d}t + 1 \right) \ \left( (1-m) \  \n{D}^{t,x} + \dfrac{m}{2} \ ( \n{D}^{ t, x+\mathrm{d}x} + \n{D}^{ t, x-\mathrm{d}x}) \right). 
\end{equation}

In case of a pulled wave, the speed $\vd$ is given by the minimal speed of the problem linearised at low drive allele numbers at the front of the wave. \lk{To seek stationary solutions in a moving reference frame, we perform the change of variable $z = x-\vd t$ and introduce the traveling wave profile $\N{D}$:}

\begin{equation}\label{eq:NDdf_profil}
    \N{D}^{z - \vd  \ \mathrm{d}t} = \left( \left(  (1-sh)   (1+c)   -  1 \right) \ \mathrm{d}t + 1 \right) \ \left( (1-m) \  \N{D}^{z} + \dfrac{m}{2} \ ( \N{D}^{ z+\mathrm{d}x} + \N{D}^{z-\mathrm{d}x}) \right). 
\end{equation}

\lk{We then approximate the decreasing drive section at the front of the wave by an exponential function of the form $ \N{D}^z \approx  e^{\lDfd z}$. Dividing both sides of Equation \eqref{eq:NDdf_profil} by $e^{\lDfd z}$, we deduce:}

\begin{equation}
 \begin{aligned}
    e^{ - \lDfd  \ \vd \  \mathrm{d}t} & = \left( \left(  (1-sh)   (1+c)   -  1 \right) \ \mathrm{d}t + 1 \right) \ \left( 1-m  + m \ \dfrac{ e^{(\lDfd)  \mathrm{d}x} + e^{-(\lDfd) \mathrm{d}x}}{2} \right)\\
    & = \left( \left(  (1-sh)   (1+c)   -  1 \right) \ \mathrm{d}t + 1 \right) \ \left( 1-m  + m \ \cosh(\lDfd \ \mathrm{d}x) \right)\\
 \end{aligned}
\end{equation}  

The minimal speed of the problem linearised at low drive allele numbers is given by:

\begin{equation}\label{eq:speed_dis}
    \vd = \underset{\lambda < 0}{\min} \left( \mathscr{V}(\lambda ) \right) = \underset{\lambda  < 0}{\min} \left(  \dfrac{ \log \left( \left[ \left(  (1-sh)   (1+c)   -  1 \right) \ d\!t + 1 \right] \ \left[ 1-m  + m \ \cosh(\lambda \ \ d\!x) \right] \right) }{ - \lambda \  d\!t} \right),
\end{equation}

\subsubsection{At the back of the wave}

Note that if the back of the wave can be defined as $z \to -\infty$ in the continuous model, it does not make sense any more in the discrete model. After a certain spatial site, there is no more individual on the left due to eradication: the exponential approximations do not hold any more. In the discrete stochastic model, the back of the wave corresponds to the non-empty spatial sites in the early increasing section of the wave.

At the back of the wave, we have:

\begin{equation}\label{eq:approx_db}
    \dfrac{\n{D}^{t,x}}{n^{t,x}} \sim 1 
  \ , \ \dfrac{\n{W}^{t,x}}{n^{t,x}} \ll 1 \  \ \text{and} \ \  n^{t,x} \ll 1.
\end{equation}

These approximations are also true around $x$, at the spatial sites $x+\mathrm{d}x$ and $x-\mathrm{d}x$. By definition of a traveling wave, the speed at the front and at the back of the wave is the same, in our case $ \vd $. 

\subsubsection*{Drive increasing section}

Using \eqref{eq:approx_db} in the first line of system \eqref{eq:discretDWT}, we know the solution of the following Eq.\eqref{eq:NDdb} is an approximation of $\n{D}$ at the back of the wave:\begin{equation} \label{eq:NDdb}
   \n{D}^{t+\mathrm{d}t,x}  =  \left( ( (r+1) (1-s) - 1) \ \mathrm{d}t + 1 \right) \ \left( (1-m) \  \n{D}^{t,x} + \dfrac{m}{2} \ ( \n{D}^{t,x+\mathrm{d}x} + \n{D}^{t,x-\mathrm{d}x}) \right). 
\end{equation}

\lk{To seek stationary solutions in a moving reference frame, we the perform the change of variable $z = x-\vd t$ and introduce the traveling wave profile $\N{D}$:} 

\begin{equation} \label{eq:NDdb_profil}
   \N{D}^{z - \vd \ \mathrm{d}t}  =  \left( ( (r+1) (1-s) - 1) \ \mathrm{d}t + 1 \right) \ \left( (1-m) \  \N{D}^{z} + \dfrac{m}{2} \ ( \N{D}^{z+\mathrm{d}x} + \N{D}^{z-\mathrm{d}x}) \right). 
\end{equation}

\lk{We then approximate the decreasing drive section at the front of the wave by an exponential function of the form $ \N{D}^z \approx  e^{\lDbd z}$. Dividing both sides of Equation \eqref{eq:NDdb_profil} by $e^{\lDbd z}$, we deduce:}

 \begin{equation} \label{eq:lDbd}
 e^{- \lDbd \ \vd \ \mathrm{d}t}  =  \left( ( (r+1) (1-s) - 1) \ \mathrm{d}t + 1 \right) \ \left( 1-m + \dfrac{m}{2} \ ( e^{\lDbd \mathrm{d}x} + e^{ - \lDbd \mathrm{d}x}) \right). 
\end{equation}

We solve Eq.\eqref{eq:lDbd} numerically to obtain the value of $\lDbd$.

\subsubsection*{Wild-type increasing section} 

Using \eqref{eq:approx_db} in the second line of system \eqref{eq:discretDWT}, we know the solution of the following Eq.\eqref{eq:NWdb} is an approximation of $\n{W}$ at the back of the wave:

\begin{equation} \label{eq:NWdb} 
 \n{W}^{t+\mathrm{d}t,x}  =  \left(  ( (r+1) (1-sh) (1-c)  - 1) \mathrm{d}t  + 1  \right)  \left(  (1-m) \  \n{W}^{t,x} + \dfrac{m}{2} \ ( \n{W}^{t,x+\mathrm{d}x} + \n{W}^{t,x-\mathrm{d}x})  \right). 
\end{equation}

\lk{To seek stationary solutions in a moving reference frame, we the perform the change of variable $z = x- \vd t$ and introduce the traveling wave profile $\N{W}$:} 

\begin{equation} \label{eq:NWdb_profil} 
 \N{W}^{z - \vd \ \mathrm{d}t}  =  \left(  ( (r+1) (1-sh) (1-c)  - 1) \mathrm{d}t  + 1  \right)  \left(  (1-m) \  \N{W}^{z} + \dfrac{m}{2} \ ( \N{W}^{z + \mathrm{d}x} + \n{W}^{z - \mathrm{d}x})  \right). 
\end{equation}

\lk{We then approximate the decreasing drive section at the front of the wave by an exponential function of the form $ \N{W}^z \approx  e^{\lWbd z}$. Dividing both sides of Equation \eqref{eq:NWdb_profil} by $e^{\lWbd z}$, we deduce:}

 \begin{equation}  \label{eq:lWdb}
 e^{- \lWbd \ \vd \ \mathrm{d}t}  =  \left( ( (r+1) (1-sh) (1-c)  - 1) \ \mathrm{d}t + 1 \right) \ \left( 1-m + \dfrac{m}{2} \ ( e^{\lWbd \mathrm{d}x} + e^{ - \lWbd \mathrm{d}x}) \right). 
\end{equation}

We solve Eq.\eqref{eq:lWdb} numerically to obtain the value of $\lWbd$.
 
\subsection{Correction of the speed in the discrete model due to stochastic effects} \label{subsec:corr_speed_dis}

\lk{The discrete model exhibits finite propagation speed, which differs from its continuous version for two reasons: firstly, the stepping-stone discretisation, and secondly, the stochastic effects due to finite population size. The stepping-stone discretisation correction has been addressed in the minimization problem \eqref{eq:speed_dis} which resulted in $\vd$.} We now address the correction on the stochastic effects due to finite population size. 

The wave is pulled, meaning that its speed is determined by the few drive individuals at the front of the wave. The model being stochastic, small densities exhibit relatively stochastic fluctuations in their dynamics, so that the speed of propagation in stochastic simulations is smaller than \eqref{eq:speed_dis}. This discrepancy has been  predicted analytically in \cite{brunet1997, brunet2001} (see also \cite{berard2010, mueller2011} for the mathematical proof) to be:
\begin{equation}\label{eq:speed_dis_cor}
     \vdc \approx \vd - \dfrac{\mathscr{V}''(\lDfd) \  \pi^2 \ (\lDfd)^2}{2 \ \left(\log\left(\frac{1}{K}\right)\right)^2},
\end{equation}
where $\vdc$ is the corrected discrete speed, and the function $\mathscr{V}$ is given in Eq.\eqref{eq:speed_dis}. This analytical prediction was obtained heuristically by a cut-off argument which takes into account the fact that the distribution of individuals vanishes beyond some finite range, in contrast with the continuous model where the density is positive everywhere.

In Table \ref{tab:speeds}, we compare the following quantities, respectively for $s=0.3$ and $s=0.7$ (see Table \ref{tab:parameters_stoch} for the set of all parameters): i) the continuous speed $\vc$ given by \eqref{eq:speed_lDf}; ii) the discrete speed $\vd$ given by \eqref{eq:speed_dis}; iii) the discrete speed corrected due to stochastic effects $\vdc$ given by \eqref{eq:speed_dis_cor}; and iv) the numerical speed $v_{\text{num}}$ as measured in stochastic numerical simulations. The best approximation of the speed computed numerically is indeed the corrected discrete speed, given by \eqref{eq:speed_dis_cor}. 

\renewcommand{\arraystretch}{1.2}

\begin{table}[H]
    \centering
    \begin{tabular}{|c|c|c|c|c|}
    \hline
    \multirow{2}{*}{$s$ value} &  \multicolumn{4}{c|}{\textbf{Speed}}\\
    \cline{2-5}
   & $\vc$ &  $\vd$ & $\vdc$ & $v_{\text{num}}$ \\
       \hline
    $0.3$ & $1.64$ & $1.63$ & $1.62$ & $1.61$ \\
     \hline
    $0.7$ & $1.21$ & $ 1.21$ & $1.20$ & $1.19$ \\
    \hline
    \end{tabular}
    \caption{Comparison of the speed values for \lk{a drive fitness cost} $s=0.3$ or $0.7$ and \lk{a carrying capacity} $K=10^8$; see Table \ref{tab:parameters_stoch} for the other parameter values. The continuous speed $\vc$ is given by \eqref{eq:speed_lDf}, the discrete speed  $\vd$ is given by \eqref{eq:speed_dis}, the correction of the discrete speed $\vdc$ is given by \eqref{eq:speed_dis_cor} and the numerical speed $v_{\text{num}}$ is computed numerically.}
    \label{tab:speeds} 
\end{table}

\clearpage

\section{Exclusion of the particular case of coexistence}\label{app:exlu_coex}

By definition, the coexistence case results in a final drive proportion strictly between $0$ and $1$ at the back of the wave in the continuous model \cite{klay2023}. The question raised in this article ``Is the last individual carrying a drive allele at the back of the wave surrounded by a sufficiently large number of drive alleles?'' — would then receive a negative answer and we expect that such coexistence state would very likely lead to recolonisation events in stochastic simulations, due to a relatively large number of wild-type alleles in small population areas. Therefore, we add a third condition to exclude coexistence cases: 

\begin{equation}
    (1-sh)(1-c) < (1-s). \label{eq:conditionfreq1_ann}
\end{equation}

We insist on the fact that coexistence is only a relative equilibrium among frequencies, because this can occur while the whole population goes extinct (under condition \eqref{eq:extinction}). We then observe a transition in the analytical computations and find $\lWb = \lDb$, meaning that the drive and wild-type alleles decay with the same exponential rate at the back of the wave, and that they both reach non-zero equilibrium frequencies at $-\infty$ in space. On the contrary, condition \eqref{eq:conditionfreq1_ann} denoted as $s<s_{1}$ in \cite{klay2023}, is equivalent to $\lWb > \lDb$ ensuring that the drive curve is less steep than the wild-type curve as in Figure \ref{fig:zoom}. Under condition \eqref{eq:conditionfreq1_ann}, the final drive proportion reaches $1$ (and the final wild-type proportion reaches $0$) \cite{klay2023}.

\clearpage

\section{Influence of the migration on the extinction time} \label{app:migration}

Extinction time $T^{100}_{0 \ \text{gw}}$ is supposed to approximate extinction time $T^{100}_{0}$ in a single isolated population modelled with a spatial Galton-Watson process. To be consistent in the comparison, we set the initial single population size with an exponential profile $\exp(z \ \lWbd)$ approximating the wild-type allele number at the back of the wave in the global simulation (Figure \ref{fig:zoom}). $T^{100}_{0}$ measures the time that the last spatial site with more than $100$ wild-type individuals at the back of the wave takes to go extinct in the global simulation; multiplied by the speed of the wave $v_{\text{num}}$, it gave us a good approximation of distance $L^{100}_{1}$ (Section \ref{subsubsec:dist_time}). $T^{100}_{0 \ \text{gw}}$ measures the time that the last spatial site with more than $100$ individuals in the initial exponential decay takes to go extinct in the spatial Galton-Watson simulation.

\begin{figure}[H]
\centering
\begin{subfigure}{\textwidth}
\begin{subfigure}{0.48\textwidth}
    \renewcommand\thesubfigure{(\alph{subfigure}1)}
    \caption{$s=0.3$} 
    \centering
    \includegraphics[width = 0.9\textwidth]{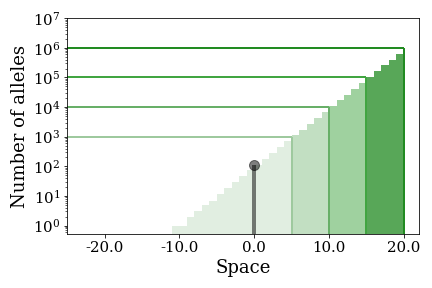}  
\end{subfigure}
\hfill
\begin{subfigure}{0.48\textwidth}
    \addtocounter{subfigure}{-1}
    \renewcommand\thesubfigure{(\alph{subfigure}2)}
    \caption{$s=0.7$} 
    \centering
    \includegraphics[width = 0.9\textwidth]{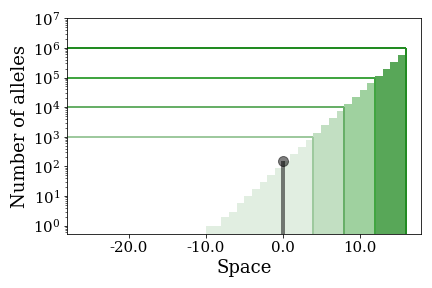}  
\end{subfigure}
\addtocounter{subfigure}{-1}
\caption{Initial conditions for the spatial Galton-Watson process, in shades of blue. The exponential profile used to initiate the simulation is cropped at different population size on the right, from $10^3$ to $10^8$, to evaluate the impact of migration from dense to less dense areas. The black line with a circle at the top represents the last spatial site with more than $100$ individuals in the initial condition. We record the extinction time $T^{100}_{0 \ \text{gw}}$ of this site, and run the simulation 500 times to obtain its statistical distribution. This distribution (multiplied by the speed $v_{\text{num}}$ to approximate distance $L^{100}_{1}$) is illustrated in Figure \ref{fig:gw_hist}.}
\label{fig:gw_ini}
\end{subfigure}
\centering
\begin{subfigure}{\textwidth}
\begin{subfigure}{0.48\textwidth}
    \renewcommand\thesubfigure{(\alph{subfigure}1)}
    \caption{$s=0.3$} 
    \centering
    \includegraphics[width = 0.9\textwidth]{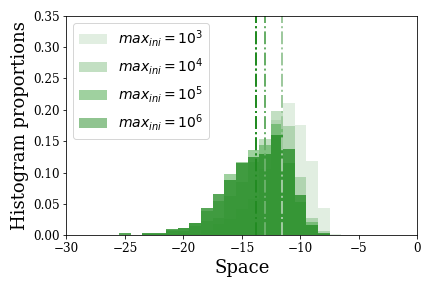}  
\end{subfigure}
\hfill
\begin{subfigure}{0.48\textwidth}
    \addtocounter{subfigure}{-1}
    \renewcommand\thesubfigure{(\alph{subfigure}2)}
    \caption{$s=0.7$} 
    \centering
    \includegraphics[width = 0.9\textwidth]{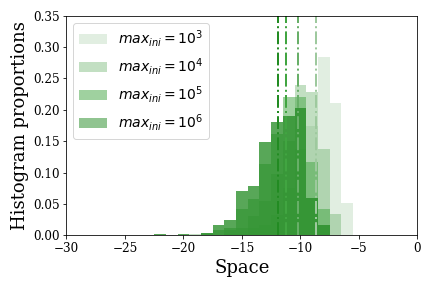}   
\end{subfigure}
\addtocounter{subfigure}{-1}
\caption{Statistical distribution of extinction time $T^{100}_{0 \ \text{gw}}$, multiplied by the speed $v_{\text{num}}$ (to approximate distance $L^{100}_{1}$). We consider the initial conditions described in Figure \ref{fig:gw_ini} in shades of blue; each blue corresponds to a different space window in which the maximum number of individuals per site is $\max_{ini}$. We record the time $T^{100}_{0 \ \text{gw}}$ at which the last spatial site with more than $100$ individuals in the initial condition goes extinct (definitively). The dashed lines are the means values of each histogram.}
\label{fig:gw_hist}
\end{subfigure}
\caption{Initial conditions and extinction times for the isolated population.}
\end{figure}

The initial condition being largely heterogeneous (because of the exponential decay), we wonder how migration from dense areas to less dense areas impact the distribution of $T^{100}_{0 \ \text{gw}}$. More precisely, we focus on how large the exponential initial condition might be for $T^{100}_{0 \ \text{gw}}$ to be a good approximation of $T^{100}_{0}$. In Figure \ref{fig:gw_ini}, we crop the initial condition on the right so that the maximum number of individuals in one spatial site ($\max_{ini}$) gradually increases from $10^3$ to $10^8$. In Figure \ref{fig:gw_hist}, we observe that the mean of the distribution $T^{100}_{0 \ \text{gw}}$, multiplied by the speed $v_{\text{num}}$ to obtain a distance, consequently increases from 2 to 3 time units. Thus, it is very important to consider a large exponential initial condition to fully capture the wild-type dynamics at the end of the wave. 

\section{Comparison with the 2D case, small carrying capacity $K=10^3$}\label{ann:2D_smallK}

We run the same simulations than in Figure \ref{fig:1D-2D} except for the carrying capacity that is taken smaller $K=10^3$. For small drive fitness cost ($s=0.3$), the outcomes stays unchanged: we observe no wild-type recolonisation event and the drive propagation leads to the full eradication of the population both in 1D and 2D. For large drive fitness cost ($s=0.7$), we observe several wild-type recolonisation and drive reinvasion events, due to the small carrying capacity value (in comparison, we observe less stochastic events for $K=10^5$ in Figure \ref{fig:1D-2D_K5s7}). The stochastic events appear to be more frequent in 2D than in 1D, similarly to what we observed for $K=10^5$ and $K=10^8$.

\begin{figure}[H]
\begin{subfigure}[t]{0.99\textwidth}
\caption{$K=10^3, s=0.3$} \label{fig:1D-2D_K3s3}
\centering
\includegraphics[width=0.18\textwidth]{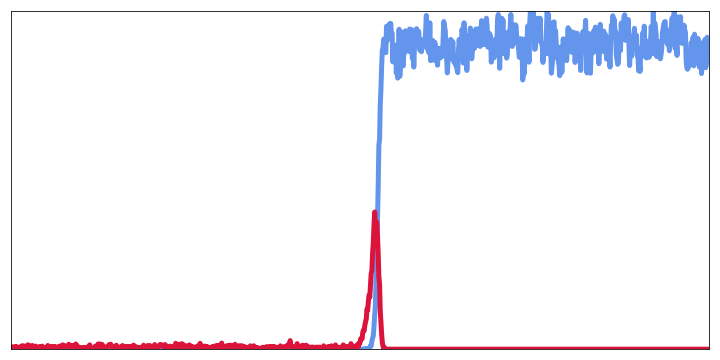} % 1D  K=10^3 et s=0.3
\includegraphics[width=0.18\textwidth]{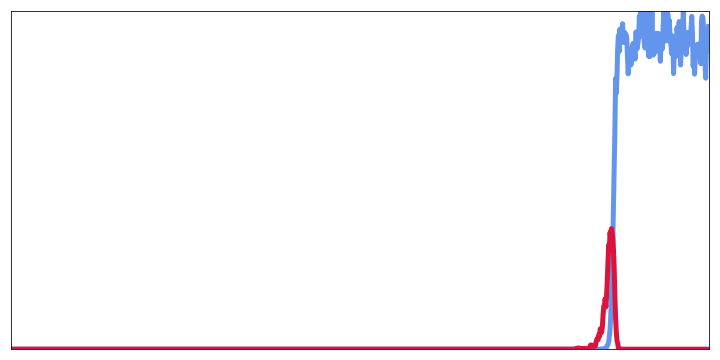}
\includegraphics[width=0.18\textwidth]{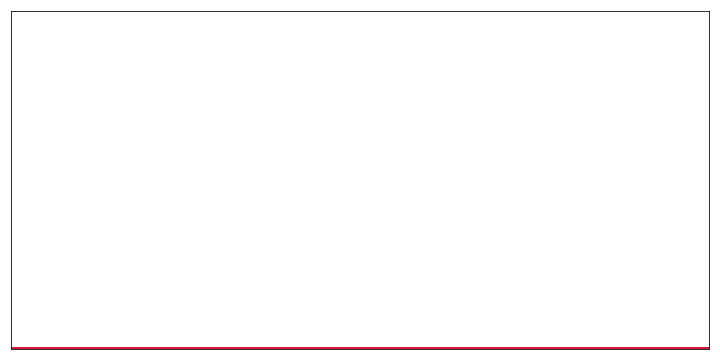}
\includegraphics[width=0.18\textwidth]{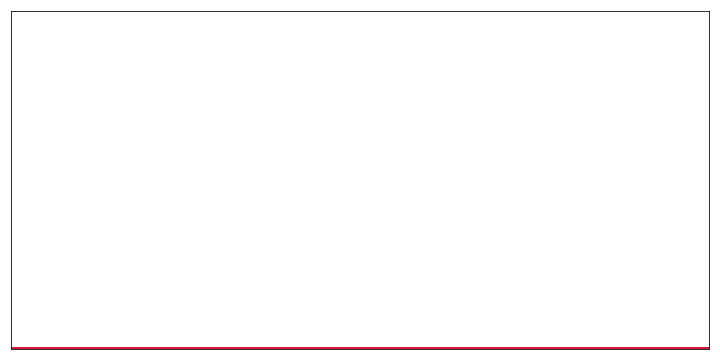}
\includegraphics[width=0.18\textwidth]{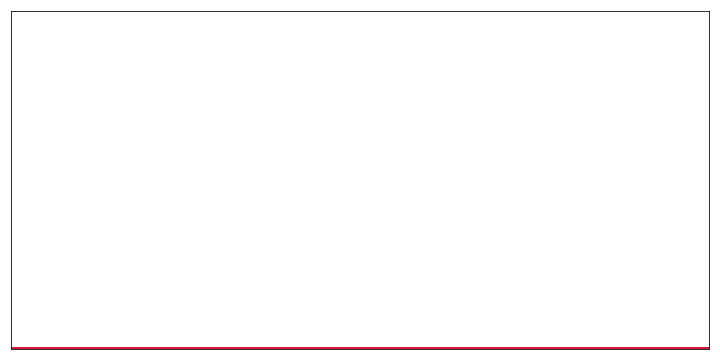}\\
\includegraphics[width=0.18\textwidth]{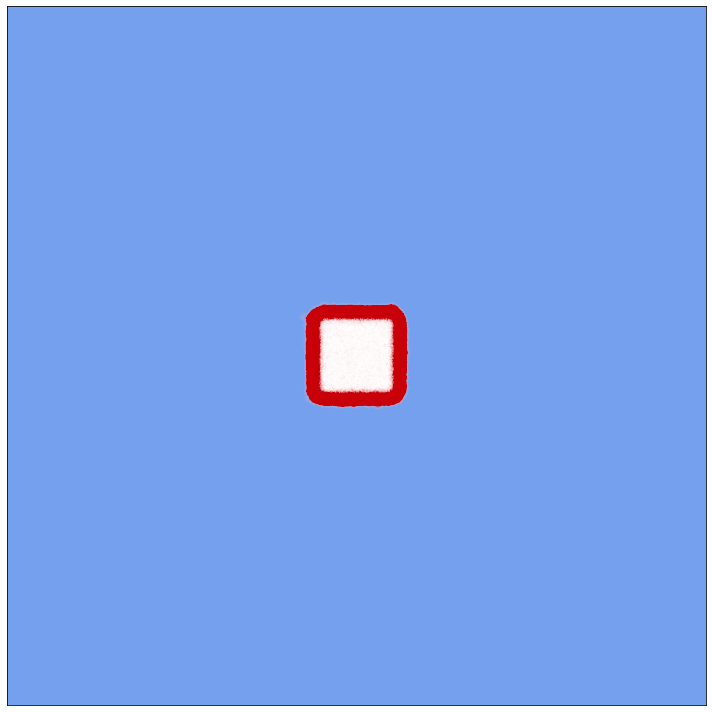} % 2D  K=10^3 et s=0.3
\includegraphics[width=0.18\textwidth]{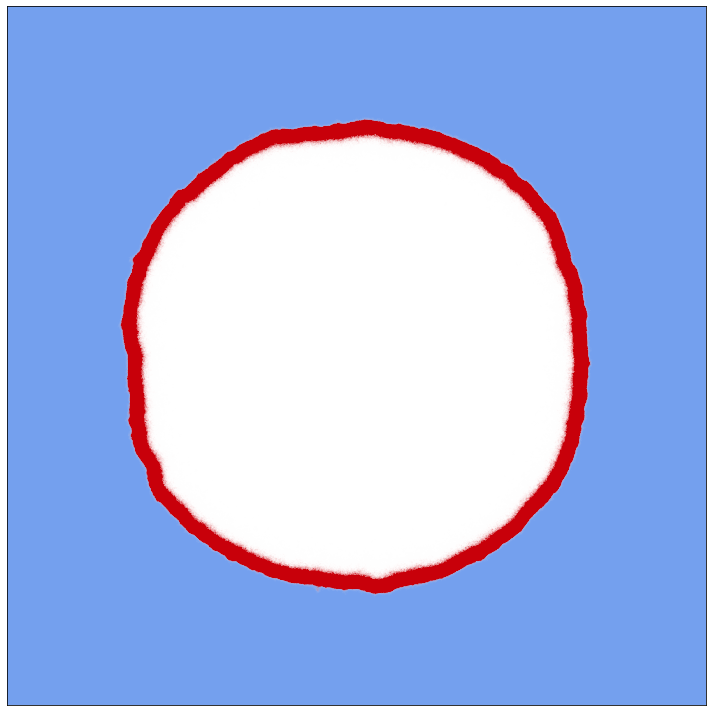}
\includegraphics[width=0.18\textwidth]{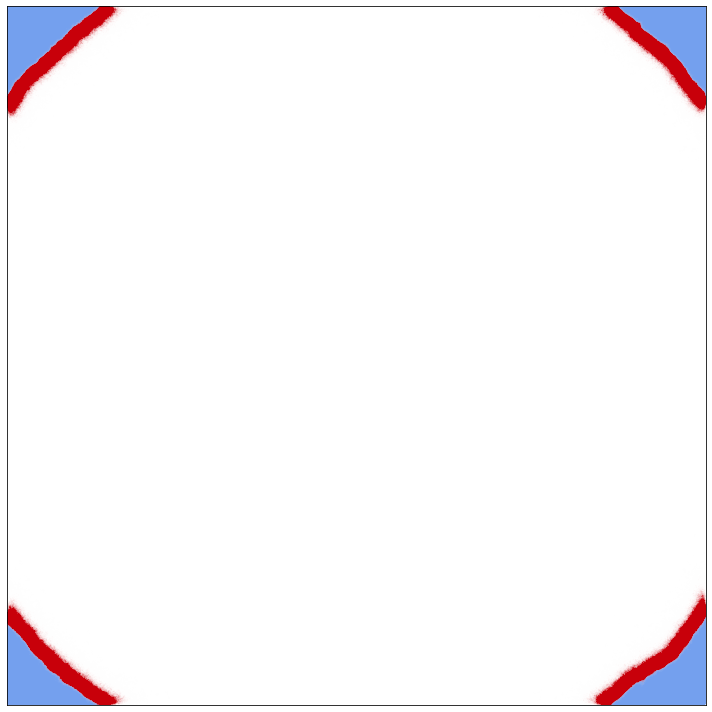}
\includegraphics[width=0.18\textwidth]{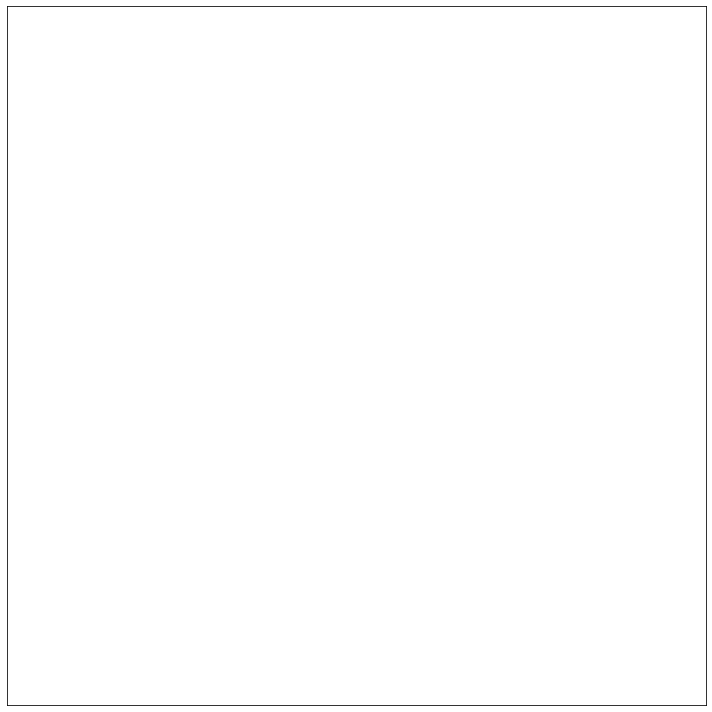}
\includegraphics[width=0.18\textwidth]{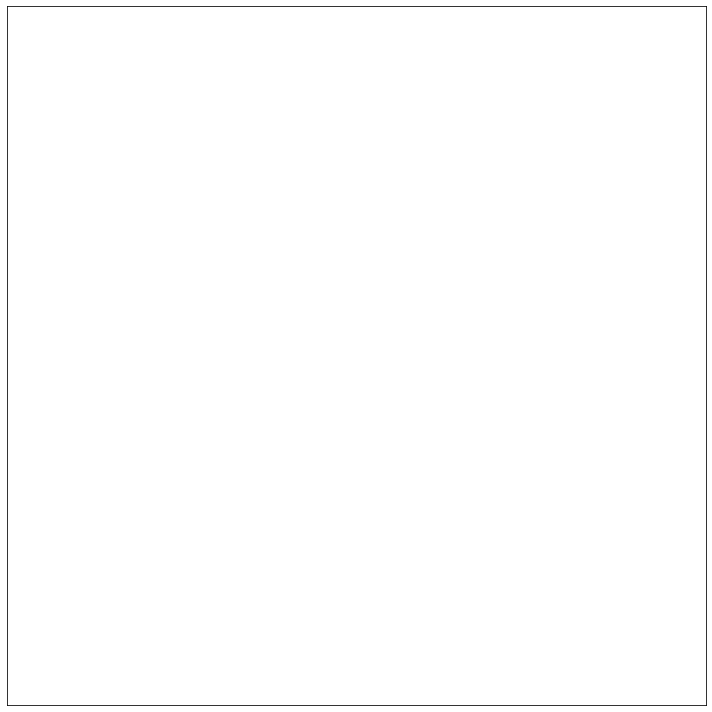}\\
\end{subfigure}
\begin{subfigure}[t]{0.99\textwidth}
\centering
\caption{$K=10^3, s= 0.7$} \label{fig:1D-2D_K3s7}
\includegraphics[width=0.18\textwidth]{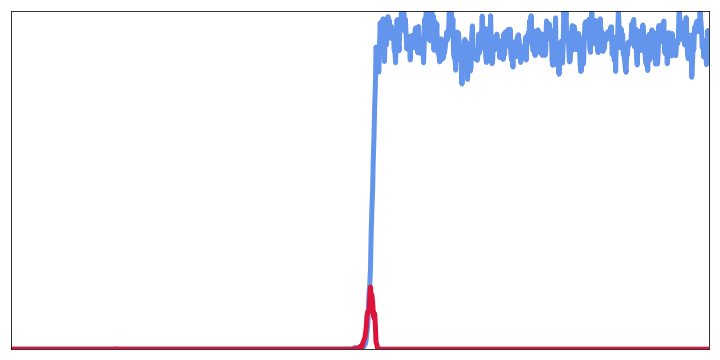} % 1D  K=10^3 et s=0.7
\includegraphics[width=0.18\textwidth]{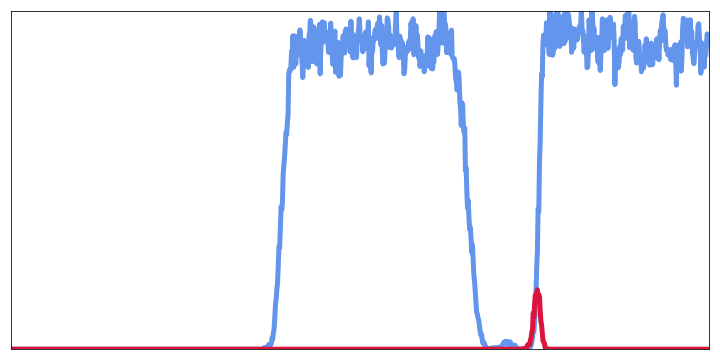}
\includegraphics[width=0.18\textwidth]{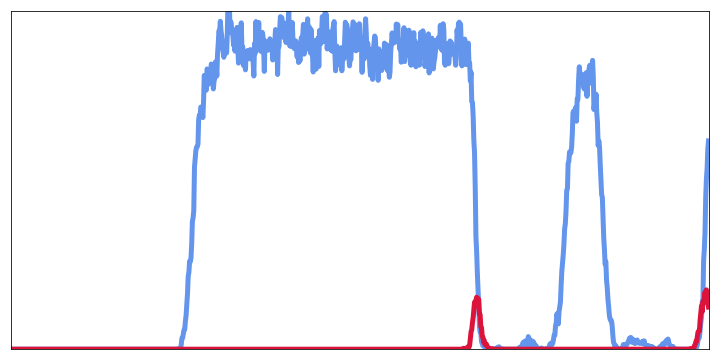}
\includegraphics[width=0.18\textwidth]{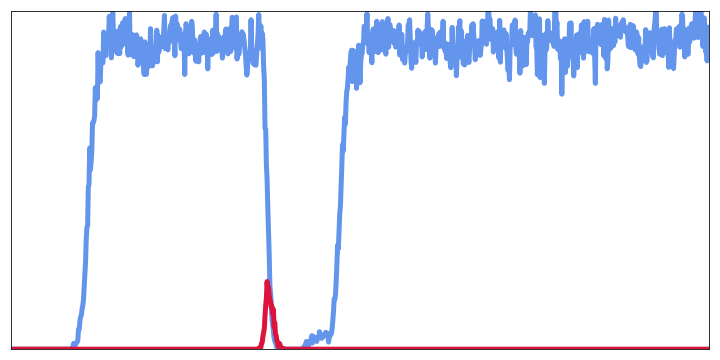}
\includegraphics[width=0.18\textwidth]{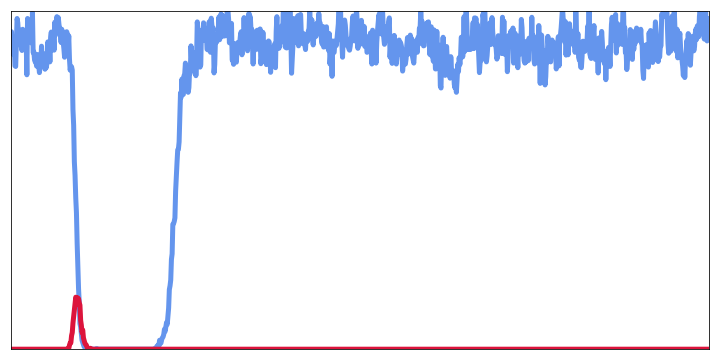}\\
\includegraphics[width=0.18\textwidth]{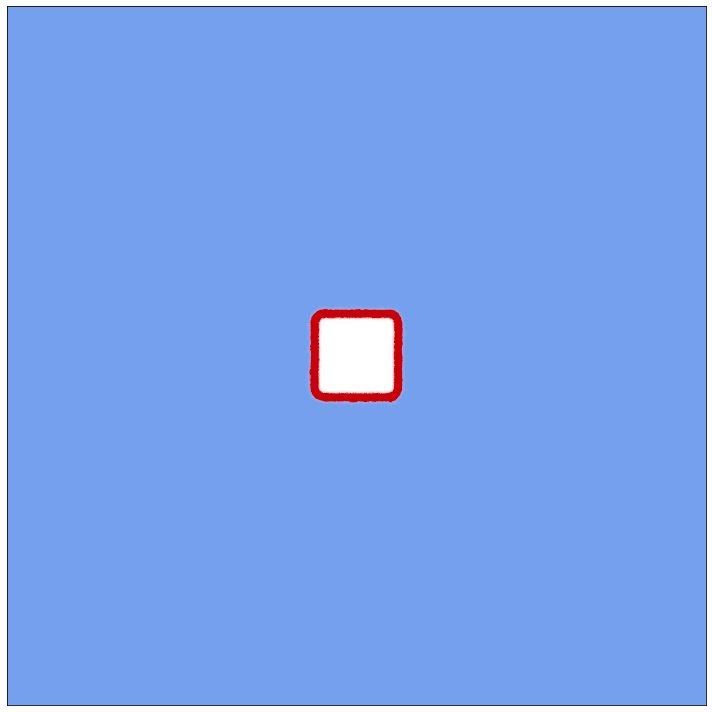} % 2D  K=10^3 et s=0.7
\includegraphics[width=0.18\textwidth]{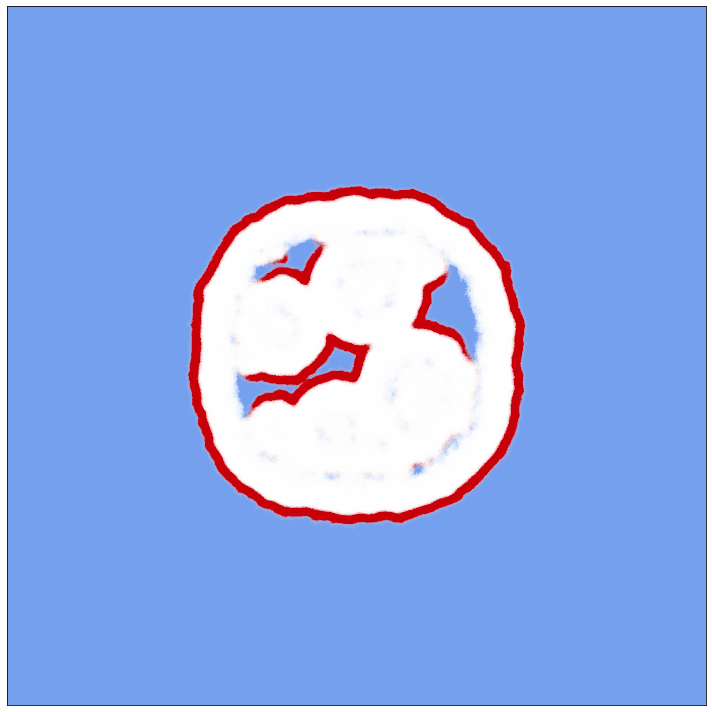}
\includegraphics[width=0.18\textwidth]{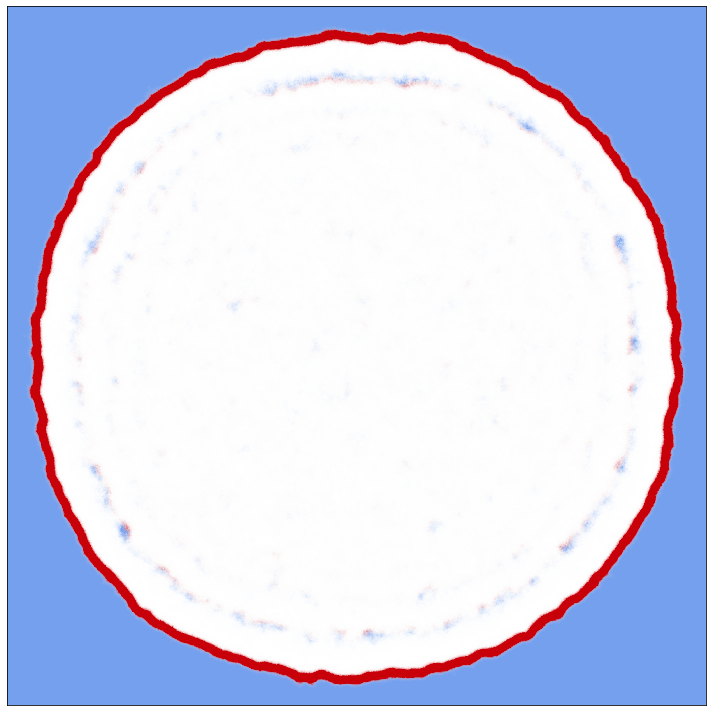}
\includegraphics[width=0.18\textwidth]{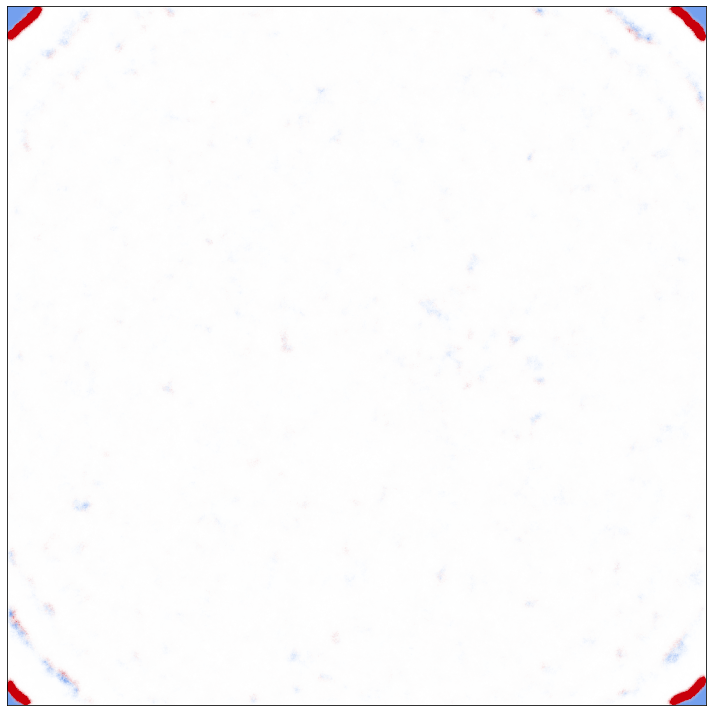}
\includegraphics[width=0.18\textwidth]{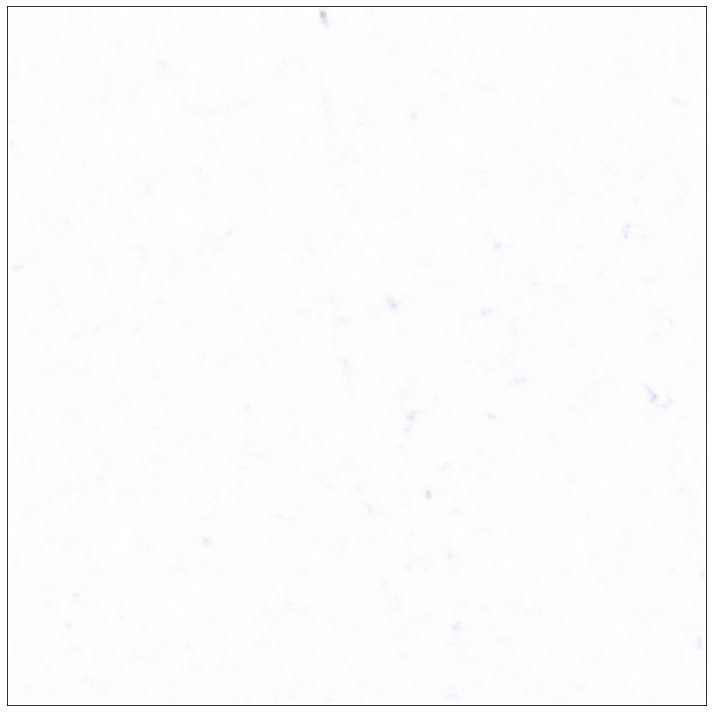}\\
\end{subfigure}
\caption{Comparison between the 1D and the 2D stochastic models for a small carrying capacity ($K=10^3$). Wild-type alleles are shown in blue and drive alleles in red. In 1D, the y-axis represents the number of alleles, while in 2D, the number of alleles is shown in shades of colour: the darker the colour, the more alleles of the corresponding type on the spatial site. For \lk{a small drive fitness cost} $s=0.3$, we observe no wild-type recolonisation events and the full eradication of the population in 1D and 2D, similarly to the numerical outcomes obtained with larger carrying capacity (Figure \ref{fig:1D-2D}(a, b)). Note that the drive presence on the first half of the domain, in the earliest 1D snapshot of panel (a), is a simulation artefact due to the initial condition (Figure \ref{fig:CI_stoch}). For a large drive fitness cost $s=0.7$, reducing the carrying capacity results in a higher number of wild-type recolonisation and drive reinvasion events (in comparison with Figure \ref{fig:1D-2D_K5s7}). Once again here, these stochastic events appear to be more frequent in 2D than in 1D.}\label{fig:1D-2D-K-2}
\end{figure}

\newpage

\section{Exploring chasing dynamics}  \label{ann:chasing}

\begin{figure}[H]
\begin{subfigure}[t]{0.99\textwidth}
\caption{$r=0.05$}
\centering
\includegraphics[width=0.18\textwidth]{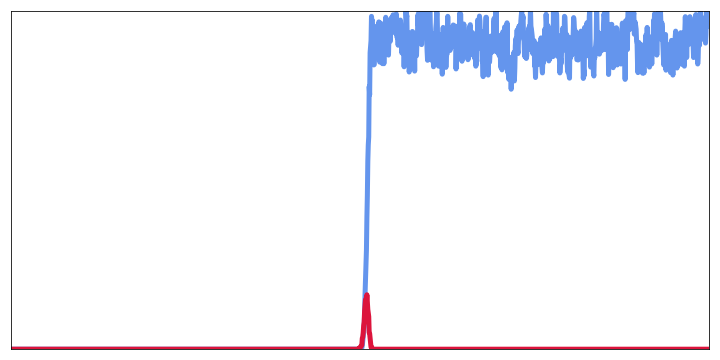} 
\includegraphics[width=0.18\textwidth]{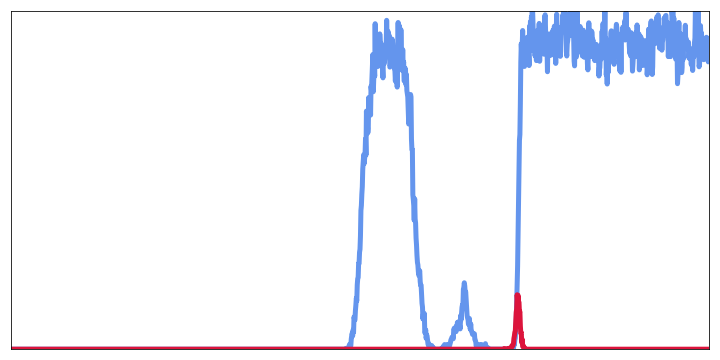} 
\includegraphics[width=0.18\textwidth]{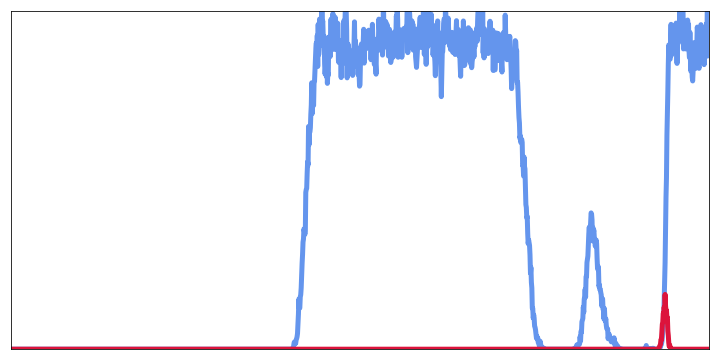} 
\includegraphics[width=0.18\textwidth]{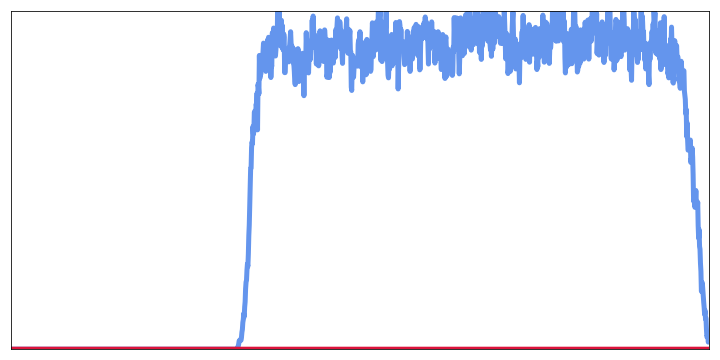} 
\includegraphics[width=0.18\textwidth]{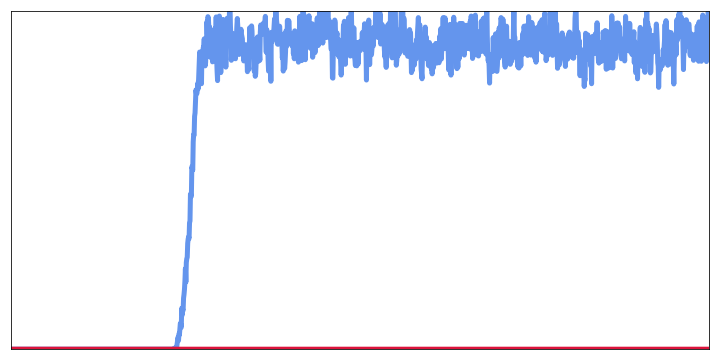} \\
\end{subfigure}
\begin{subfigure}[t]{0.99\textwidth}
\caption{$r=0.1$}
\centering
\includegraphics[width=0.18\textwidth]{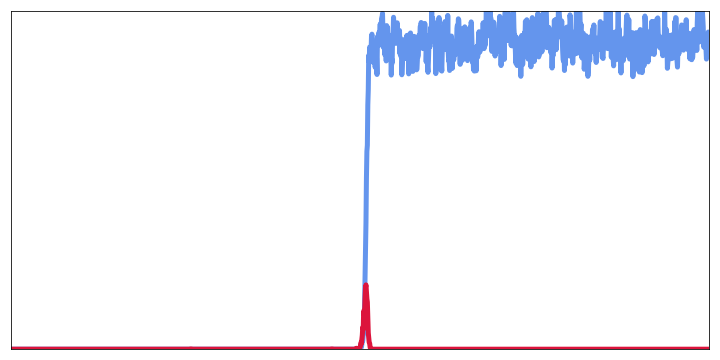} 
\includegraphics[width=0.18\textwidth]{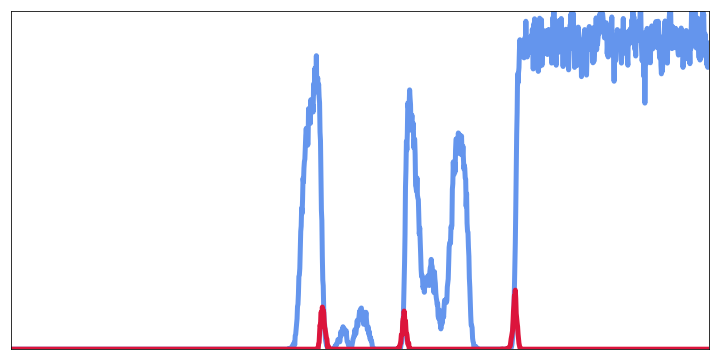} 
\includegraphics[width=0.18\textwidth]{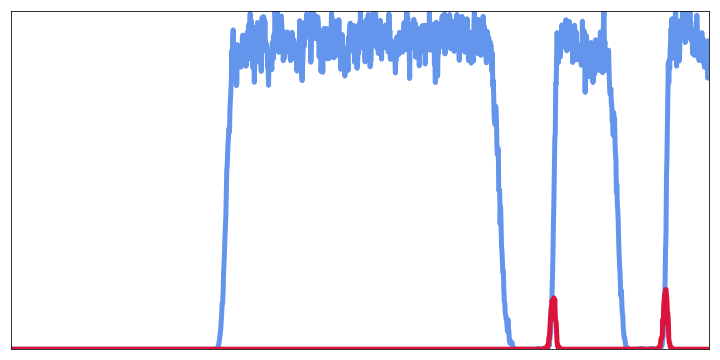} 
\includegraphics[width=0.18\textwidth]{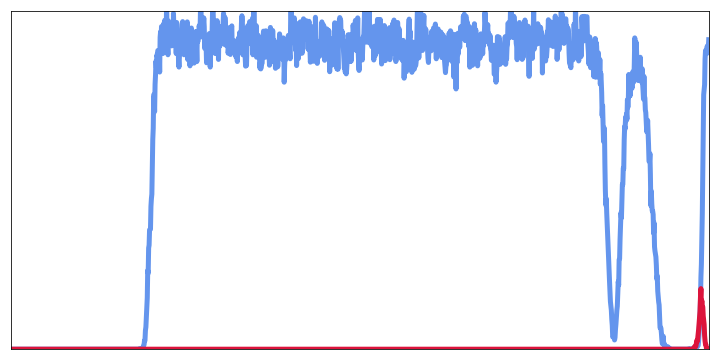} 
\includegraphics[width=0.18\textwidth]{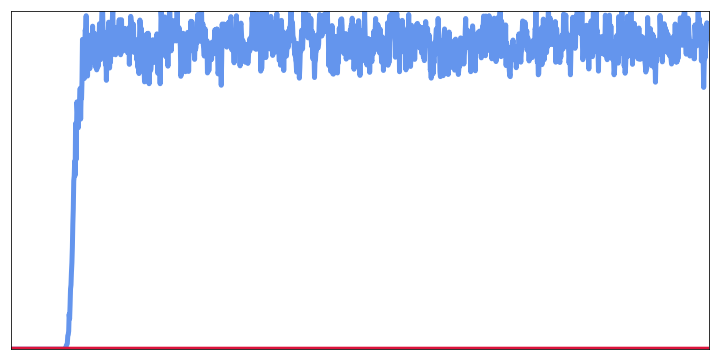} \\
\end{subfigure}
\begin{subfigure}[t]{0.99\textwidth}
\caption{$r=0.2$}
\centering
\includegraphics[width=0.18\textwidth]{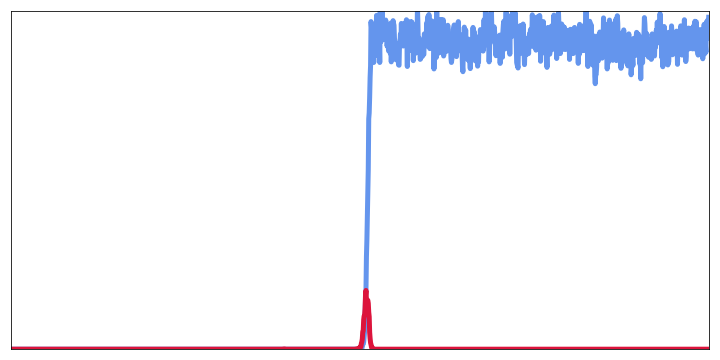} 
\includegraphics[width=0.18\textwidth]{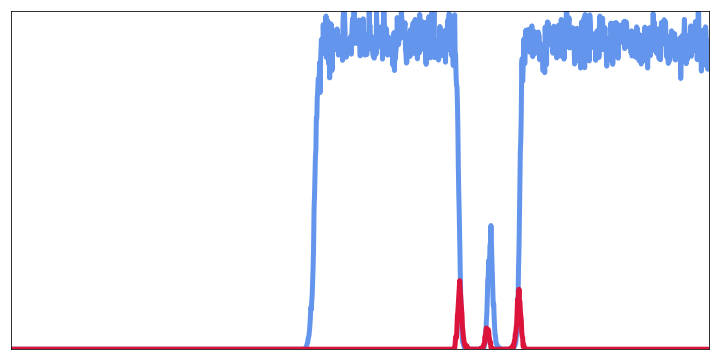} 
\includegraphics[width=0.18\textwidth]{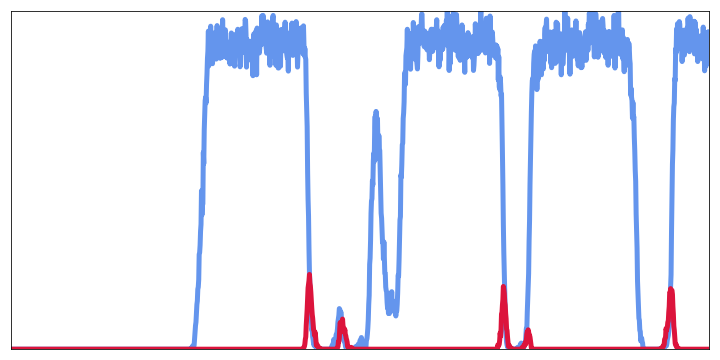} 
\includegraphics[width=0.18\textwidth]{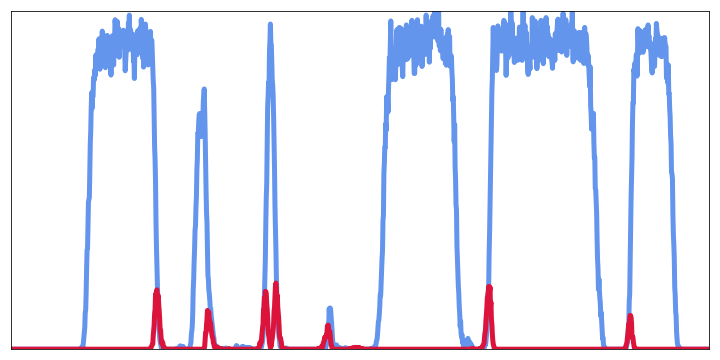} 
\includegraphics[width=0.18\textwidth]{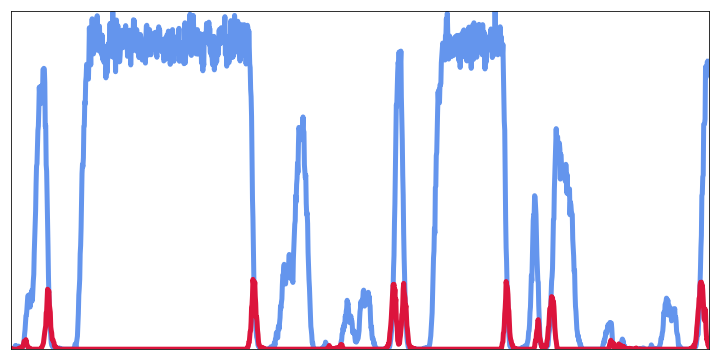} \\
\end{subfigure}
\caption{Simulations in 1D for $K=10^3$ and $s=0.7$. As the value of $r$ increases, drive reinvasion events are more and more frequent.} \label{fig:r_1D}
\end{figure}

\begin{figure}[H]
\begin{subfigure}[t]{0.99\textwidth}
\caption{$r=0.02$}
\centering
\includegraphics[width=0.18\textwidth]{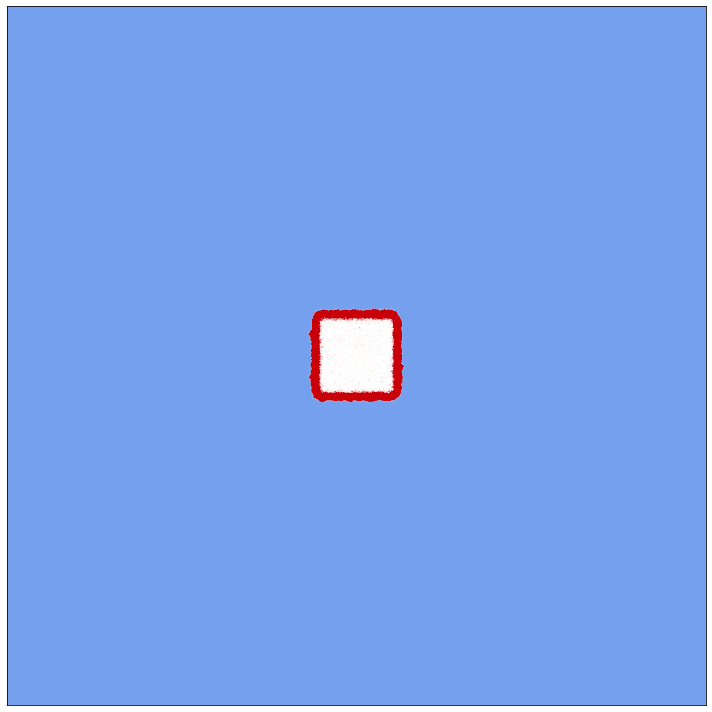} 
\includegraphics[width=0.18\textwidth]{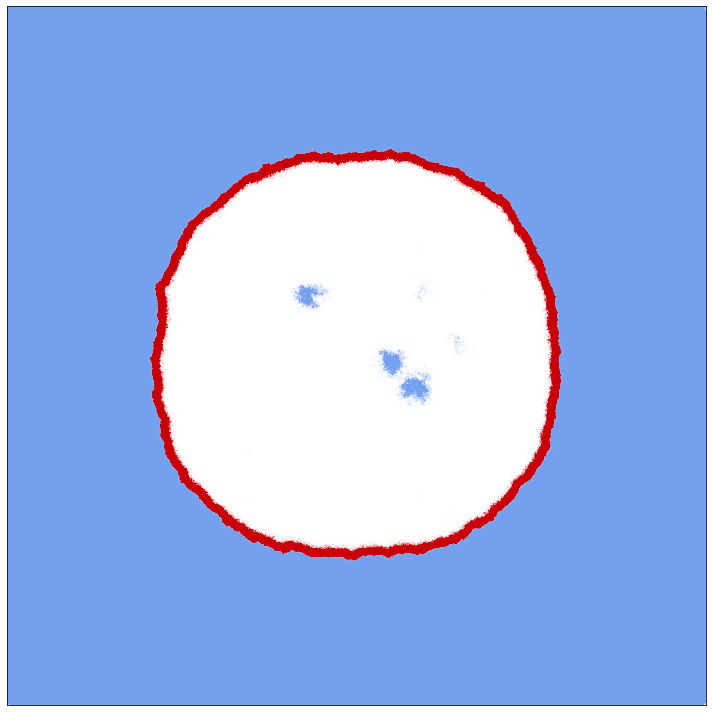} 
\includegraphics[width=0.18\textwidth]{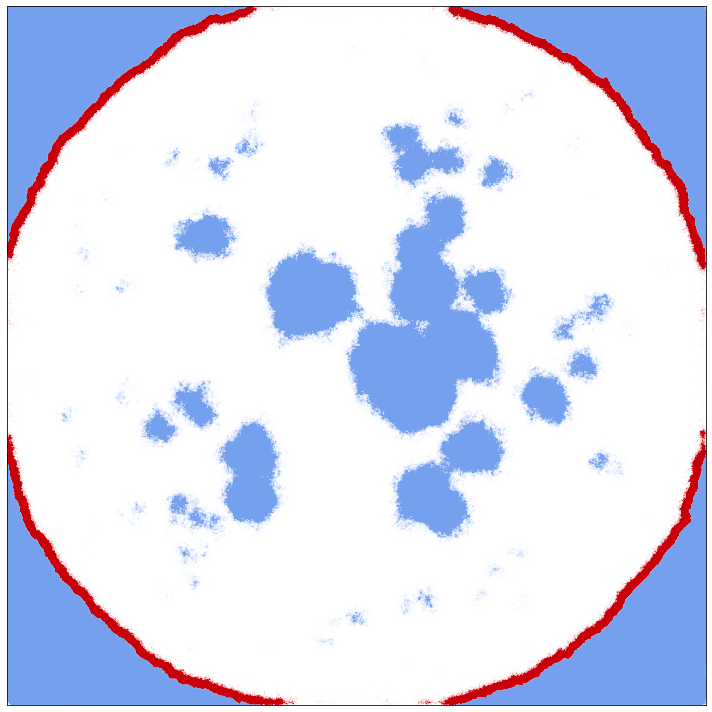} 
\includegraphics[width=0.18\textwidth]{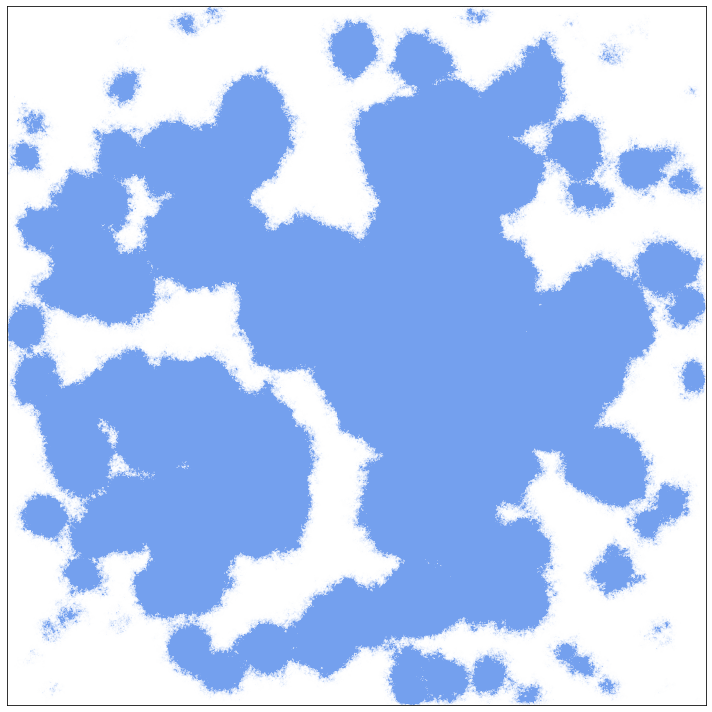} 
\includegraphics[width=0.18\textwidth]{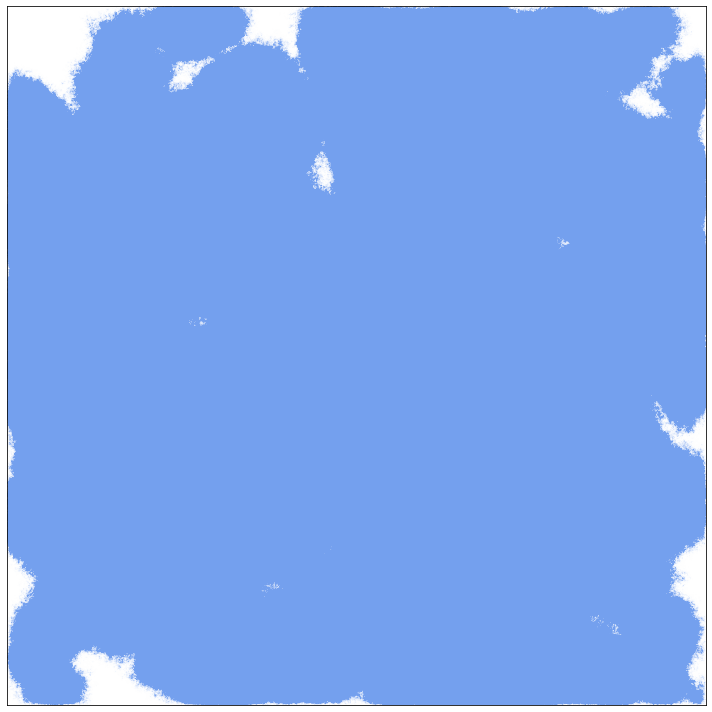} \\
\end{subfigure}
\begin{subfigure}[t]{0.99\textwidth}
\caption{$r=0.05$}
\centering
\includegraphics[width=0.18\textwidth]{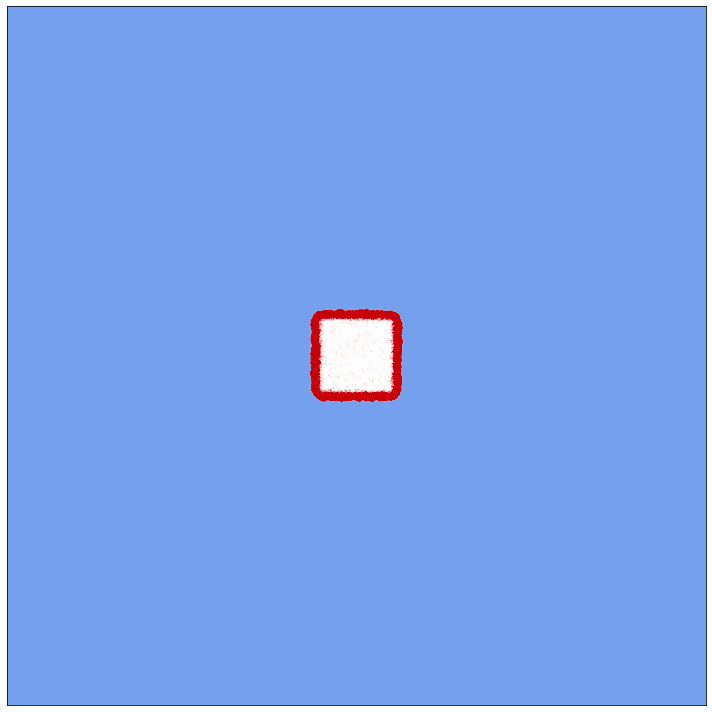} 
\includegraphics[width=0.18\textwidth]{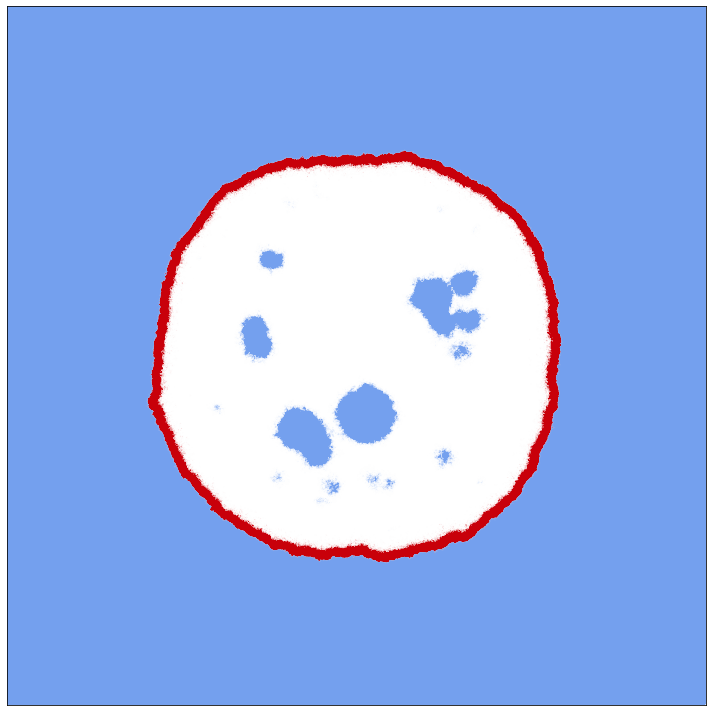} 
\includegraphics[width=0.18\textwidth]{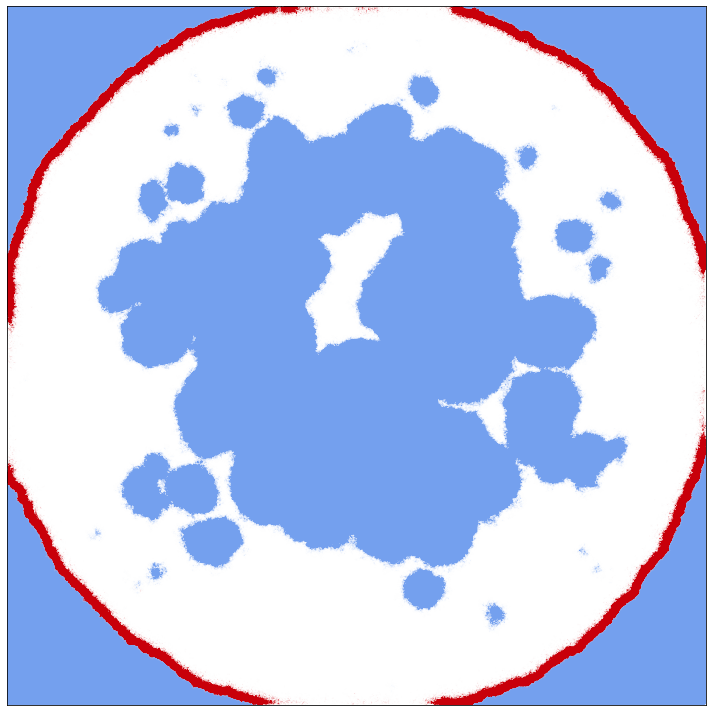} 
\includegraphics[width=0.18\textwidth]{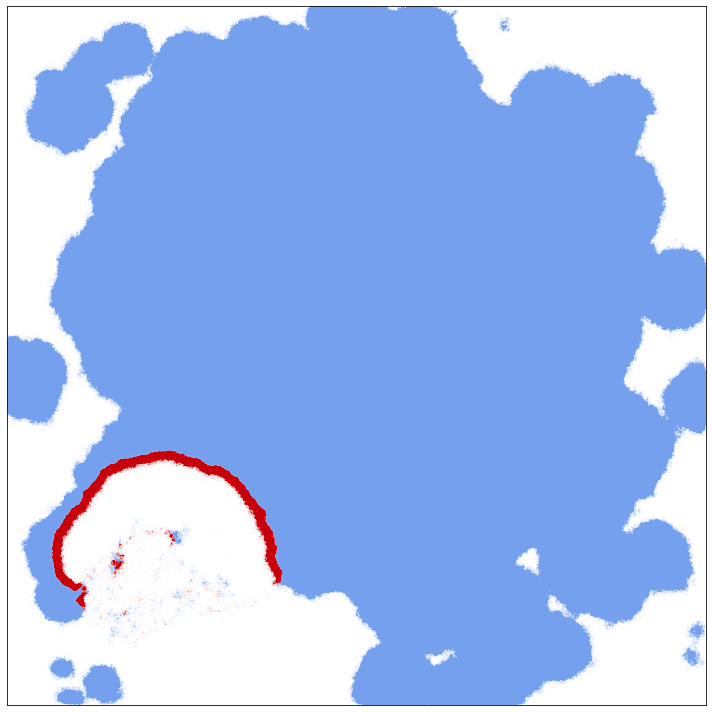} 
\includegraphics[width=0.18\textwidth]{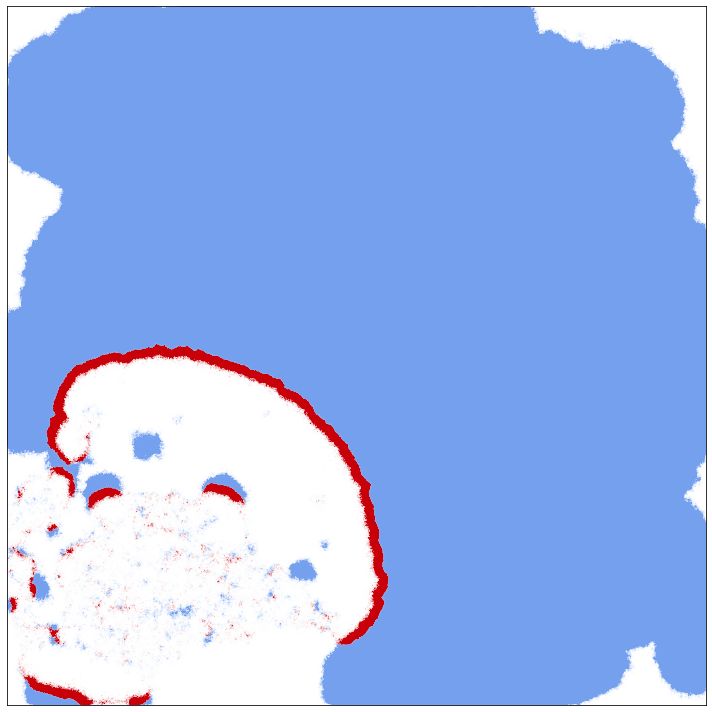} \\
\end{subfigure}
\begin{subfigure}[t]{0.99\textwidth}
\caption{$r=0.1$}
\centering
\includegraphics[width=0.18\textwidth]{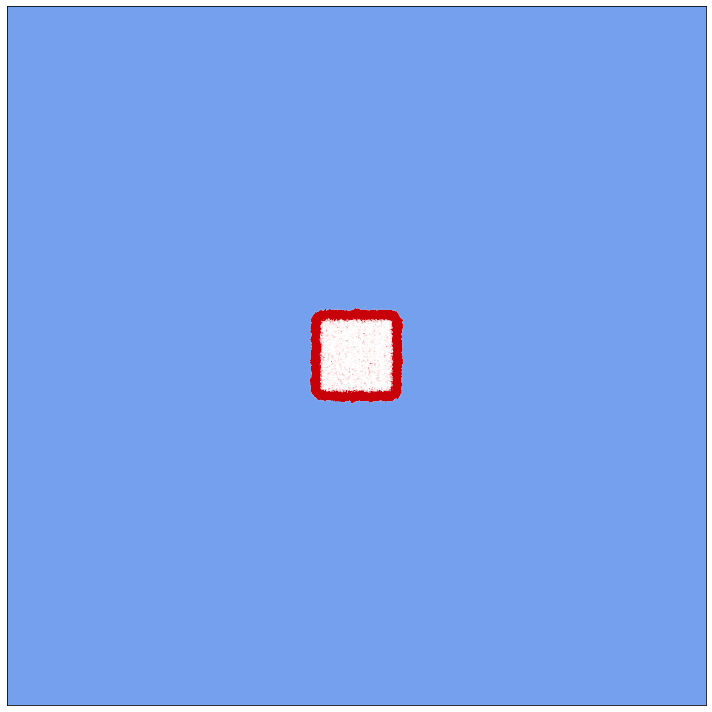} 
\includegraphics[width=0.18\textwidth]{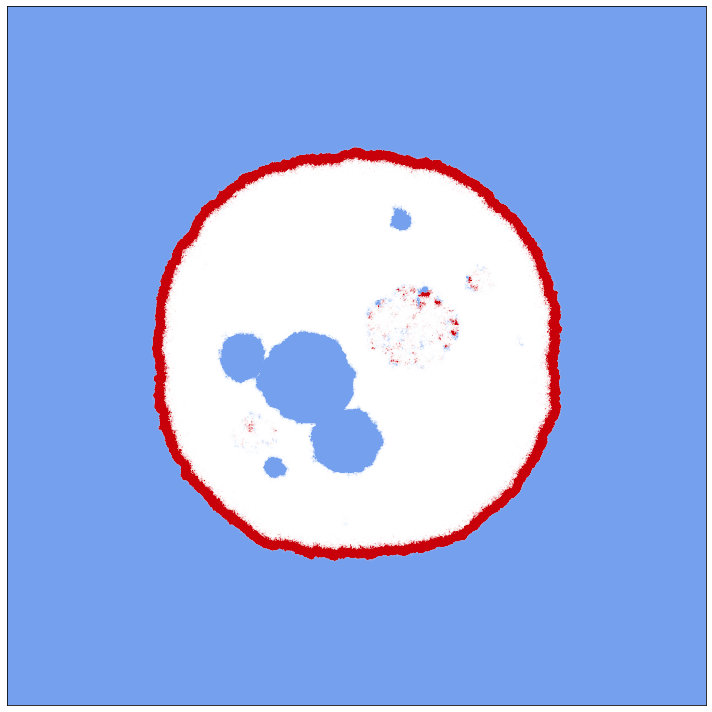} 
\includegraphics[width=0.18\textwidth]{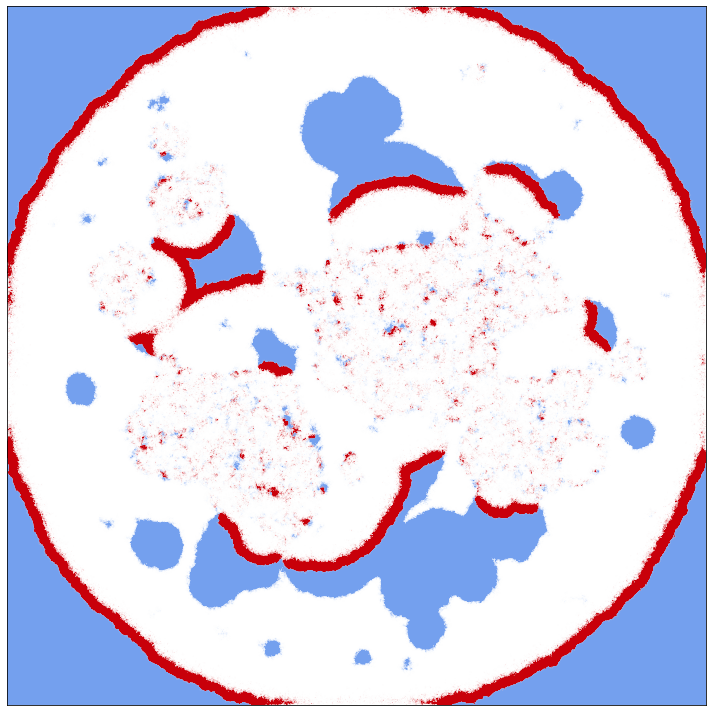} 
\includegraphics[width=0.18\textwidth]{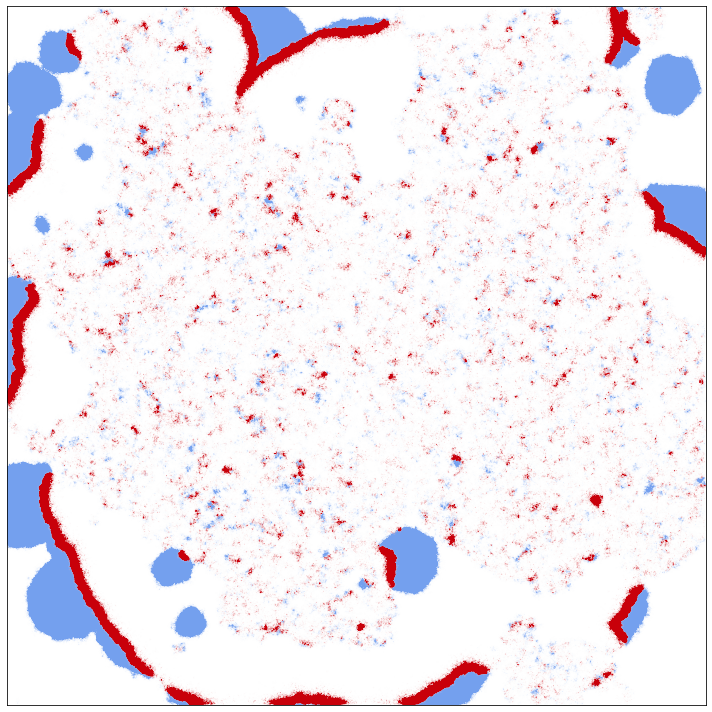} 
\includegraphics[width=0.18\textwidth]{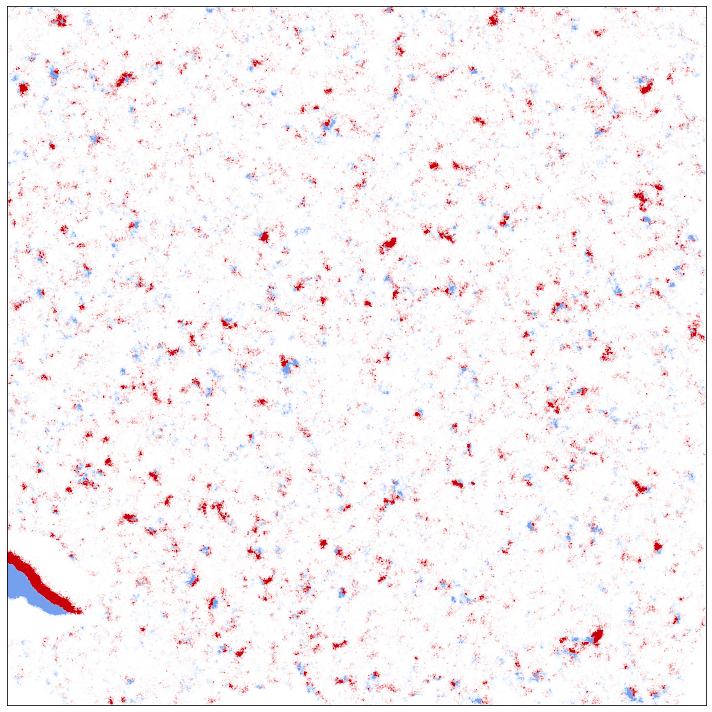} \\
\end{subfigure}
\caption{Simulations in 2D for $K=10^2$ and $s=0.3$. As the value of $r$ increases, drive reinvasion events are more and more frequent.} \label{fig:r_2D}
\end{figure}

We test various values of intrinsic growth rate $r$ in order to appreciate its influence on the probability of drive reinvasion events after a wild-type recolonisation. Both the simulations in 1D (Figure \ref{fig:r_1D}) and the simulations in 2D (Figure \ref{fig:r_2D}) illustrate the concept described in Section \ref{sec:WT_VS_chasing}: as $r$ increases, the wild-type recolonisation wave and the drive eradication wave get closer, increasing the probability to observe drive reinvasion events. In the following example, both the wild-type recolonisation probability ($p_A$) and the drive reinvasion probability ($p_B$) are high and relatively close for high values of $r$, which leads to infinite chasing dynamics (Figure \ref{fig:r_1D}(c) and \ref{fig:r_2D}(c)).

\clearpage

\section*{Acknowledgements}
This work is funded by ANR-19-CE45-0009-01 TheoGeneDrive. This project has received funding from the European Research Council (ERC) under the European Union’s Horizon 2020 research and innovation program (grant agreement No 865711).

\bibliographystyle{elsarticle-num}
\bibliography{biblio.bib}

%\printbibliography

\end{document}